\documentclass[10pt,aps,prl,showpacs,twocolumn,superscriptaddress,floats,floatfix]{revtex4-1}
\usepackage{amssymb,amsmath}
\usepackage{graphics}
\usepackage{color}
\usepackage{epstopdf}
\usepackage{epsfig}
\usepackage{rotating}
\usepackage{float}
\usepackage{bm}
\usepackage{soul}
\usepackage{hyperref}
\usepackage{breakurl}
\usepackage{natbib}
\usepackage[stable,perpage]{footmisc}
\usepackage{xr}
\usepackage{verbatim}
\newcommand{\rem}[1]{}

\begin{document}
\title{Two-point spectral model for variable-density homogeneous turbulence}
\author{Nairita Pal\footnote{nairitap2009@gmail.com}}
\affiliation{Applied Mathematics and Plasma Physics (T-5) and Center
for Nonlinear Studies, Theoretical Division, Los Alamos National Laboratory, NM 87545, USA}
\author{Susan Kurien\footnote{skurien@lanl.gov}}
\affiliation{Applied Mathematics and Plasma Physics (T-5), Theoretical Division, Los Alamos National Laboratory, NM 87545, USA}
\author{Timothy Clark\footnote{ttc@unm.edu}}
\affiliation{Department of Mechanical Engineering, University of New Mexico, Albuquerque NM USA}
\author{Denis Aslangil\footnote{denis.aslangil@lanl.gov}}
\affiliation{Department of Mechanical Engineering and Mechanics, Lehigh University, Bethlehem, Pennsylvania 18015, USA}
\author{Daniel Livescu\footnote{livescu@lanl.gov}}
\affiliation{Los Alamos National Laboratory, Los Alamos, NM 87544, USA}

\date{\today}

\begin{abstract}
We present a study of buoyancy-driven variable-density homogeneous turbulence, using a two-point spectral closure model. We compute the time-evolution of the spectral distribution in wavenumber $k$ of the correlation of density and specific-volume $b(k)$, the mass flux $\bm{a}(k)$, and the turbulent kinetic energy $E(k)$, using a set of coupled equations. Under the modeling assumptions, each dynamical variable has two coefficients governing spectral transfer among modes. In addition, the mass flux $\bm{a}(k)$ has two coefficients governing the drag between the two fluids. 
 Using a prescribed initial condition for $b(k)$ and starting from a quiescent flow, we first evaluate the relative importance of the different coefficients used to model this system, and their impact on the statistical quantities. We next assess the accuracy of the model, relative to Direct Numerical simulation of the complete hydrodynamical equations, using $b$, ${\bm a}$ and $E$ as metrics. We show that the model is able to capture the spectral distribution and global means of all three statistical quantities at both low and high Atwood number for a set of optimized coefficients. The optimization procedure also permits us to discern a minimal set of four coefficients which 
are sufficient to yield reasonable results while pointing to the mechanisms that dominate the mixing process in this problem.

\end{abstract}

\maketitle
\section{Introduction}

The mixing of fluids with different densities is an important process in many practical applications such as oceanic or atmospheric flows, combustion and inertial confinement fusion (ICF).
Variable-density flows are those in which fluctuations of the density from its mean value are large.
To predict the effects of such large density fluctuations on the mean flow in complex systems, we require efficient computational models that are accurate, and also economical to run~\cite{townsend1980structure,wilcox,hanjalic1994advanced}.
In the case of constant density turbulence there has been significant progress
in model development \cite{hanjalic1980multiple,pope1994relationship}. There are well-known efforts in the  literature on two-point (spectral) models for constant density turbulence in which energy transfer is described in wave number space using the Eddy Damped Quasi Normal Closure approach by Cambon \cite{cambon1979modelisation, cambon1981spectral, godeferd1994detailed}, and by Bertoglio \cite{bertoglio1987simplified}. Variable-density flows have been studied extensively
in experiments \cite{Gerashchenko2015, Ranjan2016,Charonko2017}, or using Direct Numerical Simulations \cite{cabot2006reynolds,livescu2008}. Analytical models for such flows are mainly limited to single-point closure models~\cite{besnard1985turbulence, banerjee2010development,Schwarzkopf2016,elghobashi1983two}, in which turbulence variables are studied as a function of a single space-point. The latter suffer the drawback of being incapable of capturing transients and scale generation \cite{clark1999two}, a fundamental feature of turbulence. 

In this paper, we work with a two-point spectral closure model for constant-density turbulent flow originally  developed in \cite{besnard96}, and modified in ~\cite{Steinkamp1999a,Steinkamp1999b} for variable density flows. The advantage of a model based on two-point correlations is it's ability to capture the evolution of scales with time. As a result, one does not need to specify an extra equation for the dissipation as is needed in one-point models~\cite{wilcox,schwarzkopf2011application}. Our work bears some resemblance to~\cite{dupuy2018equations}, although the said work is focussed only on the kinetic energy evolution.
In our work, the initial condition is specified by the 
two-point correlation of density fluctuation with specific-volume fluctuation defined as a distribution in wavenumber space. This quantity $b(k)$ in turn drives a mass flux $\bm{a}(k)$, defined as the two-point correlation of the Favre-averaged velocity with the density fluctuation, through a modulation of the pressure gradient. Consequently the mass flux drives the generation of Reynolds stress and conversion of potential energy into turbulent kinetic energy $E(k)$. This coupled system is spatially homogeneous which allows us to isolate the variable density component of the model without the complications that may have been introduced by, say, inhomogeneity as in the classical inhomogeneous Rayleigh-Taylor (RT) problem. It must be noted, that our test problem is indeed the RT instability problem, posed in a manner that homogenizes it in space.

The closure assumptions for the spectral model truncate the hierarchy of equations at the level of the triple-correlations. For the variable-density case, this introduces two coefficients for each dynamical variable, expressing nonlinear spectral transfer. Additionally, a spectral drag governing the breakup of fluid elements in scale is introduced following \cite{youngs1992two}, giving rise to two more coefficients in the evolution of the mass flux $\bm{a}$.

In the first part of our study, for a prescribed artificial initial $b(k)$ following \cite{clark1995two} we show how these coefficients affect the time evolution of the integrated model variables. The spectral transfer coefficients are varied pair-wise for each evolution equation while keeping all other constants fixed to their nominal values specified by benchmark studies in \cite{Steinkamp1999a,clark1995two,Schwarzkopf2016}. We then determine suitable constants by comparison against direct numerical simulations (DNS) of the system described in \cite{livescu2007, livescu2008}, following an optimization procedure. With these optimized constants, we find that the time-evolution of the integrated mixing parameter $b$ and mass flux $a$ are well represented by the model. The integrated turbulent kinetic energy $E$ is qualitatively well-captured including the timing of the peak, but shows some deficits in the magnitude of the peak.

\section{Model equations and implementation}
We will follow the development proposed for single-fluid incompressible flow by Besnard et al~\cite{besnard96}, and subsequently adapted for variable-density flow by \cite{Steinkamp1999a,Steinkamp1999b}.  We first decompose the flow field variables, i.e., density $\rho$, velocity ${\bm u}$, and pressure $p$ into their mean and fluctuating parts as follows:

\begin{eqnarray}
\rho & = & \overline{\rho} + \rho^{\prime} \\
{\bm u} & = & \overline{\bm u} +{\bm u}^{\prime}\\
p & = & \overline{p} + p^{\prime}
\end{eqnarray}
where the overbar denotes the mean, and the primes the fluctuations about the mean.
In the case of variable-density flows, it is useful to work with the mass-weighted averages introduced
by Favre, known as Favre averages. So the Favre-averaged velocity $\tilde{{\bm u}}$ is 
\begin{equation}
\tilde{{\bm u}} = \frac{\overline{\rho {\bm u}}}{\overline{\rho}}.
\end{equation}
Let ${\bm u}^{\prime\prime}$ denote the fluctuation about this Favre averaged velocity $\tilde{\bm u}$. Then we have
\begin{equation}
{\bm u} = \tilde{{\bm u}} + {\bm u}^{\prime\prime}.
\end{equation}
Then, for two arbitrary points ${\bm x}_1$ and ${\bm x}_2$ in space, the mass-weighted Reynolds stress tensor is defined as,
\begin{equation}
R_{ij}({\bm x}_1,{\bm x}_2)= \frac{1}{2}\overline{[\rho({\bm x}_1)+\rho({\bm x}_2)]u_i^{\prime\prime}({\bm x}_1)u_j^{\prime\prime}({\bm x}_2)},
\end{equation}
the turbulent mass flux is defined as
\begin{equation}
a_i({\bm x}_1,{\bm x}_2)= -\overline{u_i^{\prime\prime}\rho({\bm x}_1)\upsilon({\bm x}_2)}\label{a_def},
\end{equation}
and the density-specific volume covariance is defined as
\begin{equation}
b({\bm x}_1,{\bm x}_2)=-\overline{\rho^{\prime}({\bm x}_1)\upsilon^{\prime}({\bm x}_2)}.
\end{equation} 
Subscripts $i$ and $j$ indicate Cartesian components, the specific volume is $\displaystyle\upsilon({\bm x})=\frac{1}{\rho({\bm x})}$ and its fluctuations $\upsilon'(\bm{x})$ are defined with respect to the mean specific-volume. The model is further developed in spectral space for which we require Fourier transformed variables. It is useful to rewrite the arguments in terms of position $\bm{x}=\frac{1}{2}(\bm{x}_1+\bm{x}_2)$, and scale $\bm{r}=\bm{x}_1-\bm{x}_2$ and Fourier transform so that $\bm{k}$ is the wavevector associated with scale $\bm{r}$, so that
\begin{eqnarray}
 R_{ij}({\bm x},{\bm k})&=&\int R_{ij}({\bm x},{\bm r})e^{-i{\bm k}\cdot{\bm r}}{\rm d}{\bm r},\\
a_i({\bm x},{\bm k})&=&\int a_i({\bf x},{\bm r})e^{-i{\bf k}\cdot{\bm r}}{\rm d}{\bm r},\\
b({\bm x},{\bm k})&=&\int b({\bm x},{\bm r})e^{-i{\bm k}\cdot{\bf r}}{\rm d}{\bm r}
\end{eqnarray}
To simplify further, we average over the sphere in $\bm{k}$- space to obtain
\begin{eqnarray}
R_{ij}({\bm x},k)&=&\int R_{ij}({\bm x},{\bm k})\frac{k^2{\rm d}\Omega_k}{4\pi},\\
a_i({\bm x},k)&=&\int a_i({\bm x},{\bm k})\frac{k^2{\rm d}\Omega_k}{4\pi}, \\
b({\bm x},k)&=&\int b({\bm x},{\bm k})\frac{k^2{\rm d}\Omega_k}{4\pi}.
\end{eqnarray}
where ${\rm d}\Omega_k=\sin\theta~{\rm d}\theta~{\rm d}\phi$ for $0\leq \theta \leq \pi$; $0\leq \phi \leq 2\pi$.
Henceforth we will use $R_{ij}$, $a_i$ and $b$ to denote the spectral quantities, and will drop their respective arguments.
Following Steinkamp {\emph {et.~al}}~\cite{Steinkamp1999a} we write the mass and momentum conservation equations for variable-density flows driven by gravity in the $y$-direction as follows: 
\begin{widetext}
\begin{eqnarray}
\frac{\partial\overline{\rho}}{\partial t} + \frac{\partial \overline{\rho} \tilde{u}_y}{\partial y} & = & \kappa \displaystyle \frac{\partial^2 \overline{\rho}}{\partial^2 y}-\kappa k^2 {\overline{\rho}}\label{drhodt}\\
\frac{\partial \overline{\rho}\tilde{u}_y }{\partial t}+\frac{\partial \rho \tilde{u}_y\tilde{u}_y}{\partial y} & = &
-\frac{\partial \overline{p}}{\partial y}+\overline{\rho}g-\frac{\partial R_{yy}}{\partial y}+\nu \displaystyle \frac{\partial^2 \overline{\rho} \tilde{u}_y}{\partial^2 y}-\nu k^2 \overline{\rho}\tilde{u}_y
\label{dpdy}
\end{eqnarray}
\end{widetext}
From Eq.~\ref{a_def} we note that $a_y=-\tilde{u}_y$. If we write the equations for the fluctuating density and velocity fields, and take the proper convolutions~\cite{clark1995two,besnard96}, we obtain the evolution equations for the Reynolds stress $R_{ij}$, mass flux $a_i$, and density-specific volume covariance $b$. These equations contain triple correlations of the velocity and density fluctuations which represent the turbulence cascade in $k$ space.
Based on the diffusion approximation model proposed by Leith~\cite{leith67}, we model these triple correlation terms as nonlinear advection and diffusion in $k$ space~\cite{clark1995two,besnard96,leith67}.
We modify Steinkamp's set of equations for statistically homogeneous variable-density flow, keeping the gravity direction same, and arrive at the following set of equations~\cite{Steinkamp1999a,Steinkamp1999b}:

\begin{widetext}
\begin{eqnarray}
\frac{\partial R_{nn}}{\partial t} & = &2a_y\frac{\partial \overline{p}}{\partial y}+\frac{\partial}{\partial k}\left[k\Theta^{-1}\left[-C_{r1}R_{nn}+C_{r2}k\frac{\partial R_{nn}}{\partial k}\right]\right]-2\nu k^{2} R_{nn}\label{main_rnn}\\
\frac{\partial a_y}{\partial t}&=&\frac{b}{\overline{\rho}}
\frac{\partial \overline{p}}{\partial y}-\left[C_{rp1}k^2\sqrt{a_na_n}+C_{rp2}\Theta^{-1}\right]a_y
\label{press}\nonumber\\
&&+\frac{\partial}{\partial k}\left[k\Theta^{-1}\left[-C_{a1}a_y+C_{a2}k\frac{\partial a_y}{\partial k}\right]\right]- (\nu + \kappa) k^2 a_y
\label{main_a}\\
\displaystyle\frac{\partial b}{\partial t}&=&\frac{\partial}{\partial k}\left[k\Theta^{-1}\left[-C_{b1}b+C_{b2}k\frac{\partial b}{\partial k}\right]\right]
-2\kappa k^2 b\label{main_b}
\end{eqnarray}
\end{widetext}
where the turbulence frequency $\Theta^{-1}=\sqrt{\int_0^{k_1}\frac{k^2R_{nn}}{\overline{\rho}}dk}$. In the equations (\ref{main_rnn}-\ref{main_b}) the dynamical variables $R_{nn}$, $a_y$ and $b$ respectively are functions of $k$. In the original papers by Steinkamp there was an additional equation for the vertical component of Reynolds stress, $R_{yy}$, since that 
was an inhomogeneous system in which $R_{yy}$ coupled directly back into both the mass flux and the energy. In our homogeneous system this mechanism is absent and it is therefore safe to omit that equation.
Equations (\ref{drhodt}) and (\ref{dpdy}) are the mass and momentum conservation laws respectively.
The first term on the right-hand side (RHS) of Eq. (\ref{main_b}) is based on a model proposed by Leith~\cite{leith67} for a nonlocal integral cascade with a wave-like part (the $C_{b1}$ term) and a diffusive part (the $C_{b2}$ term). For $C_{b1} > 0$, the wave-like cascade of $b$ is always forward (i.e., towards higher wavenumbers), and $C_{b2} > 0$ results in a forward as well as inverse cascade~\cite{Steinkamp1999a}. The cascade terms for $R_{nn}$ and $a_y$ in Eqs. (\ref{main_rnn}) and (\ref{main_a}) respectively are written in an analogous manner \cite{Steinkamp1999a}.
The drag between the fluids is described in the mass-flux equation (\ref{main_a}) by the second term on the RHS. Here $a_n$ is the component of $\bm{a}$ normal to the fluid interface. The $C_{rp1}$ term represents a drag arising between interpenetrating fluids at different scales \cite{clark1995two}. 
The $C_{rp2}$ term represents conventional drag governed by the turbulence timescale ~\cite{clark1995two}. Previous spectral models (~\cite{Steinkamp1999a, Steinkamp1999b}) neglected explicitly the viscous and diffusive effects, while our aim here is to build a model for turbulence with viscous dissipation. 
Therefore we had dissipation terms proportional to the diffusion coefficient $\kappa$ and the kinematic viscosity coefficient $\nu$. We assume Schmidt number $Sc = \nu/\kappa = 1$. We assume that the diffusion of $b(k)$ occurs in the manner of passive scalar diffusion \cite{girimaji1992modeling}.




Since we are implementing a system which is homogeneous and isotropic, only $k$-dependent terms appear in the equations. This allows us to use only one cell for the physical direction in the computational domain. 
The pressure gradient term $\displaystyle \frac{\partial \overline{p}}{\partial y}$ term in the gravity direction is independent of $y$. We calculate $\displaystyle \frac{\partial \overline{p}}{\partial y}$ directly from Eq.~\ref{dpdy}, as follows:
\begin{widetext}
\begin{equation}
\displaystyle\frac{\partial \overline{p}}{\partial y}=\frac{\overline{\rho}g+\int_0^k\left[C_{rp1}k^2\sqrt{a_na_n}+C_{rp2}\Theta^{-1}+2\nu k^2\right]a_y(k)dk}{1+\displaystyle\frac{\int_0^{k}b(k)dk}{\overline{\rho}}}
\end{equation}
\end{widetext}

The spectral model calculations presented in this paper are performed with a code using a second order MacCormack scheme~\cite{anderson1984computational} for time integration. This code is a modified version of a code used previously for studying variable-density mixing in the Rayleigh Taylor configuration~\cite{Steinkamp1999a,Steinkamp1999b}. For the purposes of code verification we also compare our results against an independent code which uses a second-order Crank-Nicolson~\cite{crank1947practical} scheme for time-advancement. In this way we can assess confidence in the accuracy of our codes. The latter code is used only for verification. The results presented are based on the code using the  second-order MacCormack scheme for time-advancement. 
 Both computer codes use an exponential grid for the wavenumber $$k = k_s \exp\left\{\frac{z}{z_s}\right\}$$ where $k_s$ and $z_s$ are scale factors and assumed to be equal to unity~\cite{besnard96}. The variables computed are, in fact $kR_{nn}$, $kR_{ij}$, $ka_i$ and $kb$.  This choice of variables results in the cascade terms retaining a conservation form when expressed in terms of $z$ rather than $k$.  Likewise, the values of the integrals of the spectral quantities are easily determined, e.g.;
\begin{widetext} 
\begin{equation}
 R_{nn}\left(t\right) = \int_0^{+\infty} R_{nn}\left(k,t\right) dk 
= \int_{-\infty}^{+\infty} R_{nn}\left(z,t\right) \frac{k_s}{z_s}\exp\left\{\frac{z}{z_s}\right\}dz. 
\end{equation}
\end{widetext}
Setting $k_s=1$ and $z_s$ gives 
$$ R_{nn}\left(t\right) = \int_{-\infty}^{+\infty} \exp\left(z\right)R_{nn}\left(z,t\right) dz, $$
where $\exp\left(z\right)R_{nn}\left(z,t\right)  = kR_{nn}\left(k,t\right)$
The explicit MacCormack methodology is nominally second-order accurate in time and space, and utilizes two-steps. Each of the two steps uses single-sided differences for the first order derivatives, and the sides at which the differences are evaluated are different for the two steps, i.e., left-side for the first step, and right-side for the second.  
The second code utilizes a Crank-Nicolson method for time-advancement, and central differences for the $z$-space derivatives.  The implicit evaluations of the cascades are decoupled for the various variables.  However, the implicit step of the Crank-Nicolson method is iterated to achieve a coupling between the variables, and to bring the nonlinear terms, i.e., the turbulent frequency term $\Theta^{-1}$, up to date.  This typically requires three or four iterations to converge.  The implicit cascade is solved using a tridiagonal solver.  The code is thus second-order accurate in time and space, and unconditionally stable at all time-steps. It should be noted that large time-steps may necessitate more iterations to converge.

The boundary conditions at $k=1$ and $k=k_{max}$ are set to Neumann (zero flux).
We initialize our model calculations with spectra for $b, a$ and $R_{nn}$ at initial time $t=t_0$. 
We provide a value for the average density which corresponds to $\displaystyle \frac{\rho_{max}+\rho_{min}}{2}$ (here $\rho_{max}$ is the maximum density and $\rho_{min}$ is the minimum density in the variable density fluid mixture). The spectral code requires the information of the mean density, and the details of the density contrast between the fluids are present only in the initial spectral distribution of $b, a_y$ and $E$. 

In presenting results we will use integrated quantities as well as spectra for analysis. The integrated quantities are $b = \int b(k) dk$, $a = \int a_y(k) dk$, and $R_{nn} = \int R_{nn}(k) dk$. $R_{nn}$ is related to the turbulent kinetic energy $E$ in the following way:
\begin{equation}
E=\frac{1}{2\overline{\rho}}R_{nn}
\label{const}
\end{equation}

The set of equations (\ref{main_rnn}-\ref{main_b}) can describe a wide variety of homogeneous variable-density flows. We will focus on two canonical types of flow. The first is described by a non-zero initial distribution of $b(k)$ \cite{clark1995two} with other variables set to zero, and will be used to benchmark the calculations against previous efforts. The second type of flow is that computed by \cite{livescu2007,livescu2008} and is initialized by $b(k)$ describing a distribution of blobs of one fluid in another with both $a$ and $R_{ij}$ set to nominally small values. The latter choice is made so that the flow reaches a turbulent state in a reasonable period of (wallclock) time. These are discussed in the next section.

\section{Results}

In this section, we present our main results, which can be broadly divided into three categories. First, we present studies to check the numerics of our variable density model implementation. We compare the results from the model calculations performed using two codes which use different schemes for time advancement. We also demonstrate convergence with respect to grid-refinement. Second, we show how the system parameters affect the time evolution of the variables under study, i.e., $a, b, E$, for a test initial condition described by the $b(k)$ spectrum used in \cite{clark1995two}, and discuss the varying trends. Third, from this study of parameters, we choose an optimum set which minimizes error with respect to the outcomes of a highly resolved low Atwood number \Big(Atwood number defined as $\displaystyle \frac{\rho_{max}-\rho_{min}}{\rho_{max}+\rho_{min}}$ \Big) DNS study of variable density buoyancy driven turbulence \cite{aslangil2018}. We use the same set of coefficients for a high Atwood number system, and show that the time evolution of $b$ and $a$ are well captured, while there is less fidelity to $E$. Overall however, the same coefficients appear to reasonably capture the multiple stages of the mixing for a broad spread in Atwood number.

\subsection{Code convergence and time-stepping accuracy}
\begin{figure*}[ht!]
\includegraphics[scale=0.5]{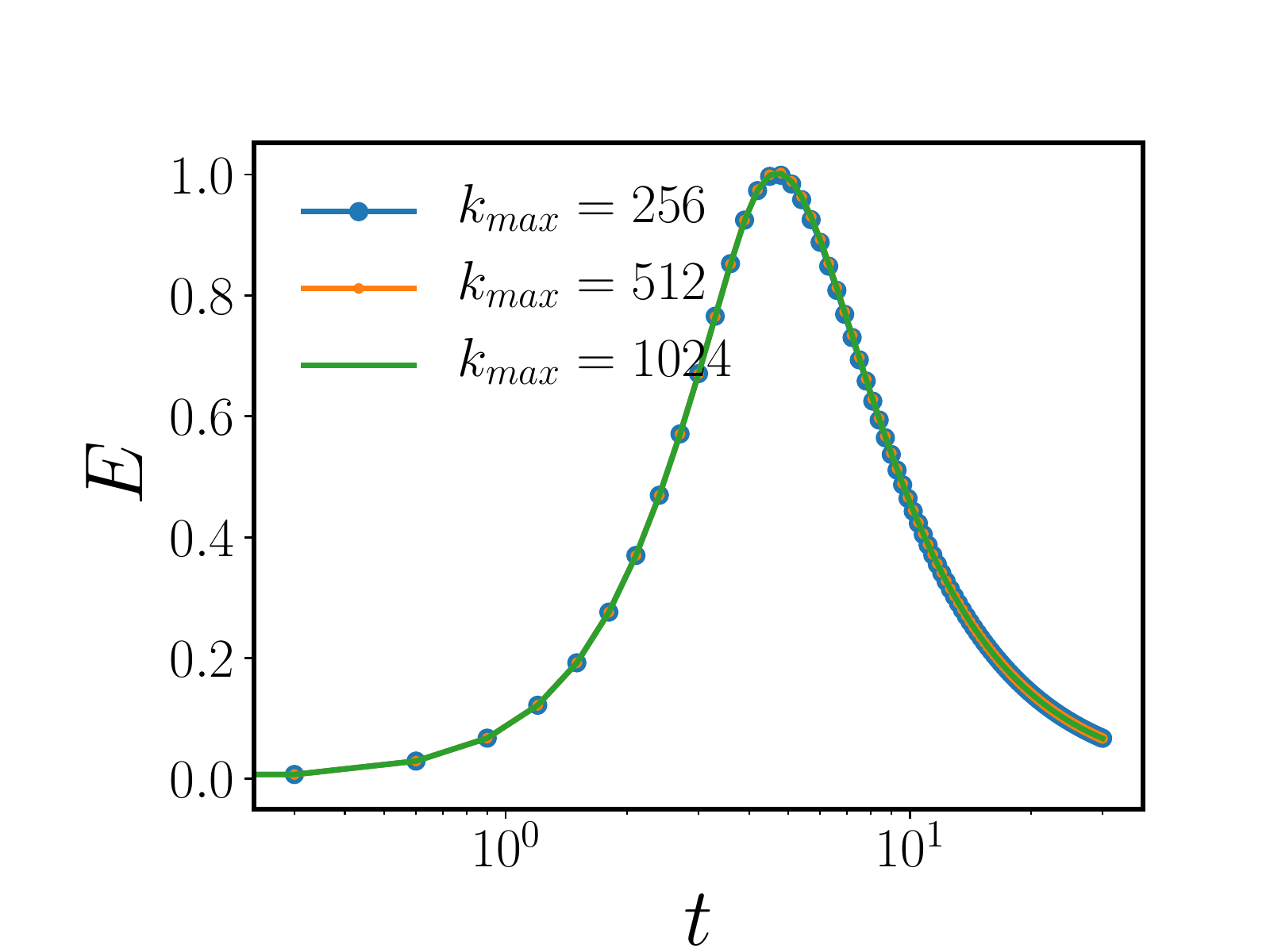}
\put(-40,135){\bf \scriptsize (a)}
\includegraphics[scale=0.5]{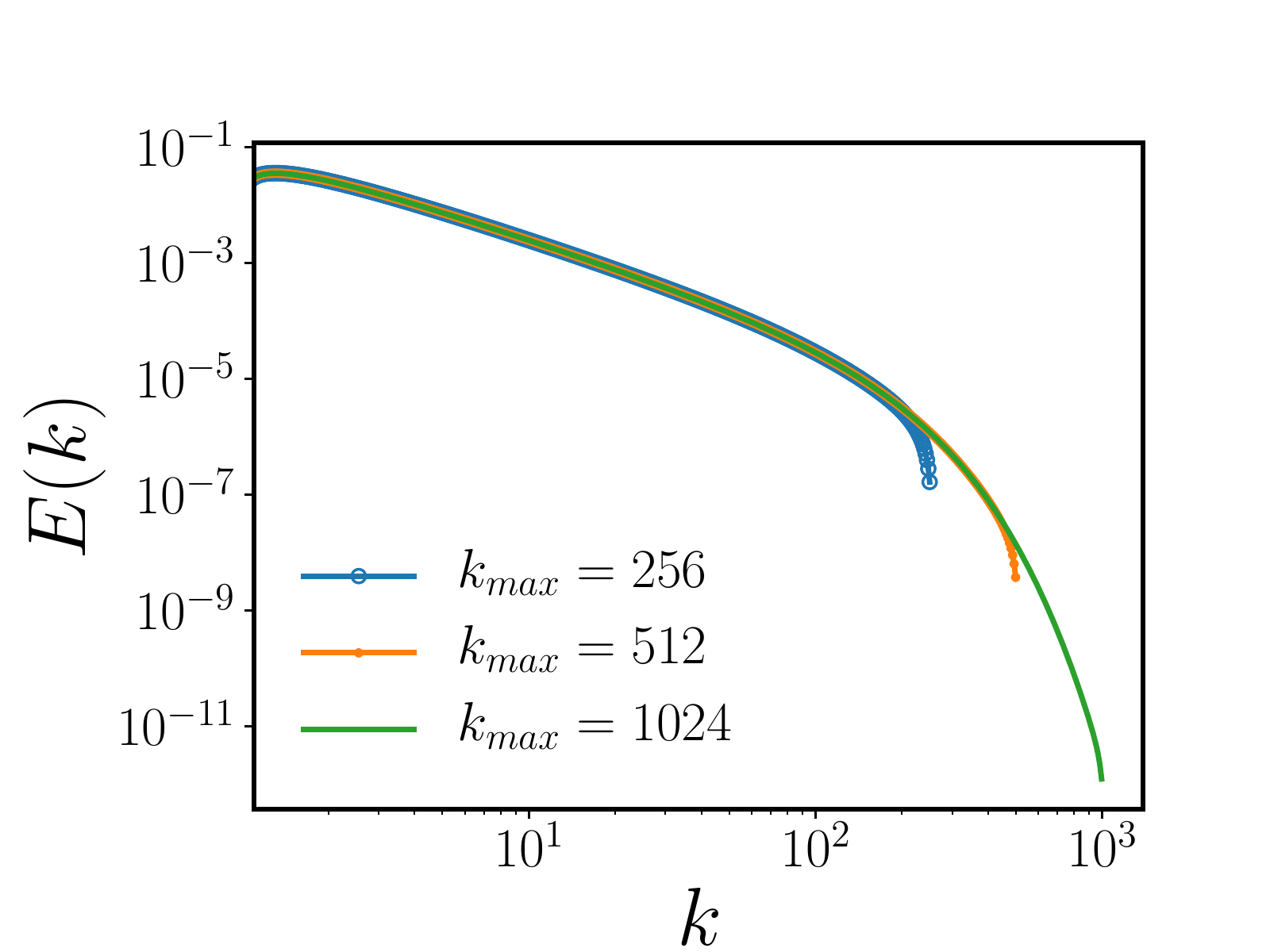}
\put(-40,135){\bf \scriptsize (b)}
\caption{[Color online] Convergence of outcomes at increasing resolution for initial Gaussian distribution for $b(k)$, and $a(k), E(k)$ set to zero at fixed viscosity $\nu = 10^{-4}$. (a) Time evolution plots of the turbulent kinetic energy $E$ at different system resolutions and (b) kinetic energy spectra at different resolutions} 
\label{conv}
\end{figure*}

\begin{table*}
{
\begin{tabular}{|l|l|l|l|l|l|l|l|l|l|l|}
\hline
$C_{b1}$&$C_{b2}$&$C_{a1}$&$C_{a2}$&$C_{r1}$&$C_{r2}$&$C_{rp1}$ & $C_{rp2}$ & $\nu$ & $\kappa$ &$k_{max}$\\
\hline
$0.12$ & $0.06$ & $0.12$ & $0.06$ &$0.12$&$0.06$&$1.0$&$1.0$ & $10^{-4}$ & $10^{-4}$ & $512$\\
\hline
\end{tabular}
}
\caption{Table showing nominal values of all coefficients and other model parameters in the code testing phase as prescribed in \cite{clark1995two}.}
\label{table1}
\end{table*}
To begin with, we test whether the spectral model code converges under different system resolutions. We choose an analytical form $b(k)=B_0{\rm e}^{-k^2}$, with $a(k)= E(k)=0$ as our initial condition. Here $B_0$ is such that $\int_0^{k_{max}}b(k)dk=1$ at $t=0$, where $k_{max}$ is the maximum number of $k$ modes, and $t$ is the time. The coefficient values used in this part of the study are listed in Table \ref{table1}. In Fig.~\ref{conv} we show that the results converge as the grid in $k$-space is refined for a fixed vertical system size of $2\pi$. As the resolution is increased from $256$ to $1024$ $k$-modes, both the global energy and the spectral distributions converge.  

\begin{figure*}[ht!]
\includegraphics[width=.35\linewidth]{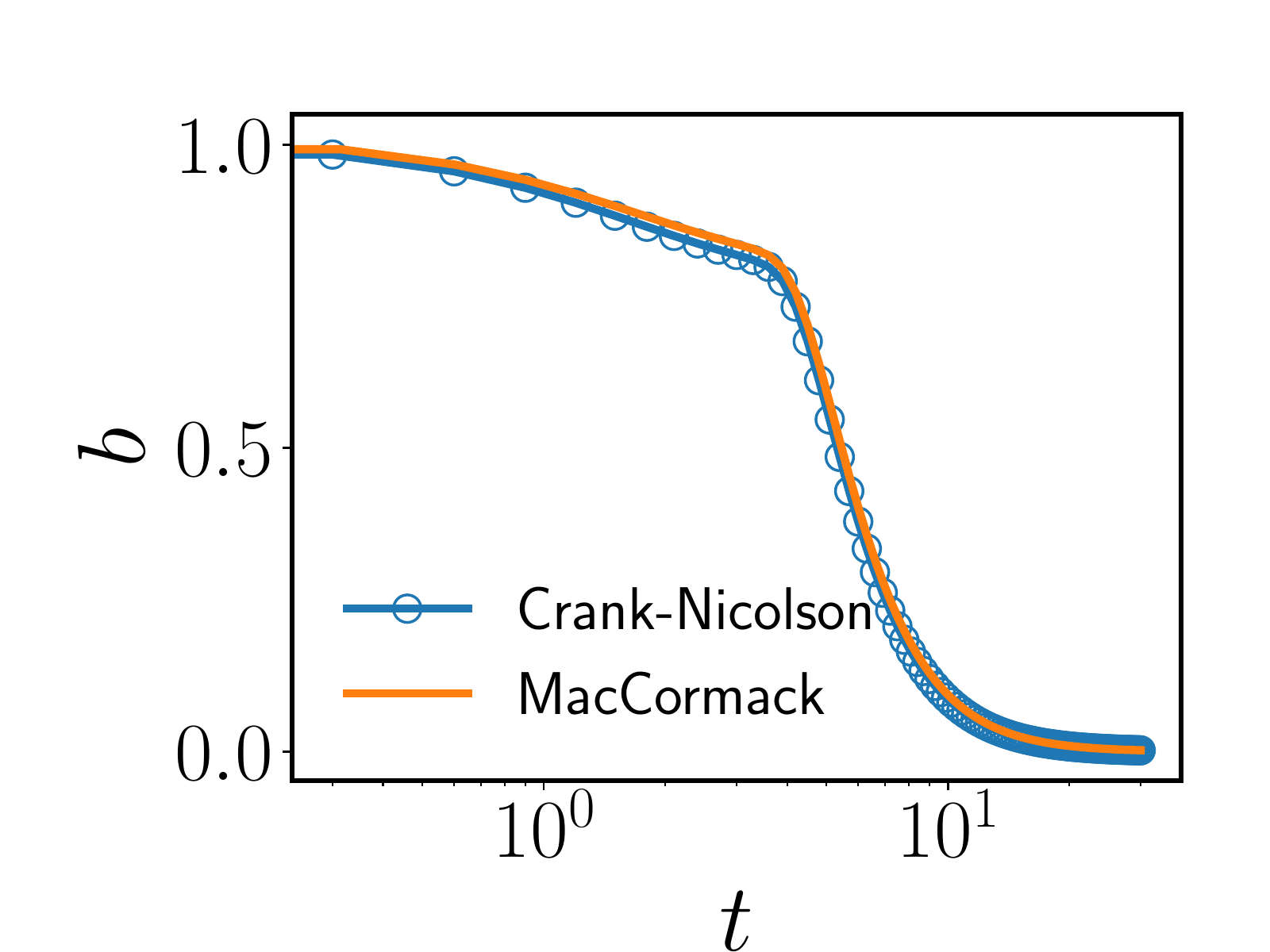}
\put(-25,110){\bf \scriptsize (a)}
\includegraphics[width=.35\linewidth]{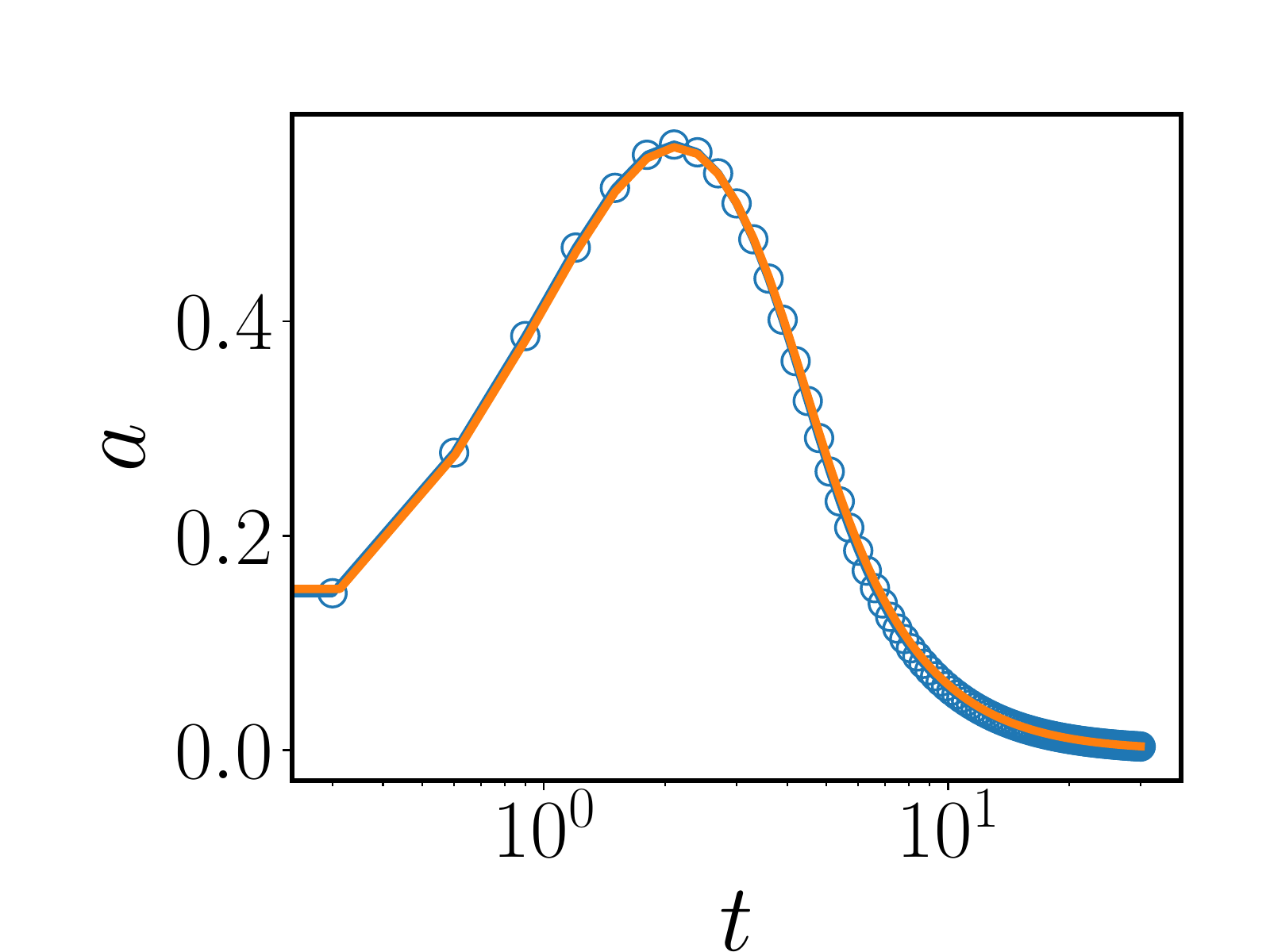}
\put(-25,110){\bf \scriptsize (b)}
\includegraphics[width=.35\linewidth]{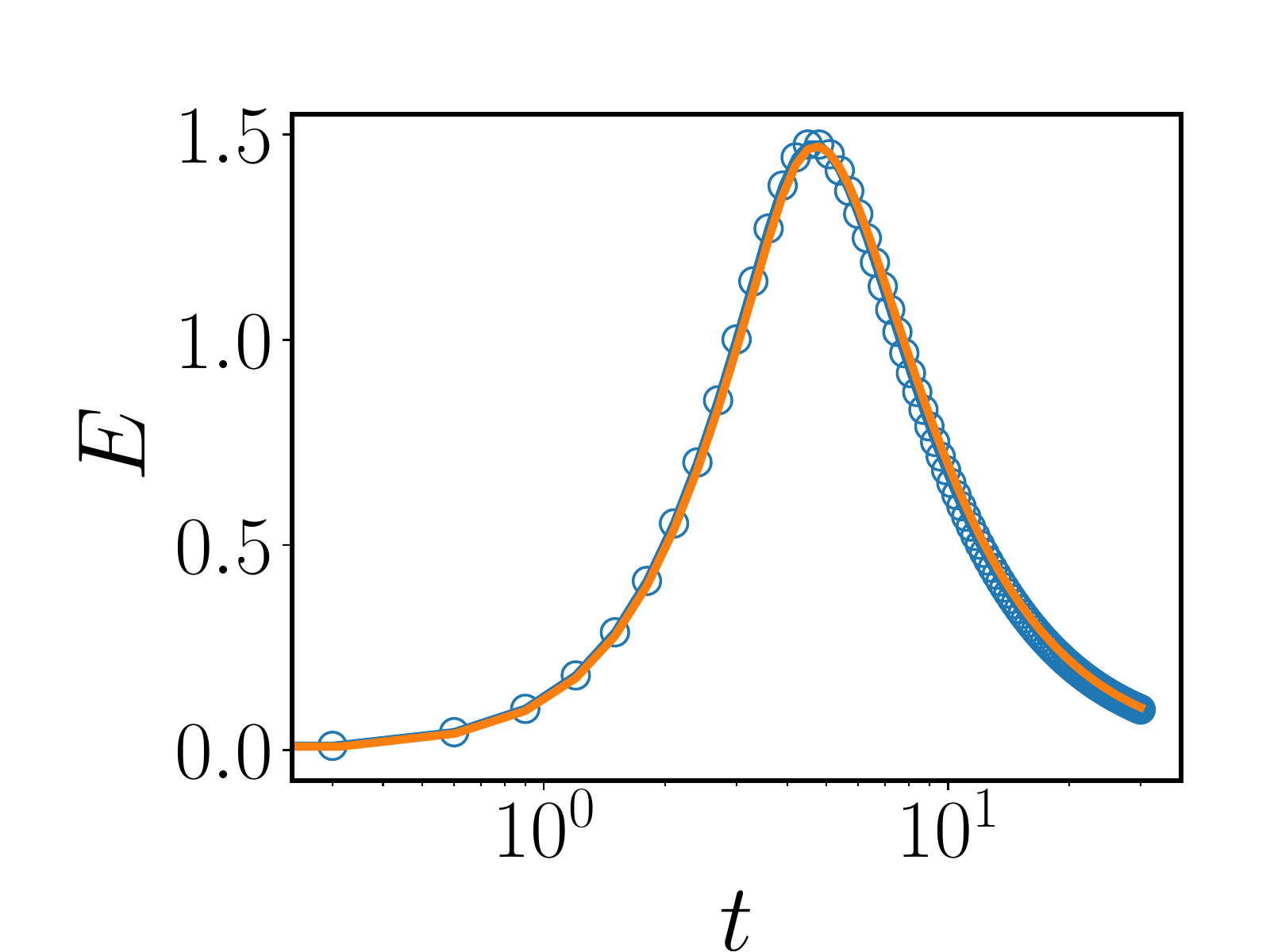}
\put(-25,110){\bf \scriptsize (c)}
\caption{[Color online] Comparison of results from codes using Crank-Nicolson (blue line with circles) and MacCormack (orange line) schemes for time integration. Plots showing time evolution of (a) the mean density-specific volume covariance $b$; (b) the mean mass flux $a$ and (c) the turbulent kinetic energy $E$. 
}
\label{test_cases}
\end{figure*}

Next we compare our results for $k_{max} = 256$ against those from a code using a Crank-Nicolson scheme for time integration for the same resolution. The results are shown in Fig.~\ref{test_cases}(a)--(c)), and demonstrate that errors due to the time-advancement scheme are not significant. 


\subsection{Coefficient variation study}
There are a total of 8 coefficients in this model. The theoretical constraint  for energy equipartition in the inviscid case $C_{r1} = 2 C_{r2}$ \cite{lee1952some,besnard1985turbulence} reduces the space to $7$ coefficients. Most of these arise from the closure approximation for the nonlinear (triple-correlation and higher order) processes. The drag terms in the evolution of $a$ were introduced in an \emph{ad hoc} fashion \cite{clark1995two} to provide an additional mechanism for the breakup of fluid structures in scale. The structure of the spectral transfer terms for $b$ and $a$ were originally written down by analogy with the arguments provided in \cite{besnard96} for the energy spectral transfer. Given the rather non-rigorous quality of these arguments, but nevertheless taking the model at face-value, it is worthwhile to assess what impact the systematic variation of these coefficients has on flow outcomes.

\begin{figure*}[ht!]
\includegraphics[width=.35\linewidth]{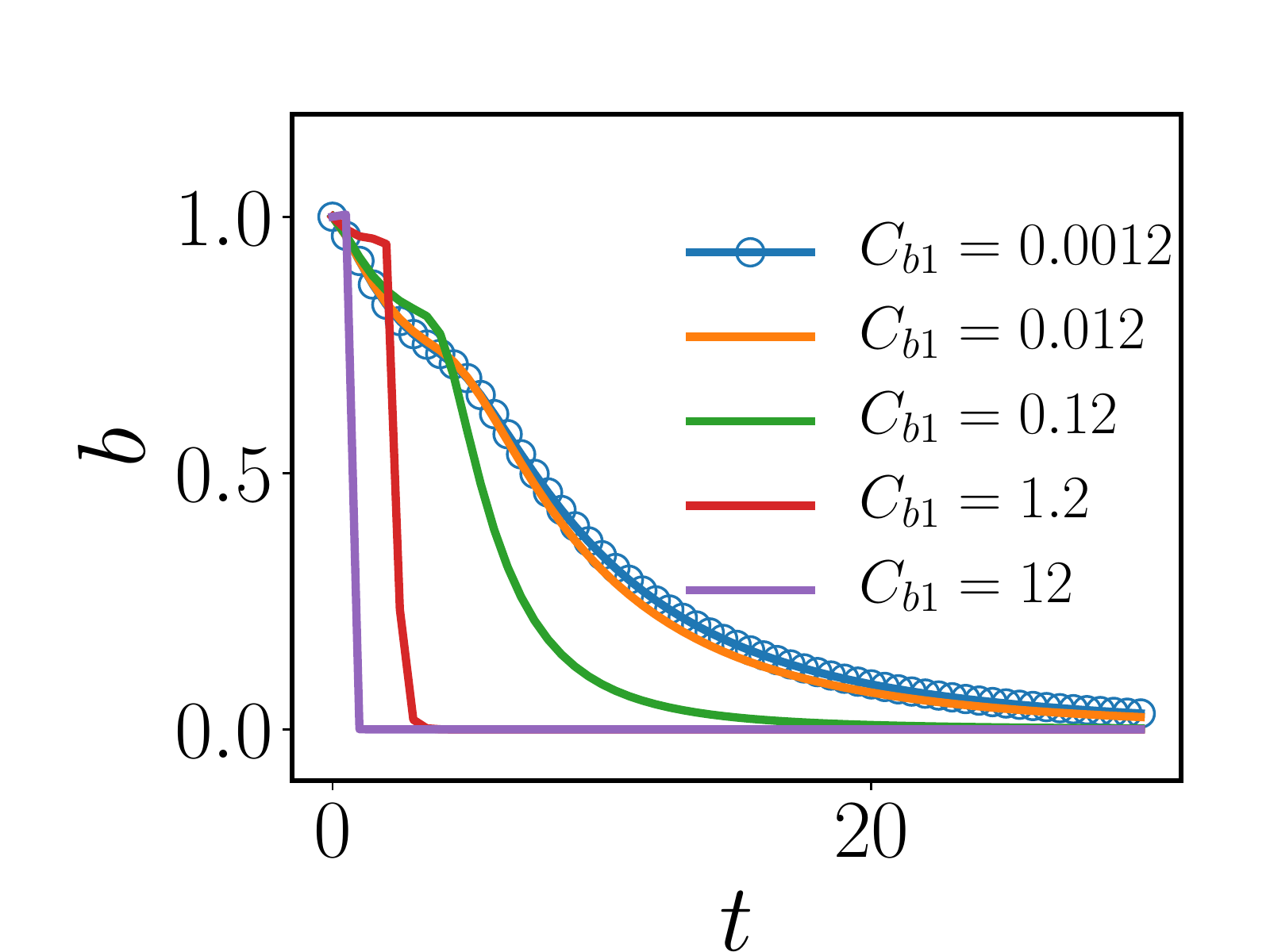}
\put(-25,110){\bf \scriptsize (a)}
\includegraphics[width=.35\linewidth]{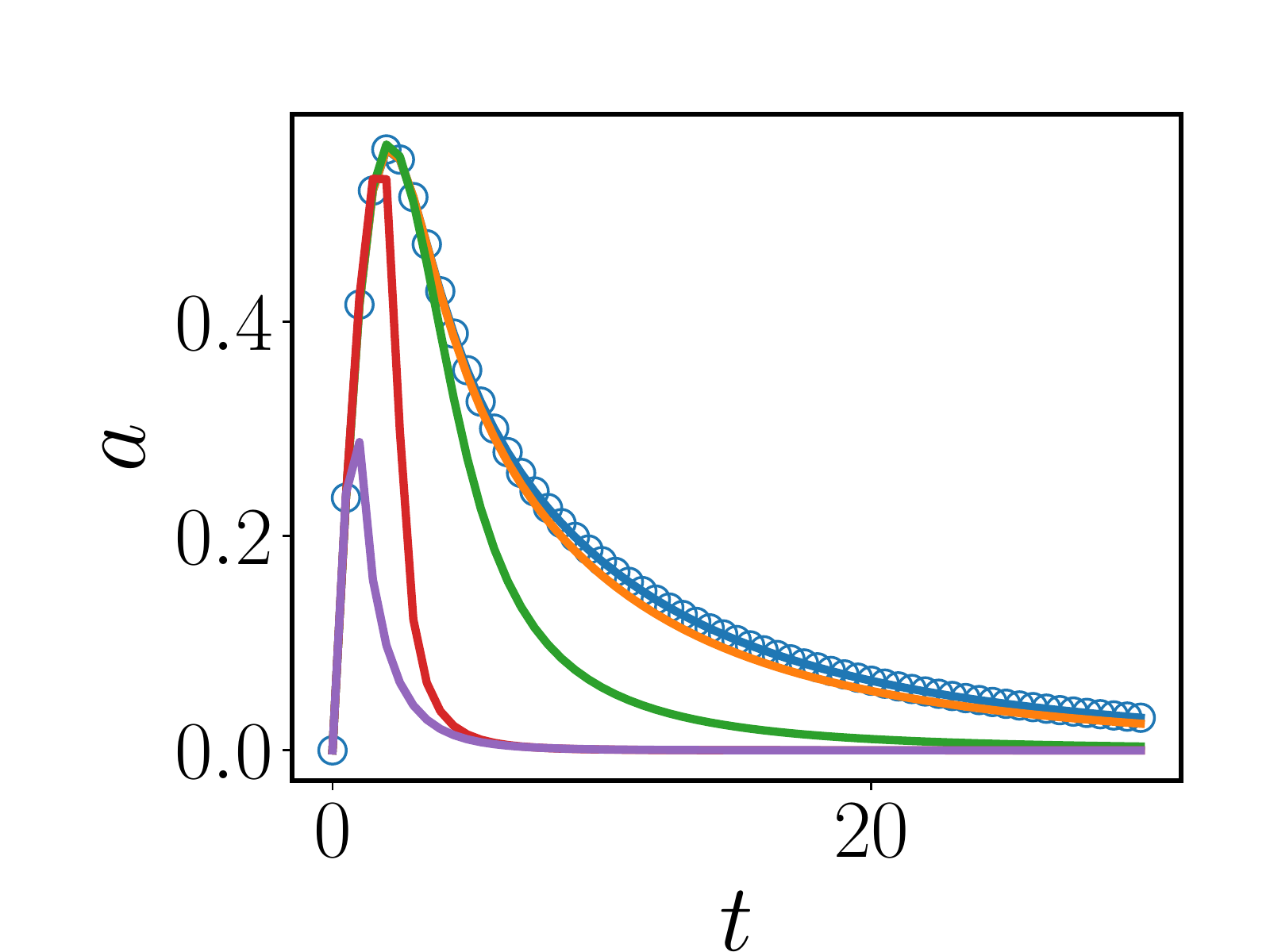}
\put(-25,110){\bf \scriptsize (b)}
\includegraphics[width=.35\linewidth]{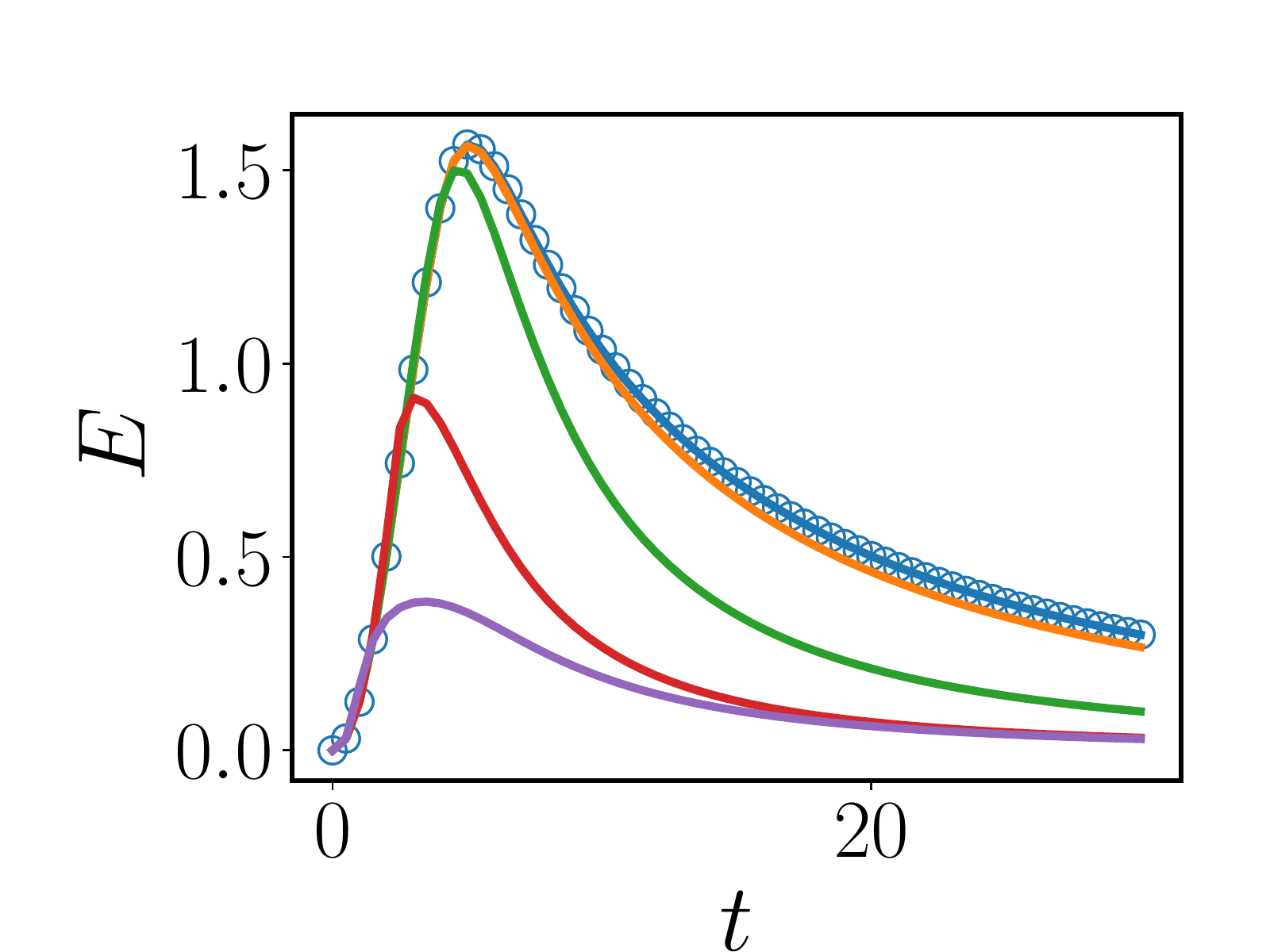}
\put(-25,110){\bf \scriptsize (c)}

\includegraphics[width=.35\linewidth]{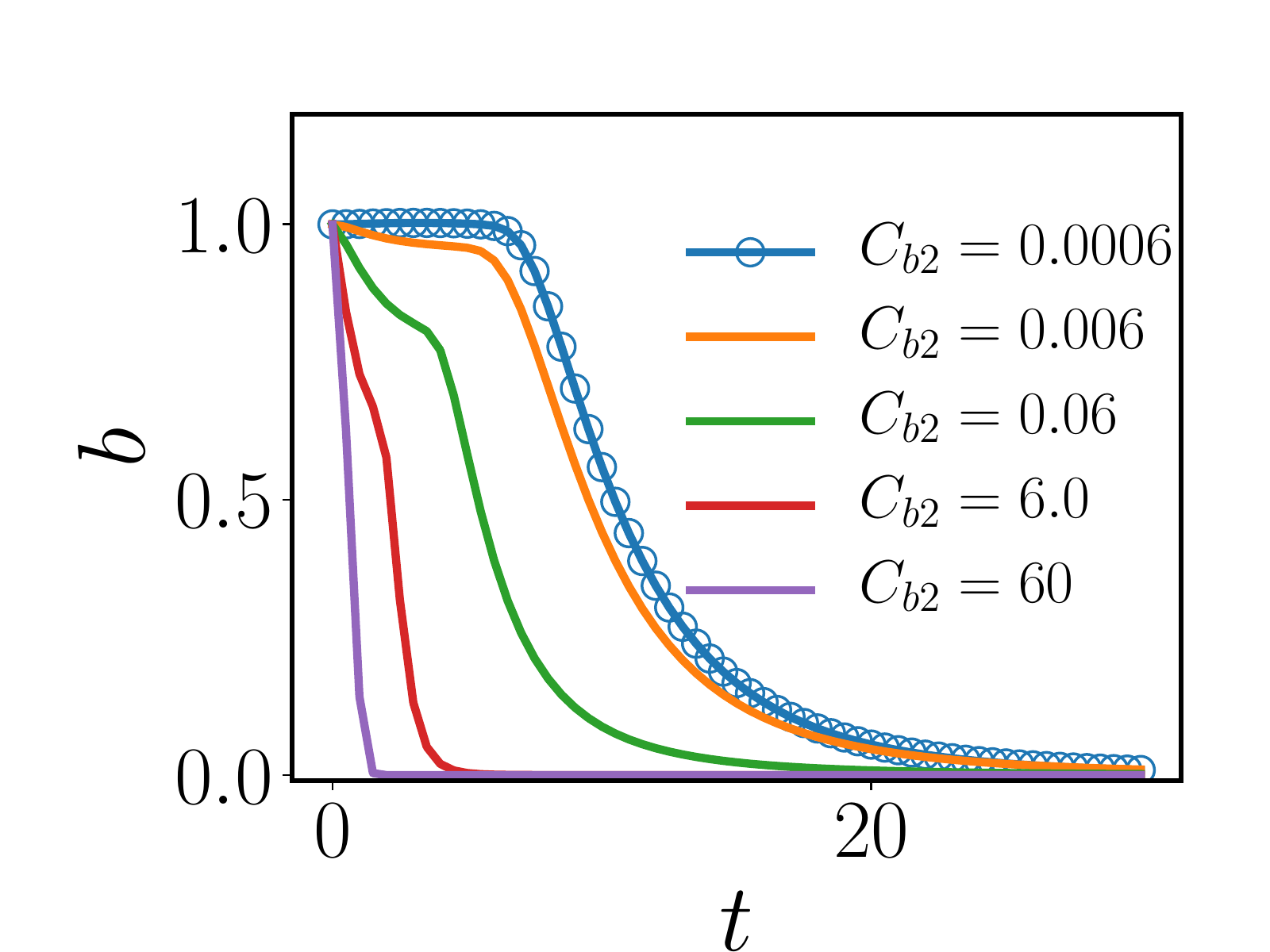}
\put(-25,110){\bf \scriptsize (d)}
\includegraphics[width=.35\linewidth]{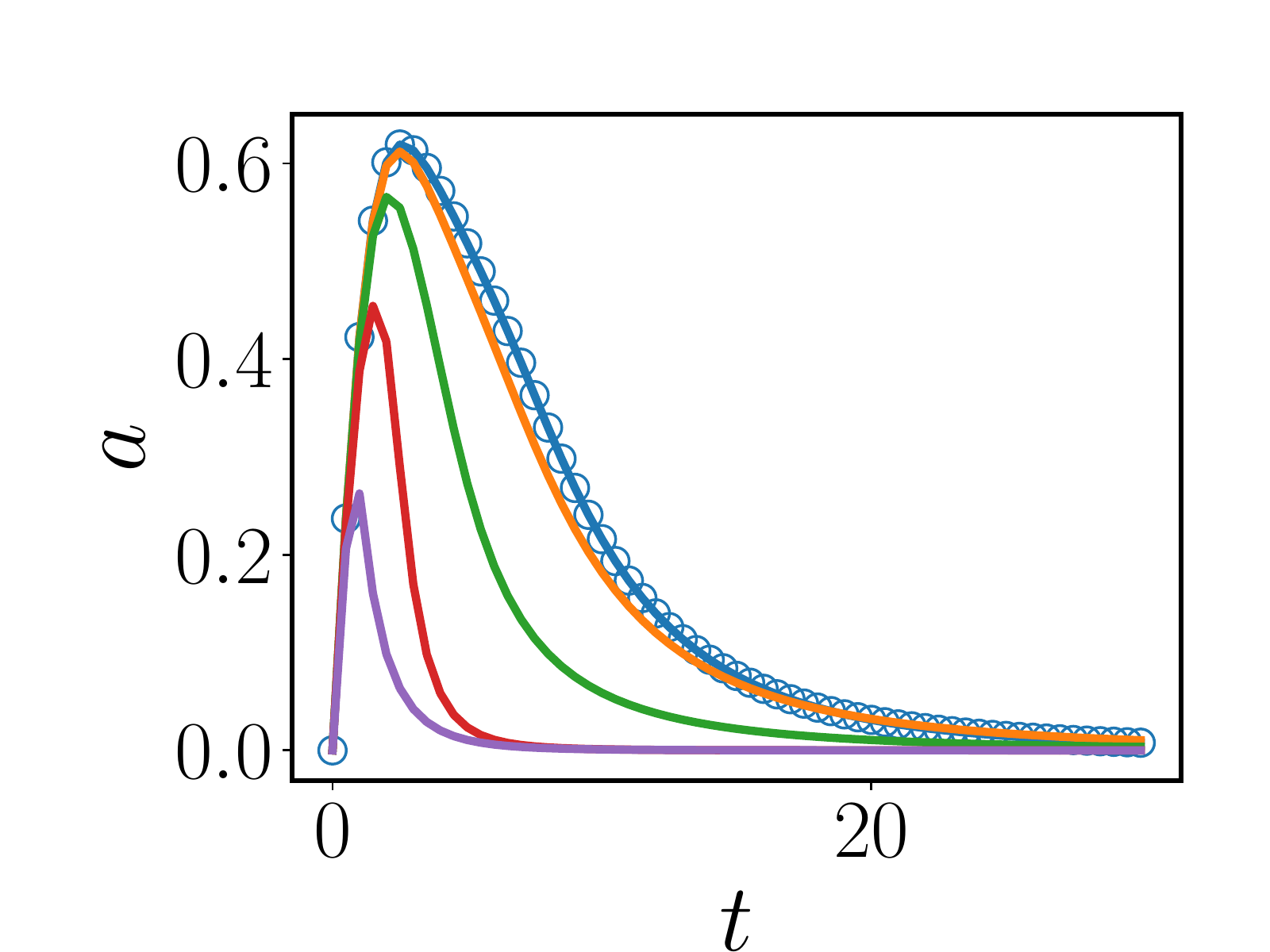}
\put(-25,110){\bf \scriptsize (e)}
\includegraphics[width=.35\linewidth]{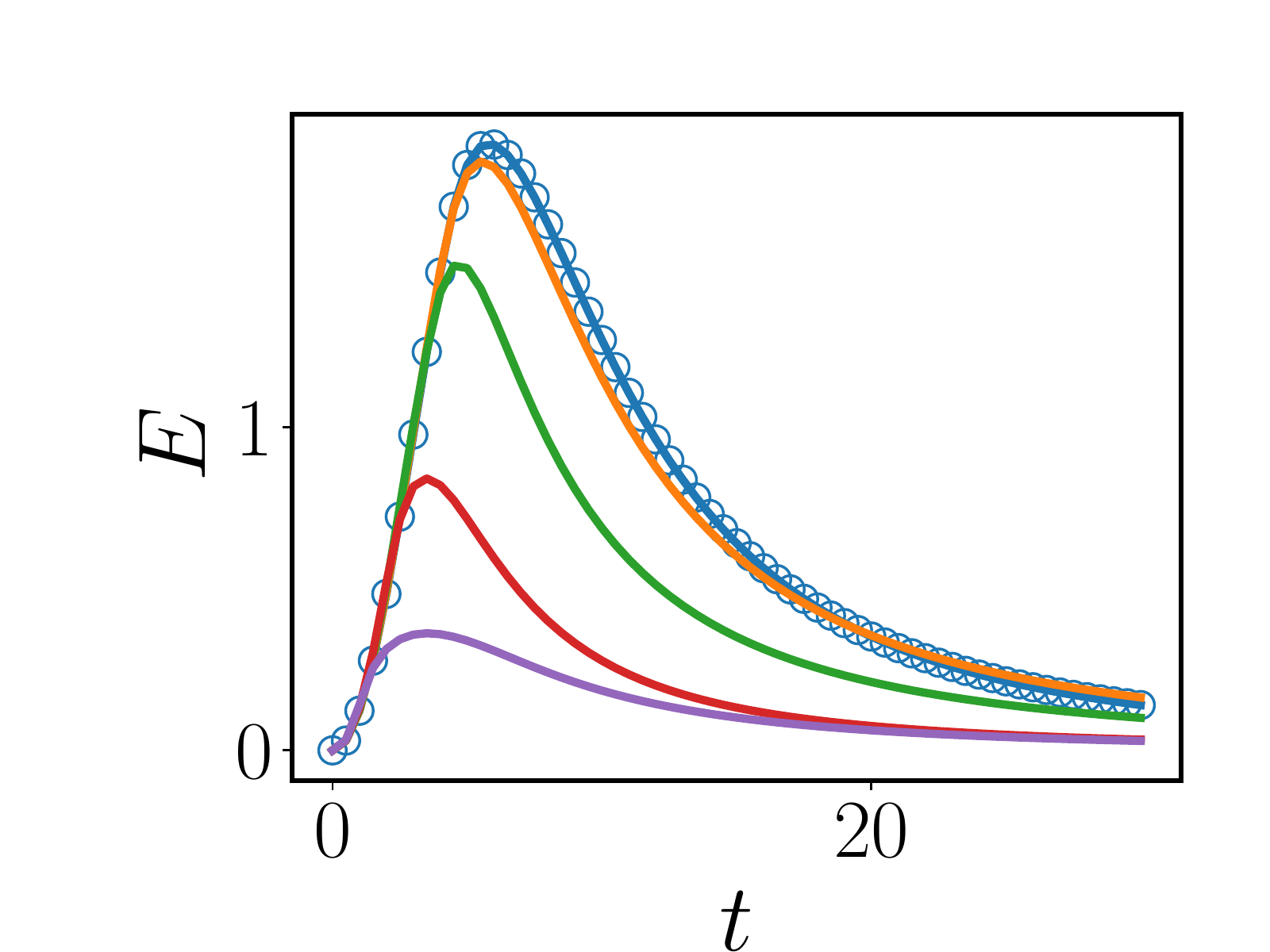}
\put(-25,110){\bf \scriptsize (f)}
\caption{[Color online] 
Time evolution plots of (a) $b$; (b) $a$; (c) $E$ for different values of $C_{b1}$. 
Time evolution plots of (d) $b$; (e) $a$; (f) $E$ for different values of $C_{b2}$.
The values of the parameters which do not vary in each case are shown in Table~\ref{table1}. 
The number of $k$ modes used is $512$.
}
\label{comp_b}
\end{figure*}

For the purposes of this study we systematically vary each coefficient while keeping the others fixed at their provisional values given in \cite{clark1995two}. We then plot the integrated quantities $b$, $a$ and $E$ as functions of time and describe how these vary relative to expected behaviors. In this test study, the same non-zero initial condition $b(k)$ as in the previous section provides the drive term for the mass flux which subsequently drives the growth of Reynolds stress $R_{nn}$. From Eq.~\ref{main_b} we note that the $b$ equation has no production term in it, since there is no production of mass in the system. The sole contribution to $b$ evolution is a redistribution in $k$-space via the terms weighted by coefficients $C_{b1}$ and $C_{b2}$. Formally, based on the terms in the model, when $C_{b1}$ is increased it should deplete $b(k)$ from the small $k$ modes, i.e., the large length scales and transfer them to the large $k$ modes where they are dissipated by viscosity. This should increase the decay rate of mean $b$ with time~(Fig.~\ref{comp_b}(a)). As $b$ decays faster, peak of $a$ is reduced, since there is less production of $a$ (through the $\displaystyle\frac{b}{\overline{\rho}} \frac{\partial \overline{p}}{\partial y}$ term). We see this is indeed the case in Fig.~\ref{comp_b}(b). The reduction in $a$ reduces $E$ because there is less production in $R_{nn}$ (due to the $a_y\displaystyle\frac{\partial \overline{p}}{\partial y}$ term). 

\begin{figure*}[ht!]
\includegraphics[width=.35\linewidth]{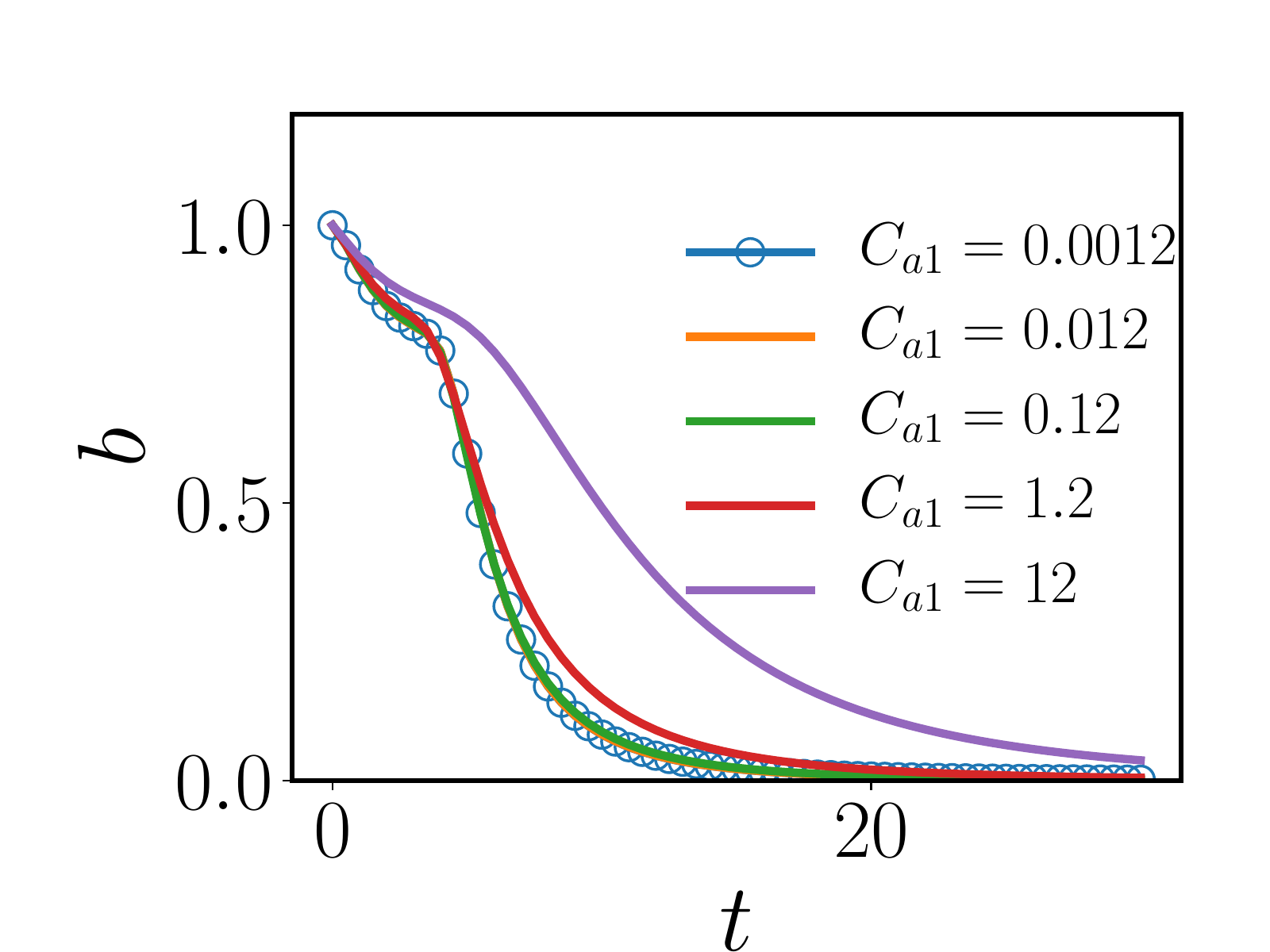}
\put(-25,110){\bf \scriptsize (a)}
\includegraphics[width=.35\linewidth]{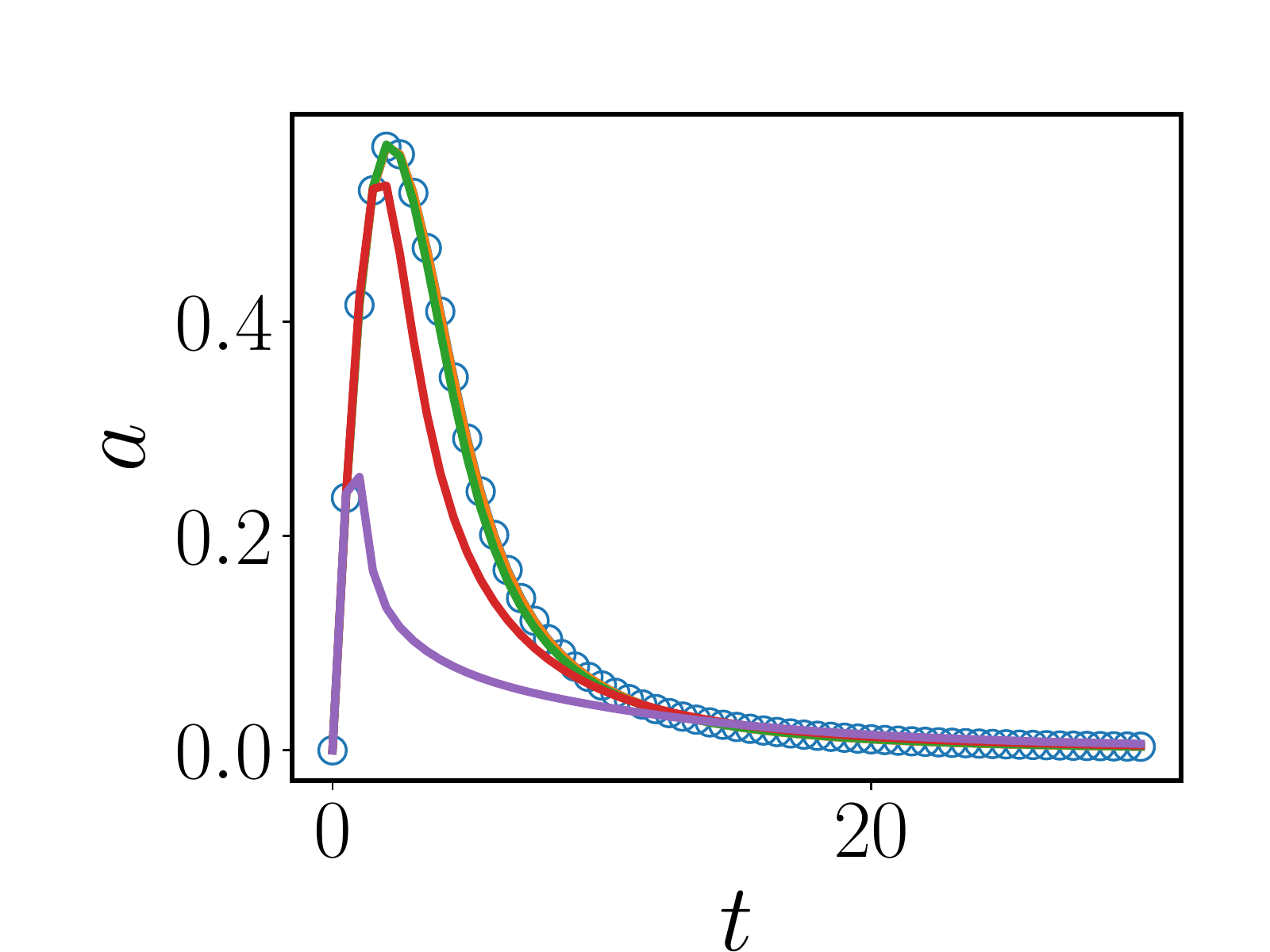}
\put(-25,110){\bf \scriptsize (b)}
\includegraphics[width=.35\linewidth]{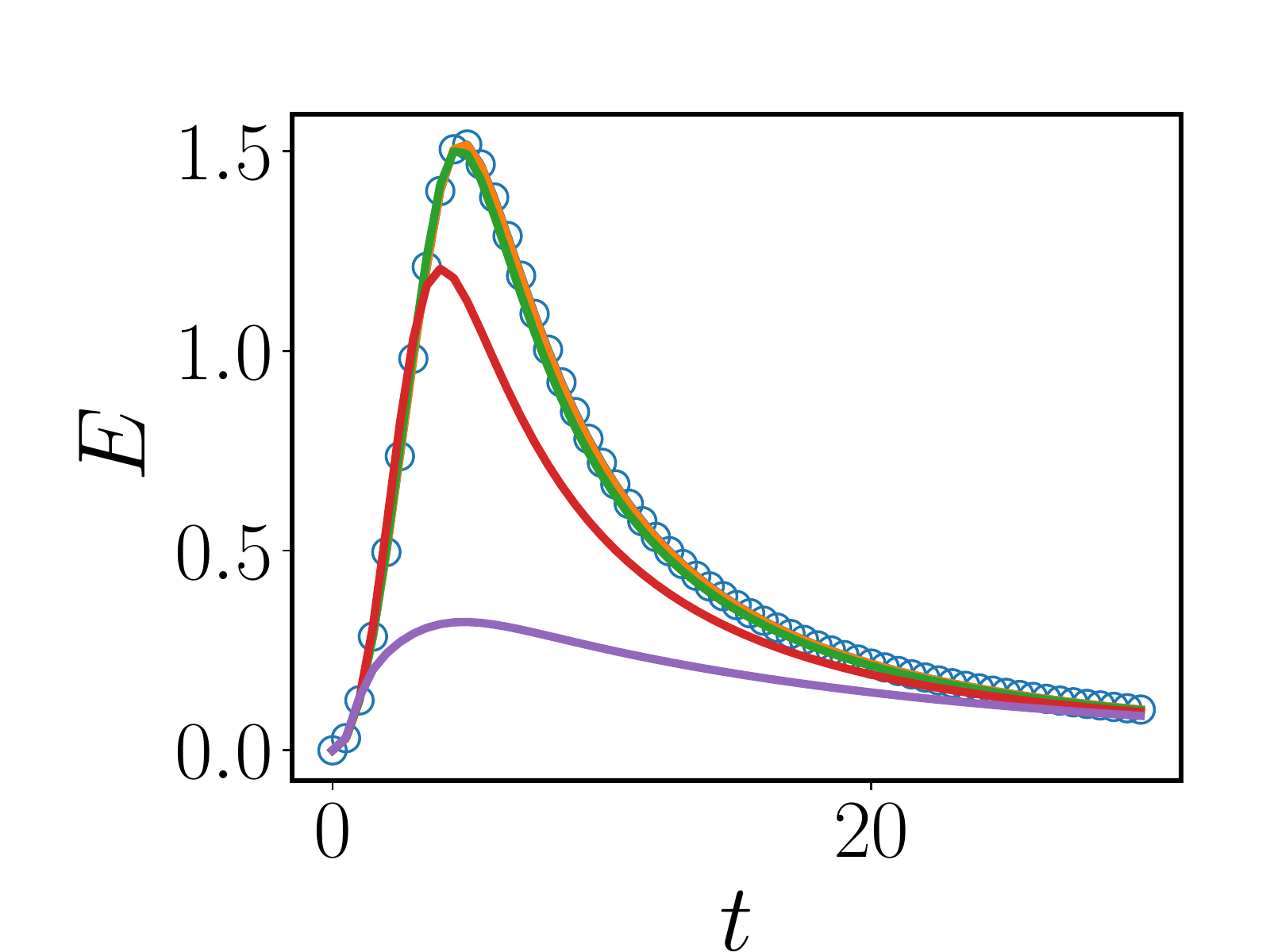}
\put(-25,110){\bf \scriptsize (c)}

\includegraphics[width=.35\linewidth]{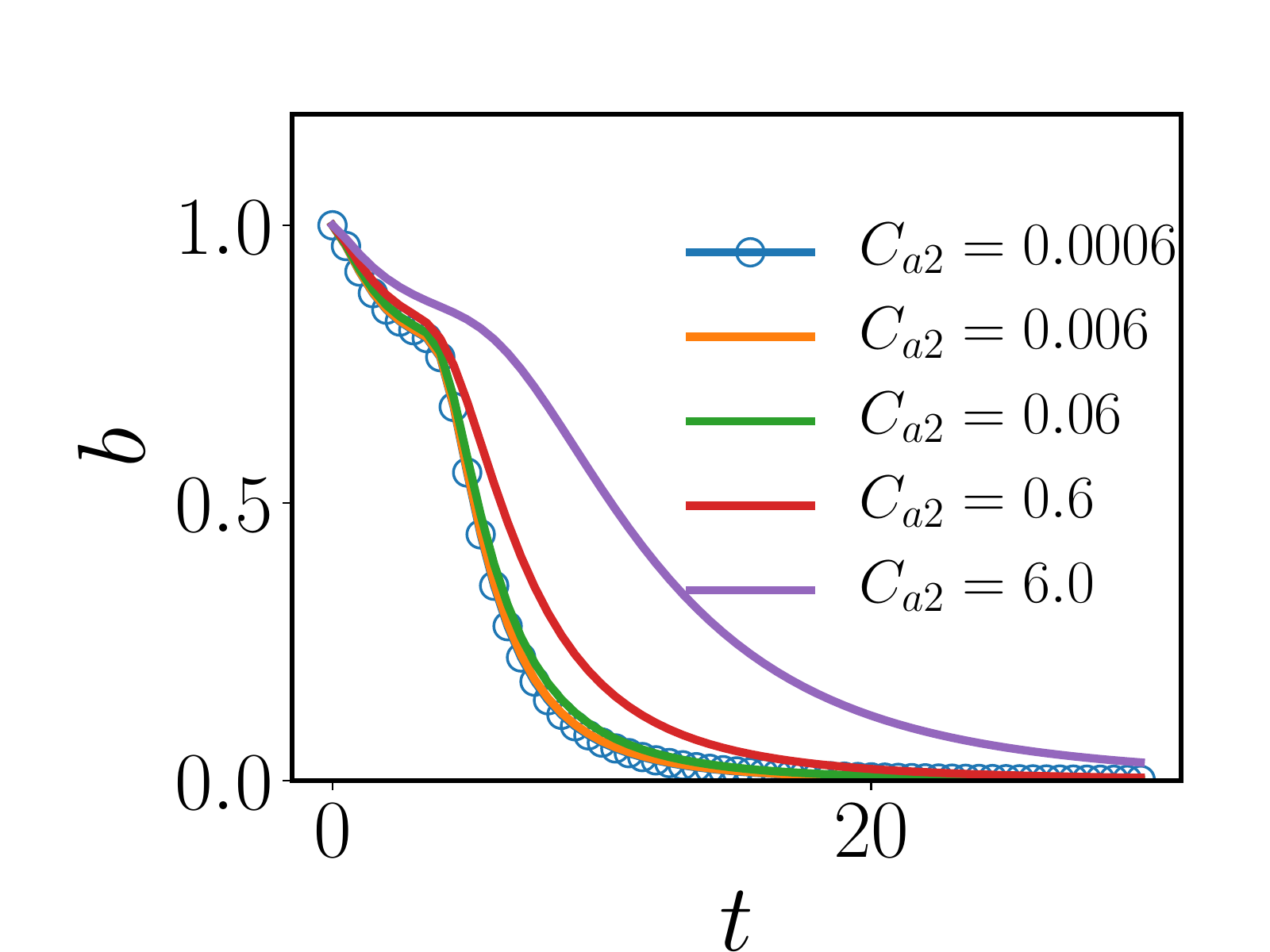}
\put(-25,110){\bf \scriptsize (d)}
\includegraphics[width=.35\linewidth]{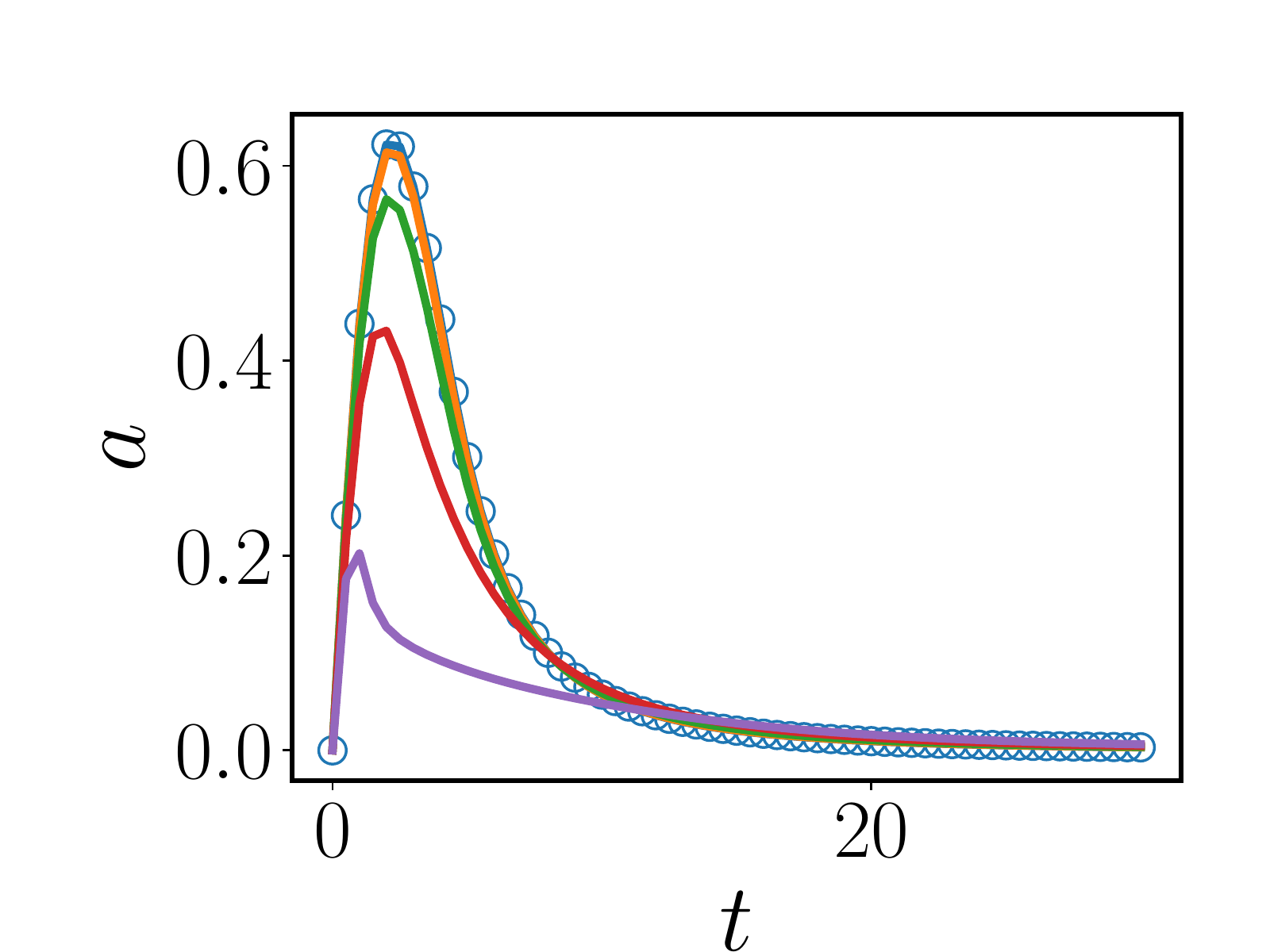}
\put(-25,110){\bf \scriptsize (e)}
\includegraphics[width=.35\linewidth]{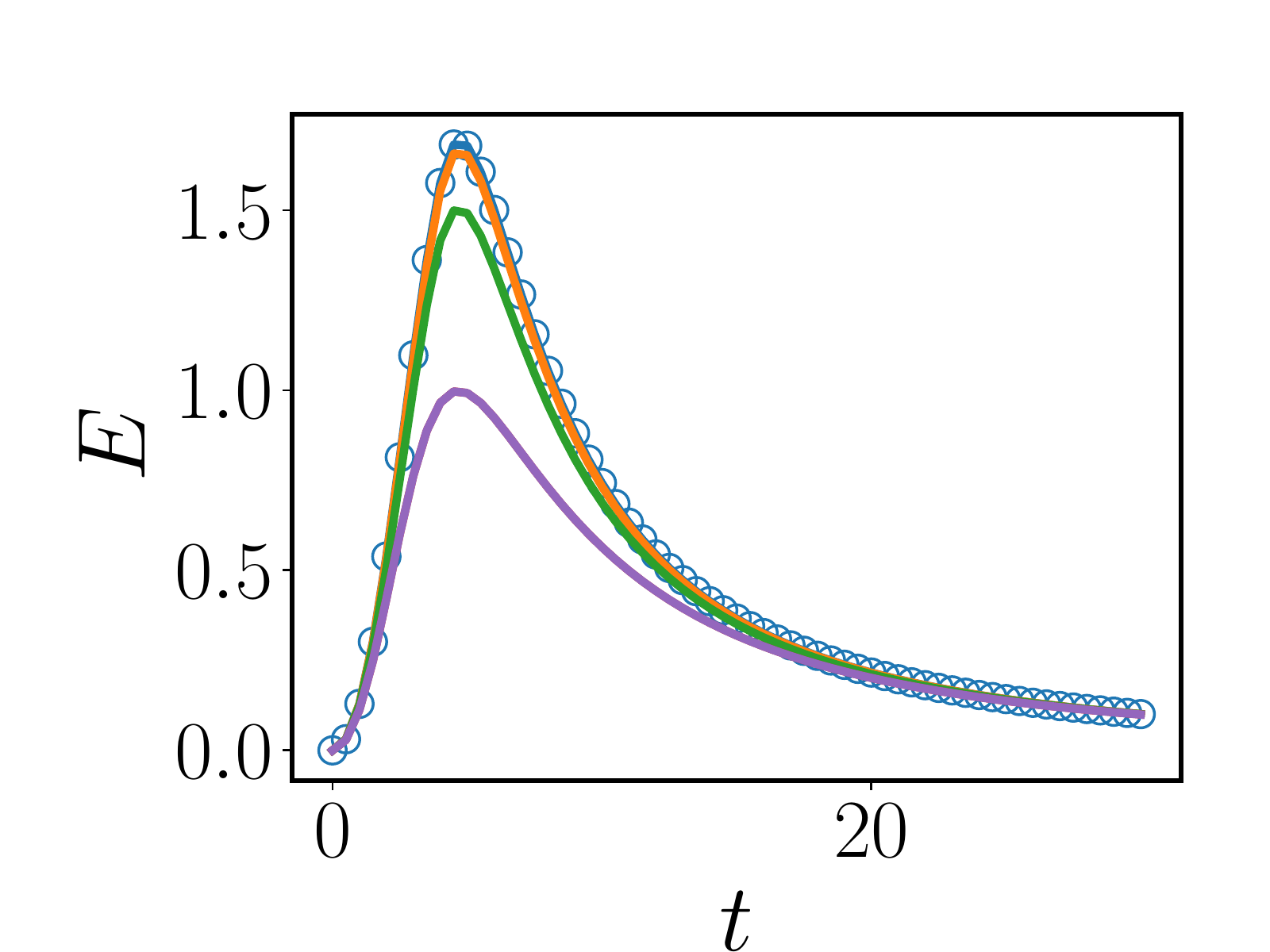}
\put(-25,110){\bf \scriptsize (f)}
\caption{[Color online] 
Time evolution plots of (a) $b$; (b) $a$; (c) $E$ for different values of $C_{a1}$.
Time evolution plots of (d) $b$; (e) $a$; (f) $E$ for different values of $C_{a2}$.
The values of the parameters which do not vary in each case are shown in Table~\ref{table1}.
}
\label{comp2}
\end{figure*}

The coefficient $C_{b2}$ multiplies a wave-like component and a diffusive component. Due to the wave-like part, an increase in $C_{b2}$ would transfer $b(k)$ from the small $k$ (large scale) modes to the large $k$ (small scale) modes, and a diffusive transfer of $b(k)$ from the large $k$ modes as well. Consistent with this interpretation, rate of decay of the mean $b$ becomes stronger as $C_{b2}$ is increased, as shown in Fig.~\ref{comp2}(d). Since $b$ is coupled to $a$, there is a corresponding decrease in the peak of $a$ as we increase $C_{b2}$ (Fig.~\ref{comp2}(e)) and a consequent reduction in the peak of $R_{nn}$, and thus $E$ (Fig.~\ref{comp2}(f)).

In Eq.~\ref{main_a} for $a_y(k)$, the spectral transfer part once again has two coefficients $C_{a1}$ and $C_{a2}$. As $C_{a1}$ is increased, more $a_y$ should be transferred from small $k$ to large $k$ modes, where it is dissipated by viscosity. 
In Fig.~\ref{comp2}(b), we see that
increase in $C_{a1}$ corresponds, as expected, to increasingly rapid decay of $a$ in time as the spectral distribution terms become more important. Since $a_y$ is coupled to $R_{nn}$ through the pressure gradient term, $E$ shows a dramatic decrease (Fig.~\ref{comp2}(c)).
However, $R_{nn}$ occurs in the inverse timescale $\Theta^{-1}$. So, quite interestingly, reduction of $R_{nn}$ reduces $\Theta^{-1}$, and thus slows the decay of $b$ as we see from Fig.~\ref{comp2}(a). $C_{a2}$ affects the decay of $a$ significantly, because, increasing $C_{a2}$ causes rapid transfer of $a_y$ from intermediate $k$ modes to large $k$ modes, where they are dissipated due to the diffusive part in the spectral transfer term.
So there is a substantial decrease in the peak of $a$
(Fig.~\ref{comp2}(e)), and the peak of $E$ (Fig.~\ref{comp2}(f)) as well. Reduction in $R_{nn}$ results in a slower decay of mean $b$ (Fig.~\ref{comp2}(d)).

\begin{figure*}
\includegraphics[width=.35\linewidth]{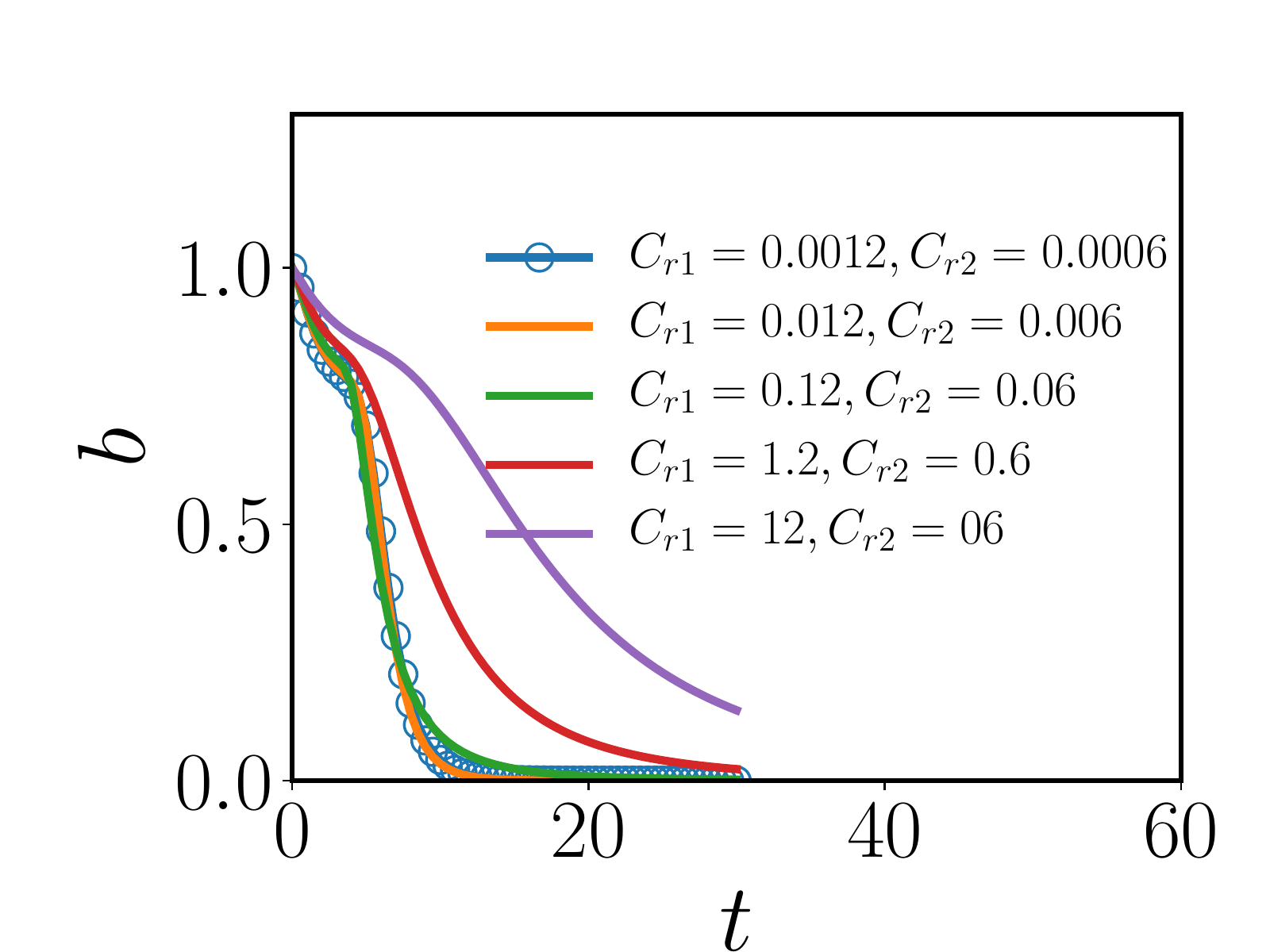}
\put(-25,110){\bf \scriptsize (a)}
\includegraphics[width=.35\linewidth]{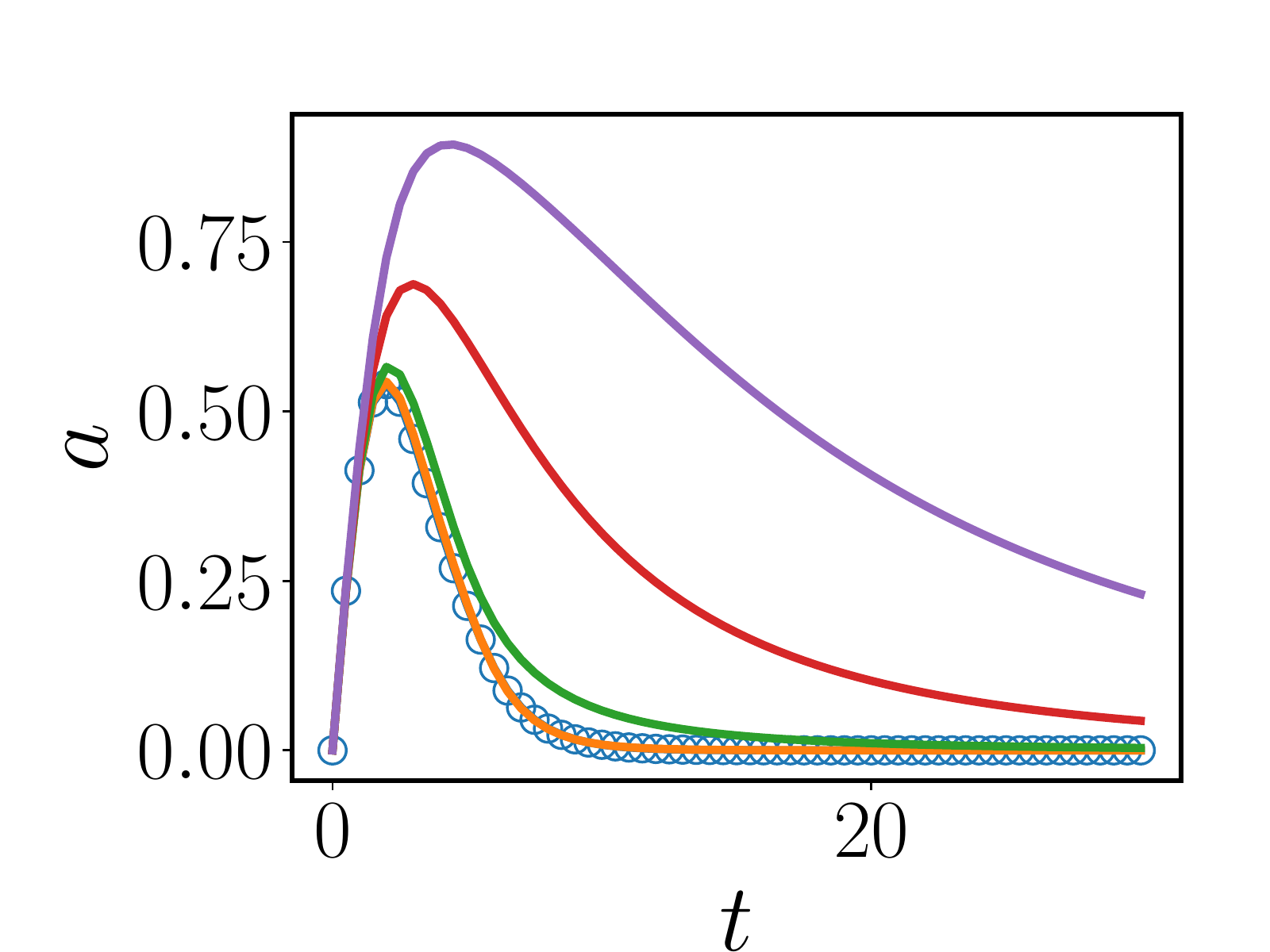}
\put(-25,110){\bf \scriptsize (b)}
\includegraphics[width=.35\linewidth]{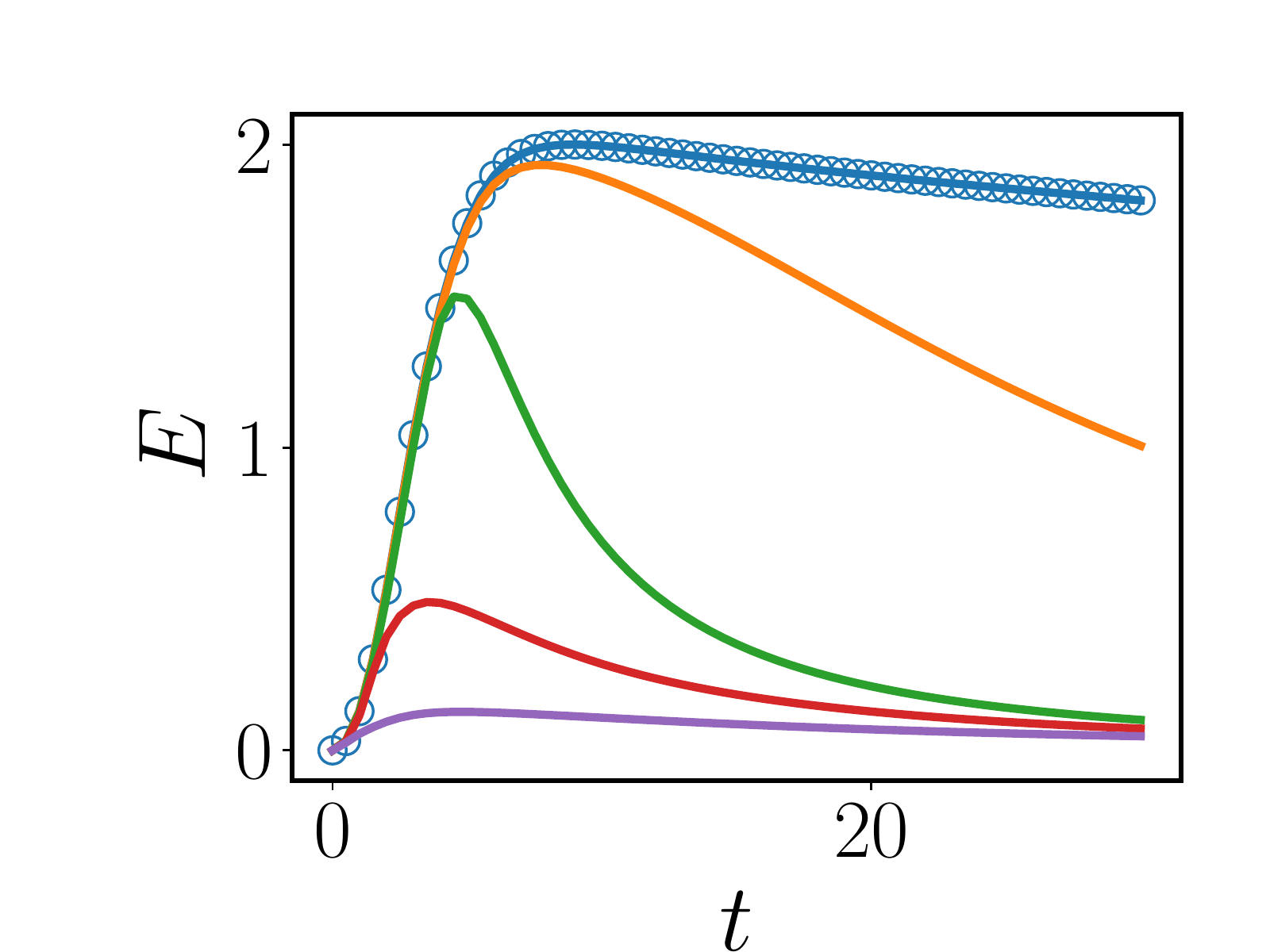}
\put(-25,110){\bf \scriptsize (c)}
\caption{[Color online] 
Time evolution plots of (a) $b$; (b) $a$; (c) $E$ for different values of $C_{r1} =  2C_{r2}$.
The values of the other parameters are shown in Table~\ref{table1}.
}
\label{comp}
\end{figure*}

Next we vary the $C_{r1}$ and $C_{r2}$ coefficients pairwise, so as to maintain the condition $C_{r1}=2C_{r2}$~\cite{lee1952some}. When we increase $C_{r1}$ and $C_{r2}$, the peak of the energy goes down, as well as the decay rates (Fig.~\ref{comp}(c)). This results in a decrease if the inverse timescale $\Theta^{-1}$. Thus $a$ and $b$ decay slowly (Fig.~\ref{comp}(a) and (b)). It is interesting to note that the shape of the energy as a function of time is very different for small values of the coefficients compared to the larger values. This variation study is the only one in our series that gives rise to this type of difference. However it is consistent with the observation that as $C_{r1}$ and $C_{r2}$ go to zero the spectral redistribution of energy ceases and only the drive and dissipation terms remain. The inflected shape of the time-series of energy as it decays (Fig. \ref{comp}(c)) is thus captured only with proper information about spectral distribution.

\begin{figure*}[ht!]
\includegraphics[width=.35\linewidth]{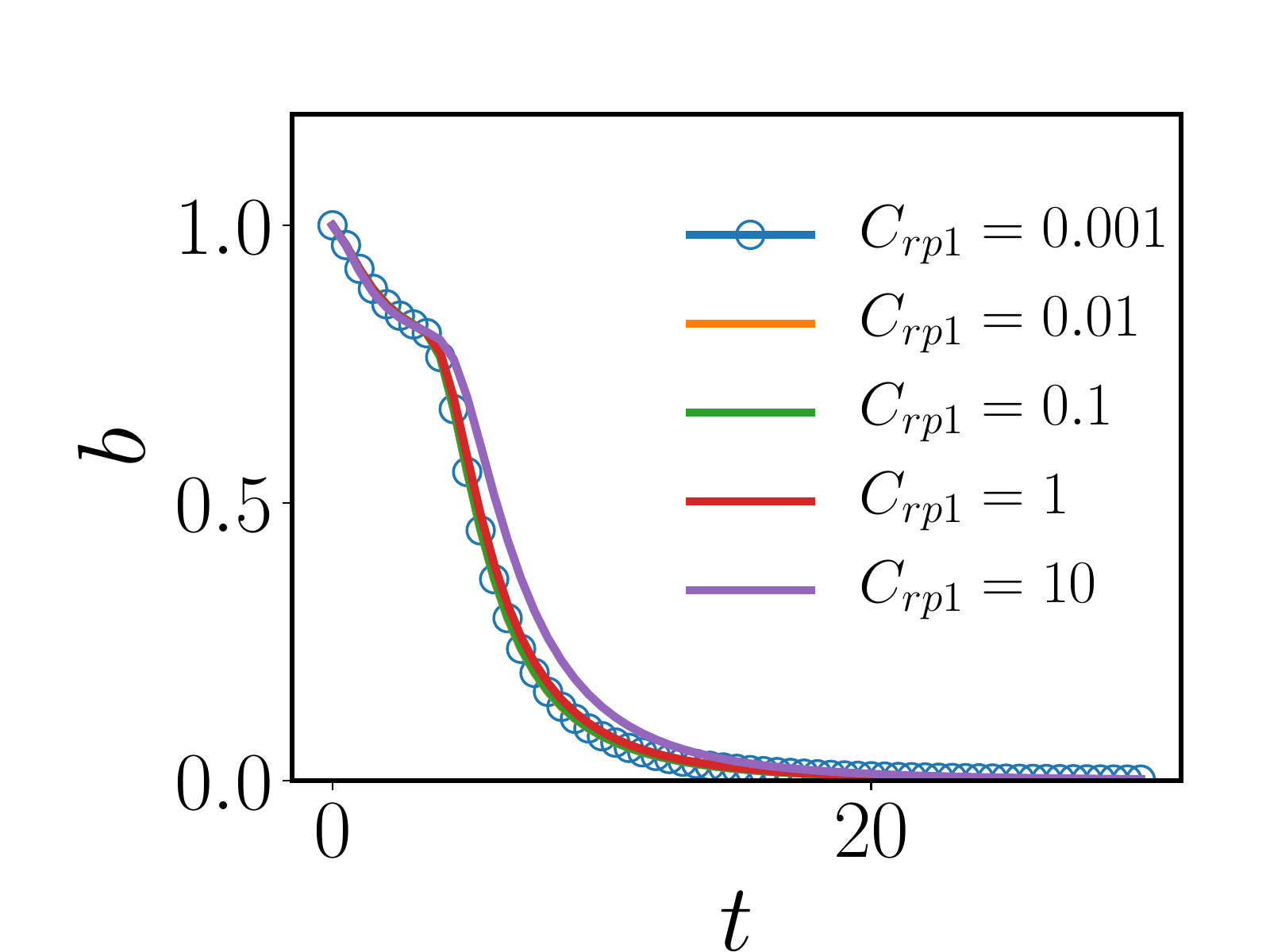}
\put(-25,110){\bf \scriptsize (a)}
\includegraphics[width=.35\linewidth]{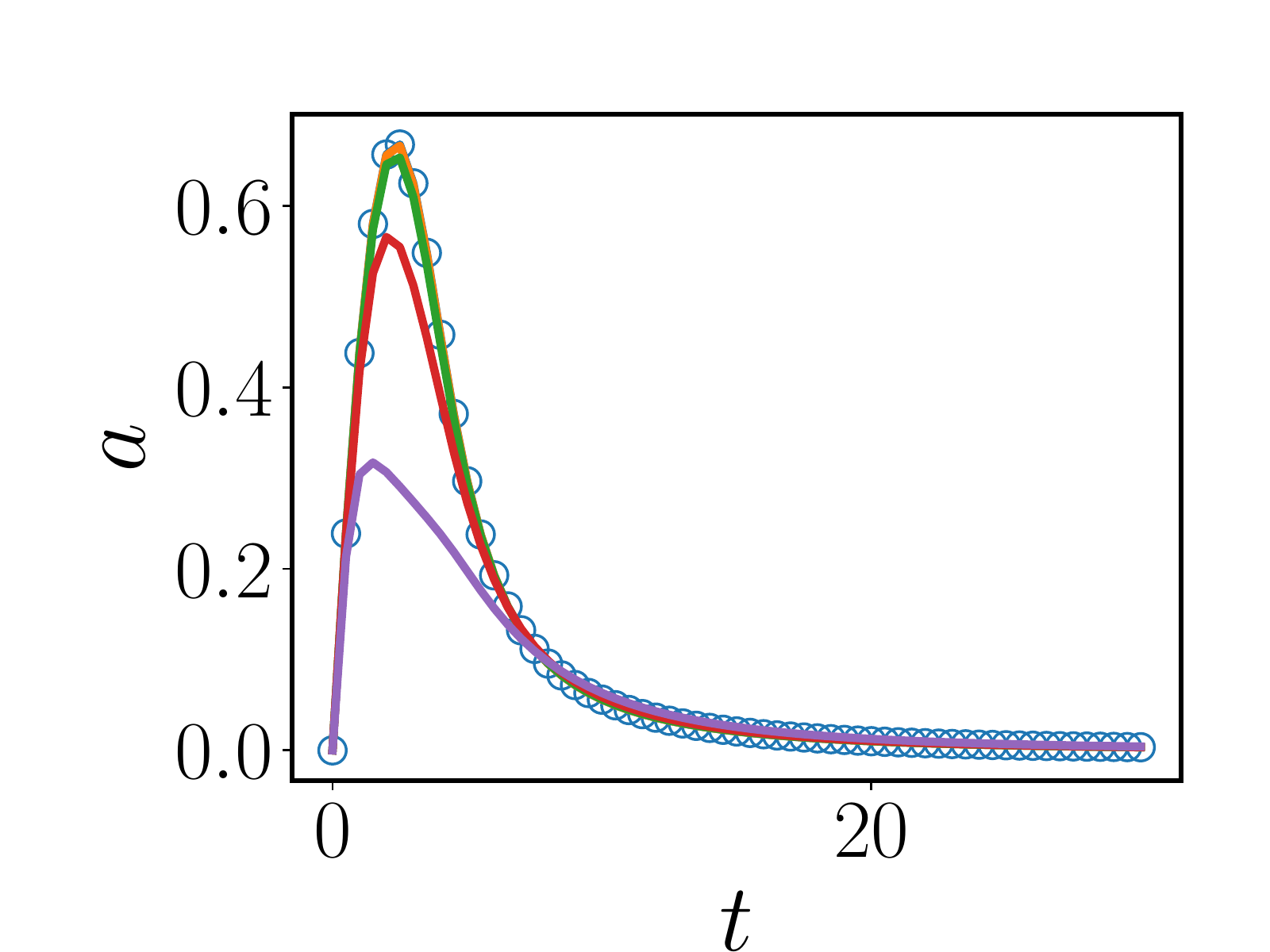}
\put(-25,110){\bf \scriptsize (b)}
\includegraphics[width=.35\linewidth]{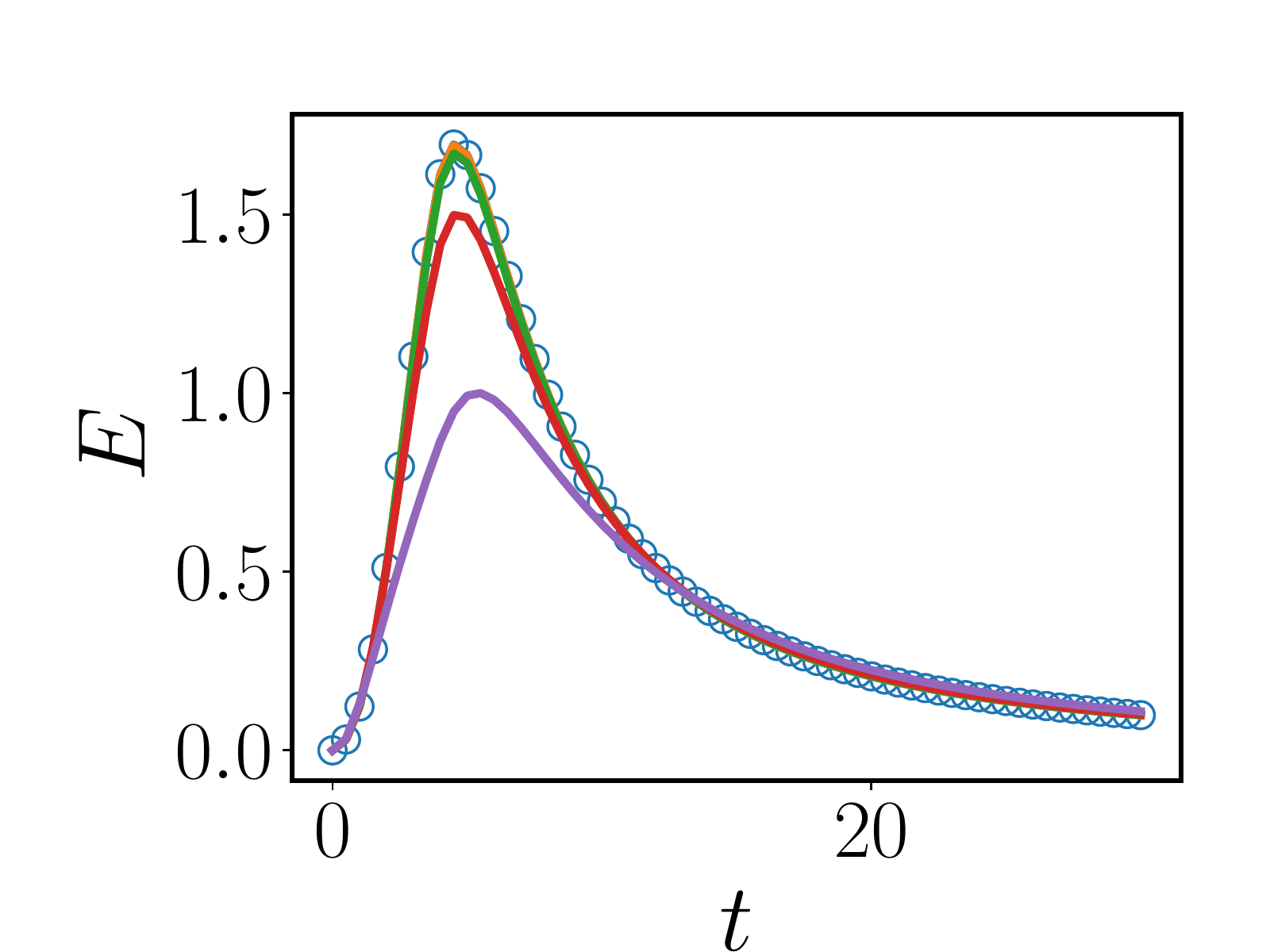}
\put(-25,110){\bf \scriptsize (c)}

\includegraphics[width=.35\linewidth]{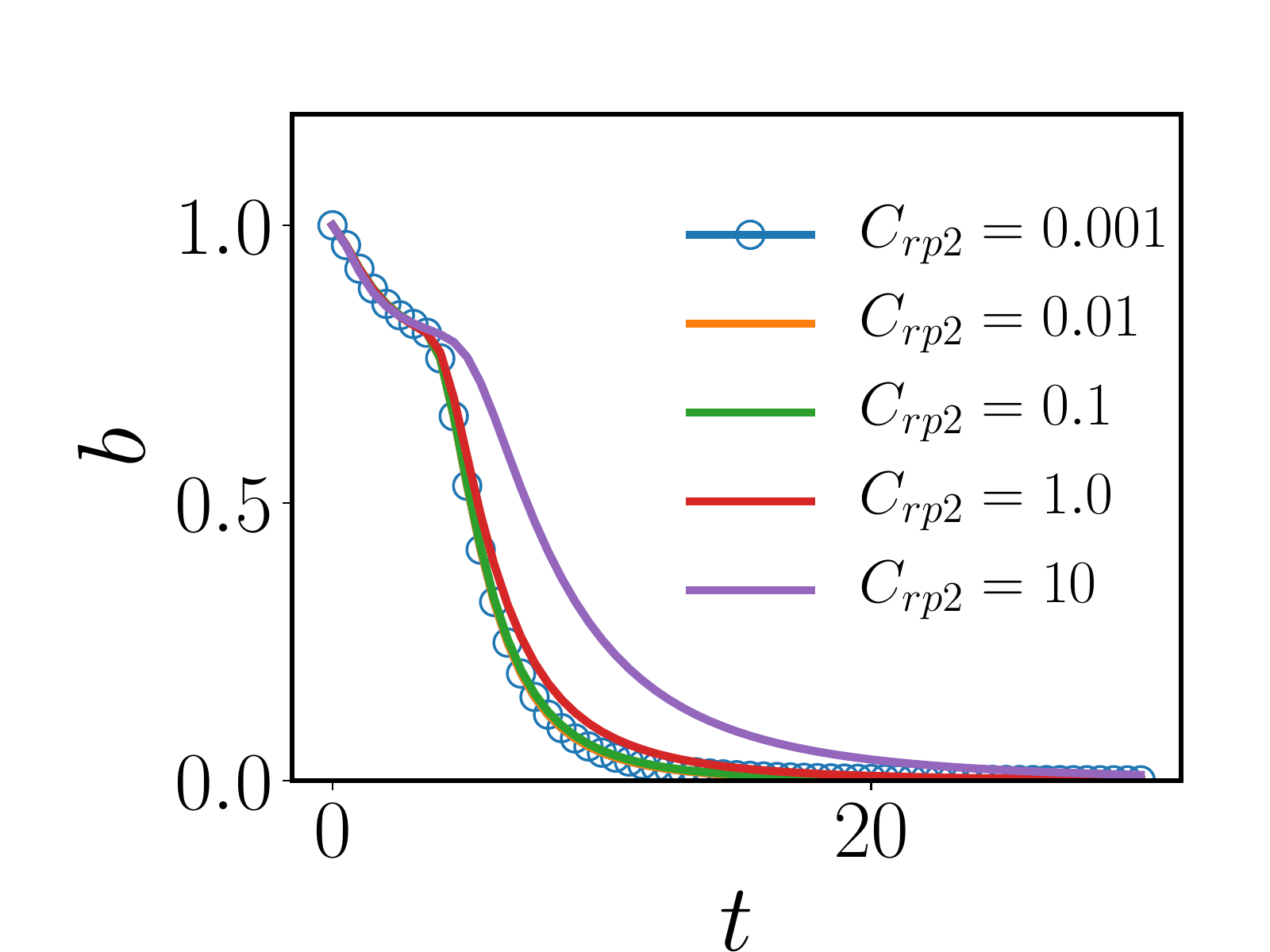}
\put(-25,110){\bf \scriptsize (d)}
\includegraphics[width=.35\linewidth]{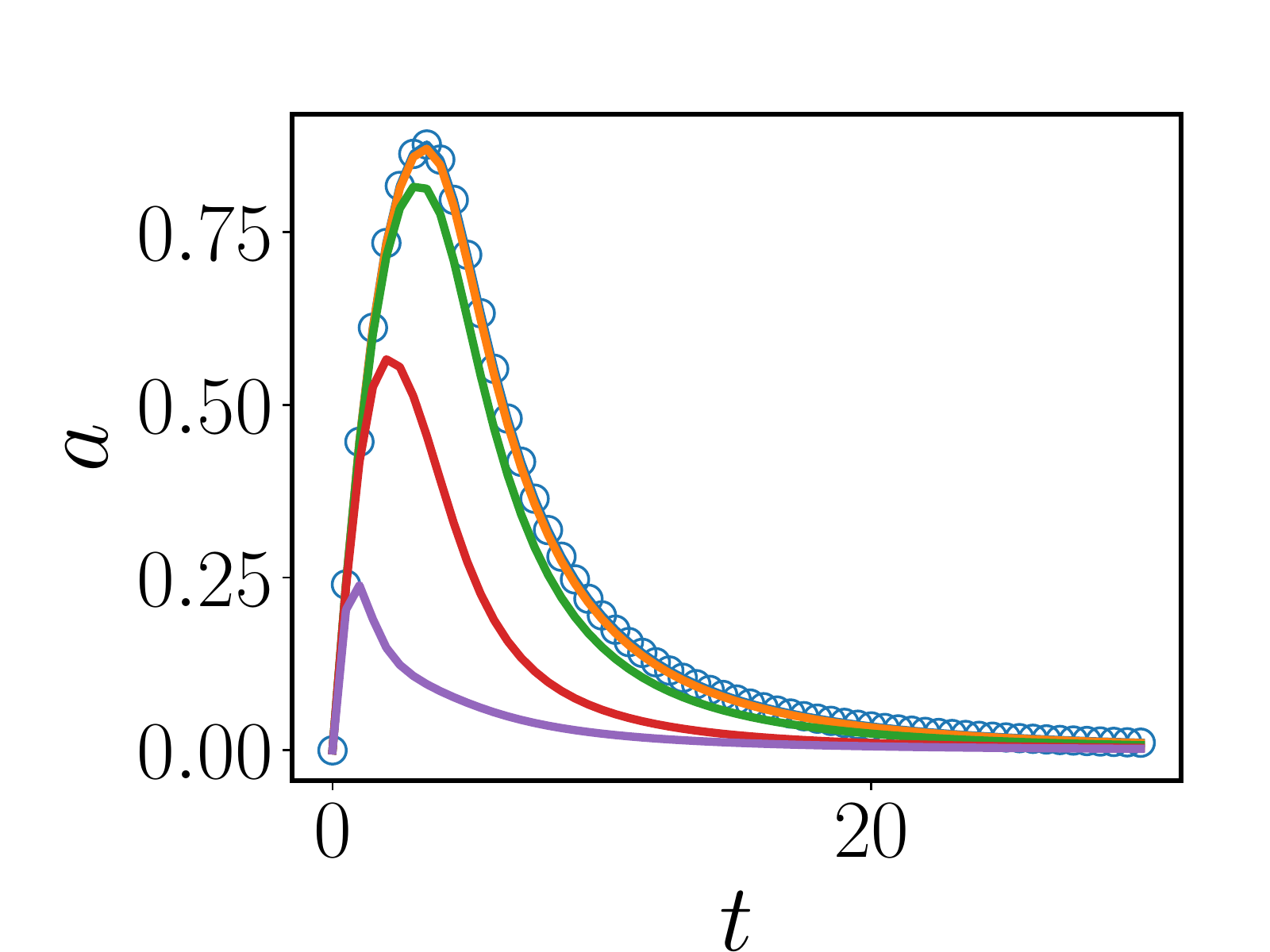}
\put(-25,110){\bf \scriptsize (e)}
\includegraphics[width=.35\linewidth]{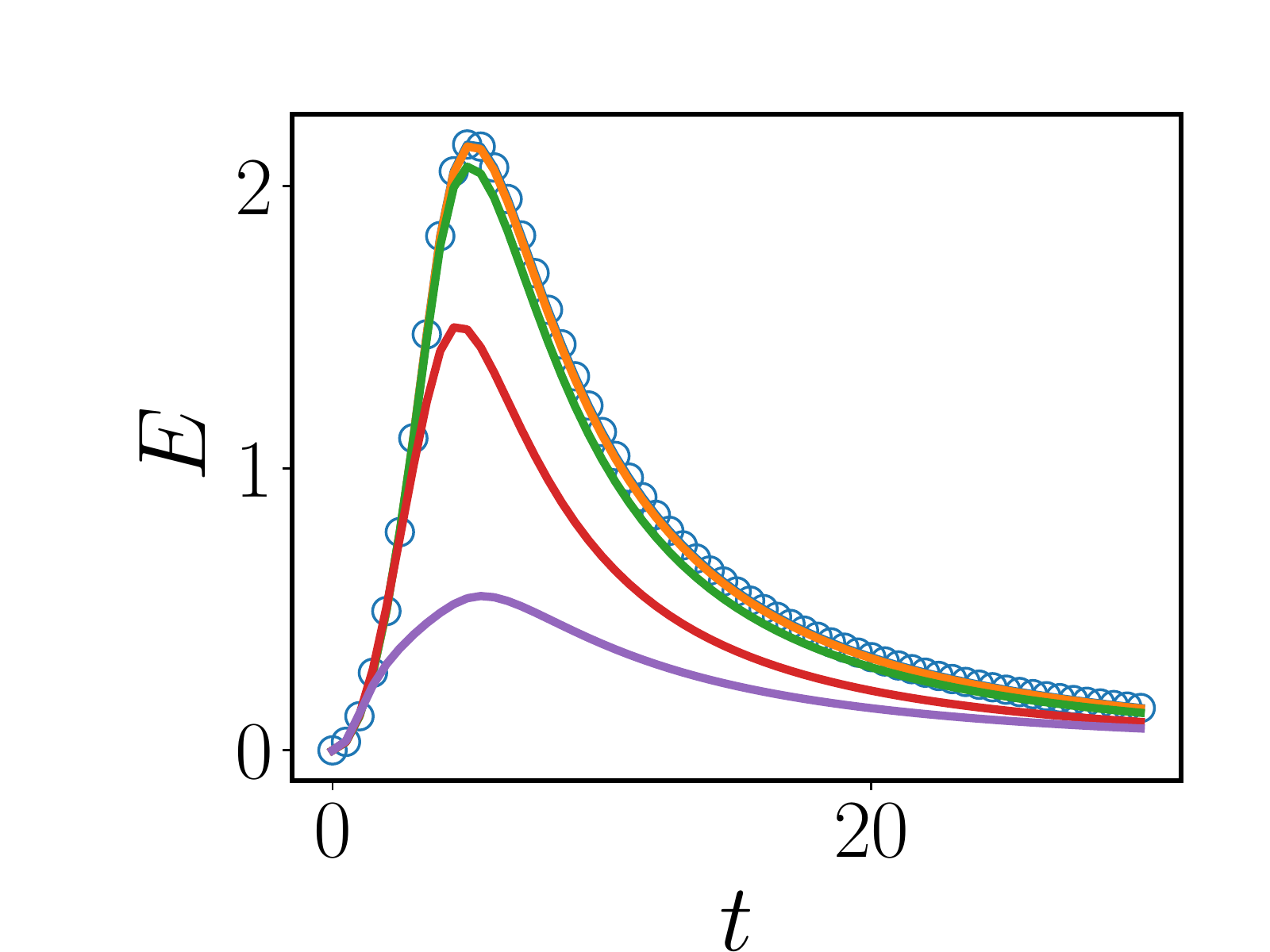}
\put(-25,110){\bf \scriptsize (f)}
\caption{[Color online] 
Time evolution plots of (a) $b$; (b) $a$; (c) $E$ for different values of $C_{rp1}$.
Time evolution plots of (d) $b$; (e) $a$; (f) $E$ for different values of $C_{rp2}$.
The values of the parameters which do not vary in each case are shown in Table~\ref{table1}.
}
\label{drag}
\end{figure*}

We also look at the effects of varying the drag coefficients $C_{rp1}$ and $C_{rp2}$ on $b,a, E$. As $C_{rp1}$ is increased, drag on $a$ increases, and thus peak of $a$ is suppressed (Fig.~\ref{drag}(b)). This reduces the production of $E$, and thus peak of $E$ is also suppressed (Fig.~\ref{drag}(c)). Reduction in $E$ reduces the turbulence frequency $\Theta^{-1}$. This slows the decay of $b$ (Fig.~\ref{comp}(a)). $C_{rp2}$ increases the decay of $a$ substantially (Fig.~\ref{comp}(e)) which in turn increases the decay of $E$ (Fig.~\ref{comp}(f)). This reduces the inverse timescale $\Theta^{-1}$, which results in a slower decay rate for $b$ (Fig.~\ref{comp}(d)).

These results are summarized in Table \ref{table_results} for convenience. With some knowledge and intuition for the impact the model coefficients we next proceed to do a more in-depth study in comparison with DNS data of variable-density buoyancy-driven turbulence.

\begin{table*}[ht!]
\resizebox{1.1\linewidth}{!}
{
\begin{tabular}{|l|l|l|l|l|l|l|l|l|l|}
\hline
 & $C_{b1}$ & $C_{b2}$ & $C_{r1}$ & $C_{r2}$&$C_{a1}$&$C_{a2}$ & $C_{rp1}$ & $C_{rp2}$ & effect\\
\hline
Effect on $b$ & $0.0012$--$12$ & $0.06$ & $0.12$ & $0.06$ & $0.12$ & $0.06$ & $1.0$ & $1.0$ & as $C_{b1}$ increases, $b$ decays faster  \\
Effect on $a$ & $0.0012$--$12.0$ & $0.06$ & $0.12$ & $0.06$ & $0.12$ & $0.06$ & $1.0$ & $1.0$ & as $C_{b1}$ increases, peak of $a$ decreases \\
Effect on $R_{nn}$ & $0.0012$--$12.0$ & $0.06$ & $0.12$ & $0.06$ & $0.12$ & $0.06$ & $1.0$ & $1.0$ & as $C_{b1}$ increases peak of $E$ decreases \\
\hline
Effect on $b$ & $0.12$ & $0.006$--$6.0$ & $0.12$ & $0.06$ & $0.12$ & $0.06$ & $1.0$ & $1.0$ & decay rate increases as $C_{b2}$ increases \\
Effect on $a$ & $0.12$ & $0.006$--$6.0$ & $0.12$ & $0.06$ & $0.12$ & $0.06$ & $1.0$ & $1.0$ & peak decreases  as $C_{b2}$ increases \\
Effect on $R_{nn}$ & $0.12$ & $0.006$--$6.0$ & $0.12$ & $0.06$ & $0.12$ & $0.06$ & $1.0$ & $1.0$ & peak decreases as $C_{b2}$ increases\\
\hline
Effect on $R_{nn}$ & $0.12$ & $0.06$ & $0.0012$--$12.0$ & $0.0006$--$6$ & $0.12$ & $0.06$ & $1.0$ & $1.0$ & as $C_{r1}$, $C_{r2}$ increases, peak decreases \\
Effect on $a$ & $0.12$ & $0.06$ & $0.0012$--$12.0$ & $0.0006$--$6$ & $0.12$ & $0.06$ & $1.0$ & $1.0$ & as $C_{r1}$, $C_{r2}$ increases, decay of $a$ decreases   \\
Effect on $b$ & $0.12$ & $0.06$ & $0.0012$--$12.0$ & $0.0006$--$6$ & $0.12$ & $0.06$ & $1.0$ & $1.0$ & as $C_{r1}$, $C_{r2}$ increases, decay of $b$ decreases  \\
\hline
Effect on $a$ & $0.12$ & $0.06$ & $0.12$& $0.06$ & $0.0012$--$12$ & $0.06$ & $1.0$ & $1.0$ & as $C_{a1}$ increases, decay of $a$ is better\\
Effect on $R_{nn}$ & $0.12$ & $0.06$ & $0.12$ & $0.06$ & $0.0012$--$12.0$ & $0.06$ & $1.0$ & $0.25$ & as $C_{a1}$ increases, peak decreases \\
Effect on $b$ & $0.12$ & $0.06$ & $0.12$ & $0.06$ & $0.0012$--$12.0$ & $0.06$ & $1.0$ & $1.0$ & as $C_{a1}$ increases, decay is slower\\
\hline
Effect on $a$ & $0.12$ & $0.06$ & $0.12$ & $0.06$ & $0.12$ & $0.006$ -- $6.0$ & $1.0$ & $1.0$ & peak decreases  as $C_{a2}$ increases.\\
Effect on $R_{nn}$ & $0.12$ & $0.06$ & $0.12$ & $0.06$ & $0.12$ & $0.006$ -- $6.0$ & $1.0$ & $1.0$ & peak decreases as $C_{a2}$ increases \\
Effect on $b$ & $0.12$ & $0.06$ & $0.12$ & $0.06$ & $0.12$ & $0.006$ -- $6.0$ & $1.0$ & $1.0$ & decay rate decreases  as $C_{a2}$ increases .\\
\hline
Effect on $a$ & $0.12$ & $0.06$ & $0.12$ & $0.06$ & $0.12$ & $0.06$ & $0.0001-10.0$ & $1.0$ & peak decreases  as $C_{rp1}$ increases.\\
Effect on $R_{nn}$ & $0.12$ & $0.06$ & $0.12$ & $0.06$ & $0.12$ & $0.06$ & $0.0001-10.0$ & $1.0$ & peak decreases  as $C_{rp1}$ increases.\\
Effect on $b$ & $0.12$ & $0.06$ & $0.12$ & $0.06$ & $0.12$ & $0.06$ & $0.0001-10.0$ & $1.0$ & decay rate decreases  as $C_{rp1}$ increases.\\
\hline
Effect on $a$ & $0.12$ & $0.06$ & $0.12$ & $0.06$ & $0.12$ & $0.06$ & $1.0$ & $0.0001$ -- $10$ & peak decreases  as $C_{rp2}$ increases.\\
Effect on $R_{nn}$ & $0.12$ & $0.06$ & $0.12$ & $0.06$ & $0.12$ & $0.06$ & $1.0$ & $0.0001$ -- $10$ & peak decreases  as $C_{rp2}$ increases.\\
Effect on $b$ & $0.12$ & $0.06$ & $0.12$ & $0.06$ & $0.12$ & $0.06$ & $1.0$ & $0.0001$ -- $10$ & decay rate decreases  as $C_{rp2}$ increases.\\
\hline
\end{tabular}
}
\caption{The parameters $C_{b1}$, $C_{b2}$, $C_{r1}$, $C_{r2}$, $C_{a1}$, $C_{a2}$, $C_{rp1}$, and $ C_{rp2}$ for different calculations. The initial condition was a Gaussian in $k$ space for $b$ field, and $a,E$ were set to zero. The system had $512$-$k$ modes. The viscosity is kept fixed at $10^{-4}$. The diffusivity of the density field is also $10^{-4}$.}
\label{table_results}
\end{table*}

\subsection{Comparison and optimization with respect to DNS data}

In this section we will attempt to optimize the coefficients for a particular problem that has been exactly computed using the equations of motion in a highly resolved Direct Numerical Simulation. The goal is to demonstrate the operation of the model for a realistic problem and assess whether and how accurately the modeling assumptions capture both integrated and spectral quantities.

\begin{figure*}[ht!]
\includegraphics[width=.5\linewidth]{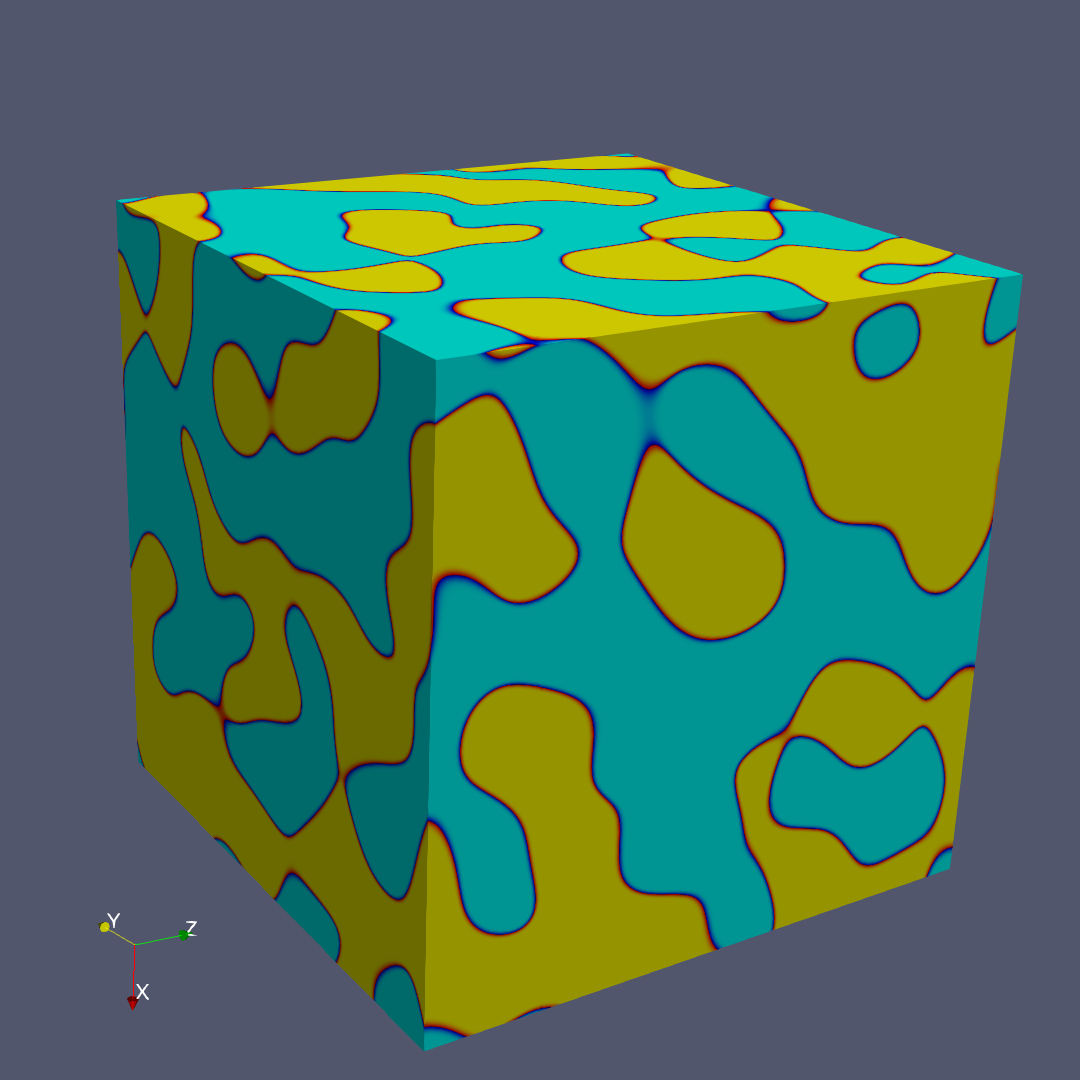}
\put(-20,240){\bf \textcolor{white}{(a)}}~
\includegraphics[width=.5\linewidth]{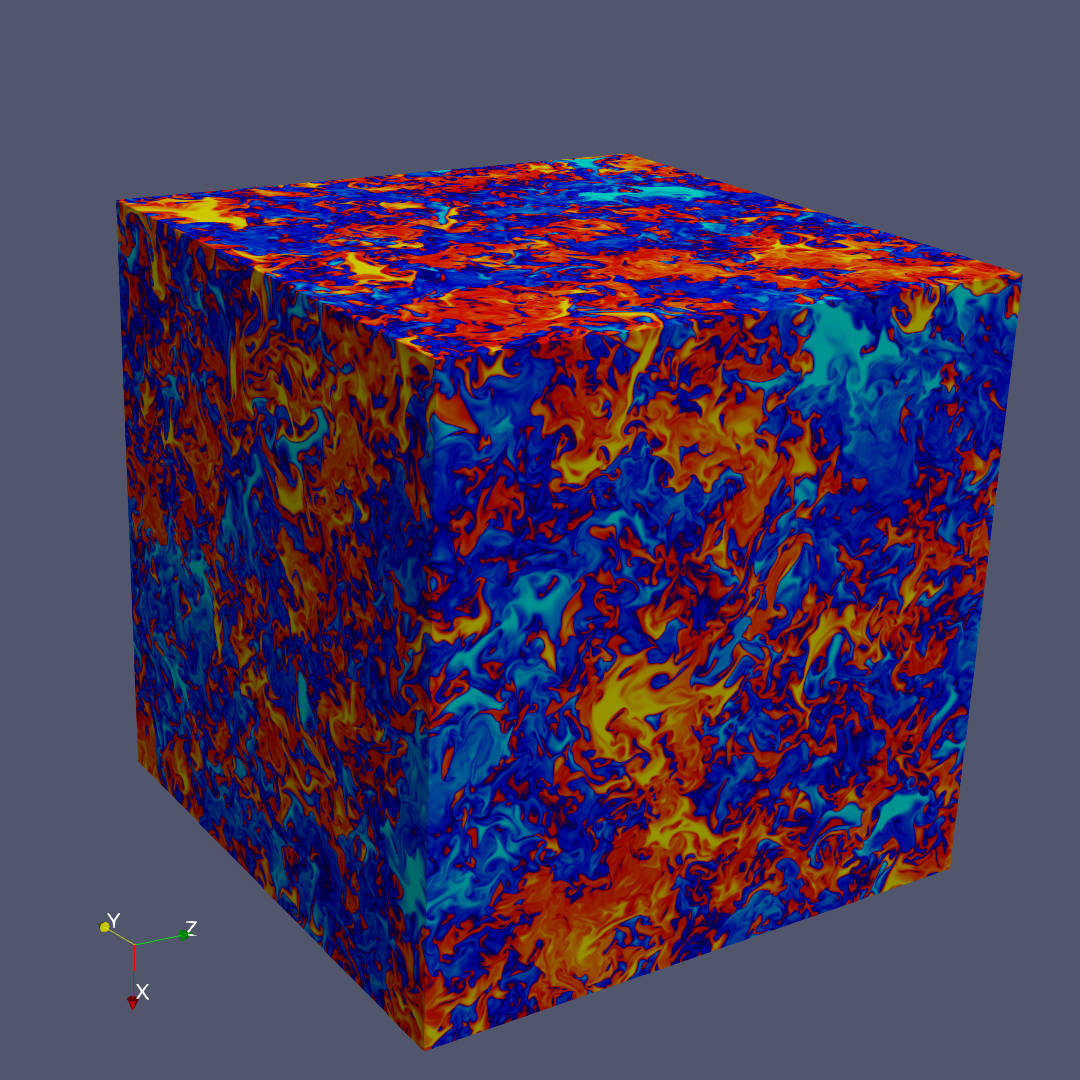}
\put(-20,240){\bf \textcolor{white}{(b)}}

\caption{3D visualization of the density field for the $At=0.05$ DNS run ($1024^3$ mesh) at a) initial time and b) turbulent kinetic energy peak time as described in Ref.~\cite{aslangil2018}.}
\label{spectra}
\end{figure*}
The DNS set-up follows the triply periodic buoyancy driven turbulence studied in Refs.~\cite{livescu2007,livescu2008}.
This flow represents a homogeneous version of the classical Rayleigh-Taylor instability and, during the growth stage, resembles the interior of the Rayleigh-Taylor mixing layer. The flow is described by the variable-density Navier-Stokes equations,
which are obtained as the incompressible (infinite speed of sound) limit of the fully compressible Navier-
Stokes equations with two miscible species with different molar masses~\cite{livescu2009cfdns,livescu2013numerical}. In this limit, the
density variations arise from compositional changes as the two species mix and lead to non-zero divergence
of velocity. The boundary conditions are triply periodic, and the two fluids are initialized
as random blobs, consistent with the homogeneity assumption. The flow starts from rest, with only
a small amount of dilatational velocity necessary to satisfy the divergence condition and turbulence
is generated as the two fluids start moving in opposite directions due to differential buoyancy forces.
However, as the fluids become molecularly mixed, the buoyancy forces decrease and at some point
turbulence starts decaying. The non-stationary evolution of turbulence, resulting from the interplay between buoyancy turbulence production and mixing, is very difficult to be captured by one-point models~\cite{Schwarzkopf2016}. 

To calibrate and test the spectral model, we use new higher resolution simulations~\cite{aslangil2018}.  Similar to Refs.~\cite{livescu2007,livescu2008}, the simulations were performed with the CFDNS code~\cite{livescu2009cfdns}, using a pseudo-spectral method. The time integration was performed with a third order predictor-corrector Adams-Bashforth-Moulton method coupled with a pressure projection method, which results in a variable coefficient Poisson equation. The solution method uses direct Poisson solvers, with no loss of accuracy. The density contrast between the fluids is obtained from the value of the Atwood number $\displaystyle At=\frac{\rho_{max}-\rho_{min}}{\rho_{max}+\rho_{min}}$. Here, we use two sets of simulations, with $At=0.05$ and $At=0.75$, on $(2\pi)^3$ domains discretized using $1024^3$ meshes. The initial density spectrum is a top-hat between wavenumbers 3 and 5, resulting in a normalized initial density integral scale of $~0.21$ and a mixing state metric $\theta \sim 0.07$. The maximum turbulent Reynolds number attained by the two cases ($At=0.05$ and $At=0.75$) are $13330$ and $2230$, resulting in Taylor Reynolds numbers using the isotropic formula of $298$ and $122$, respectively.

In order to remain as systematic as possible given the relatively large number of tuneable coefficients, we 
first assign nominal values prescribed in \cite{clark1995two,Steinkamp1999a}. The values of $C_{r1} = 0.12$ and $C_{r2} = 0.06$, their relationship constrained by \cite{lee1952some}, have the 
most prior validation due to studies of single-fluid homogeneous isotropic and anisotropic flows \cite{clark1995two, Clarkthesis}. Although we do not have have a theoretical expectation for these in the variable density case, but the single fluid values seem a reasonable place to search for an optimum. The corresponding spectral transfer coefficients for $a$ ($C_{a1}$ and $C_{a2}$) and $b$ ($C_{b1}$ and $C_{b2}$) were assigned to be identical to $C_{r1}$ and $C_{r2}$ respectively on a provisional basis in previous work, but there exists no theory or other expectation for these to the best of our knowledge. The drag coefficients for $C_{rp1}$ and $C_{rp2}$ were set to unity in \cite{clark1995two} but assigned the values of 5 and 6 respectively by \cite{Steinkamp1999a} following arguments by \cite{youngs1992two} for effective drag around bluff bodies and spheres. 
We found that the values of 5 and 6 for $C_{rp1}$ and $C_{rp2}$ respectively in the present comparison were too high in that they strongly damped the growth of $a$ relative to the DNS. Therefore we choose drag coefficients around unity as in \cite{clark1995two}.


We proceed to optimize the coefficients as follows. Each coefficient is varied, keeping all others fixed, and we can define the Pearson's $\chi^2$ test function \cite{pearson1956karl} as follows:
\begin{equation}
\chi^2=\sum \limits_{t=0}^{t_{max}} \frac{\Big(\left(b_{m}-b_{D}\right) + \left(a_{m}-a_{D}\right)/a_0 + \left(E_{m}-E_{D}\right)/E_0\Big)^2}{b_D + a_D/a_0 + E_D/E_0}
\label{chisq}
\end{equation}
where $b_{m} = b_{m}(t)$ refers to the mean density specific-volume correlation obtained from the model as a function of time, $b_D = b_D(t)$ are the corresponding values from the resolved DNS. The quantities for $a$ and $E$ are defined analogously. $a_0 = E_0 = 1$ are chosen so that they have the same dimensions as $a_m$ and $E_m$ respectively. The upper limit on time $t_{max} = 20$ in our case and is chosen so that the essential features of growth, peak and decay are sufficiently captured without weighting the results too much by the very late times in which errors are naturally minimized. 

\begin{figure*}
\includegraphics[scale = 0.35]{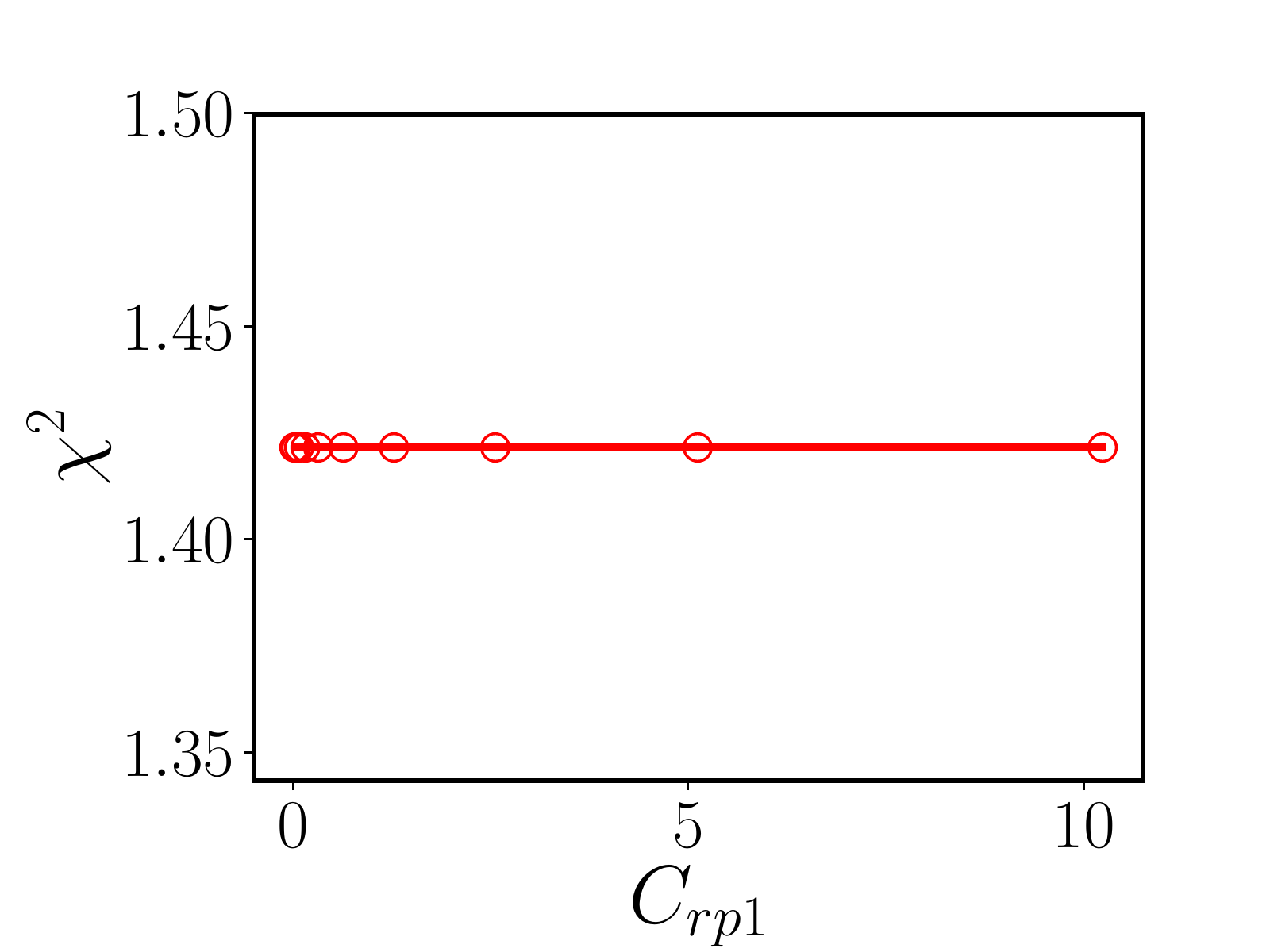}
\put(-128,100){\bf \scriptsize (a)}
\includegraphics[scale = 0.35]{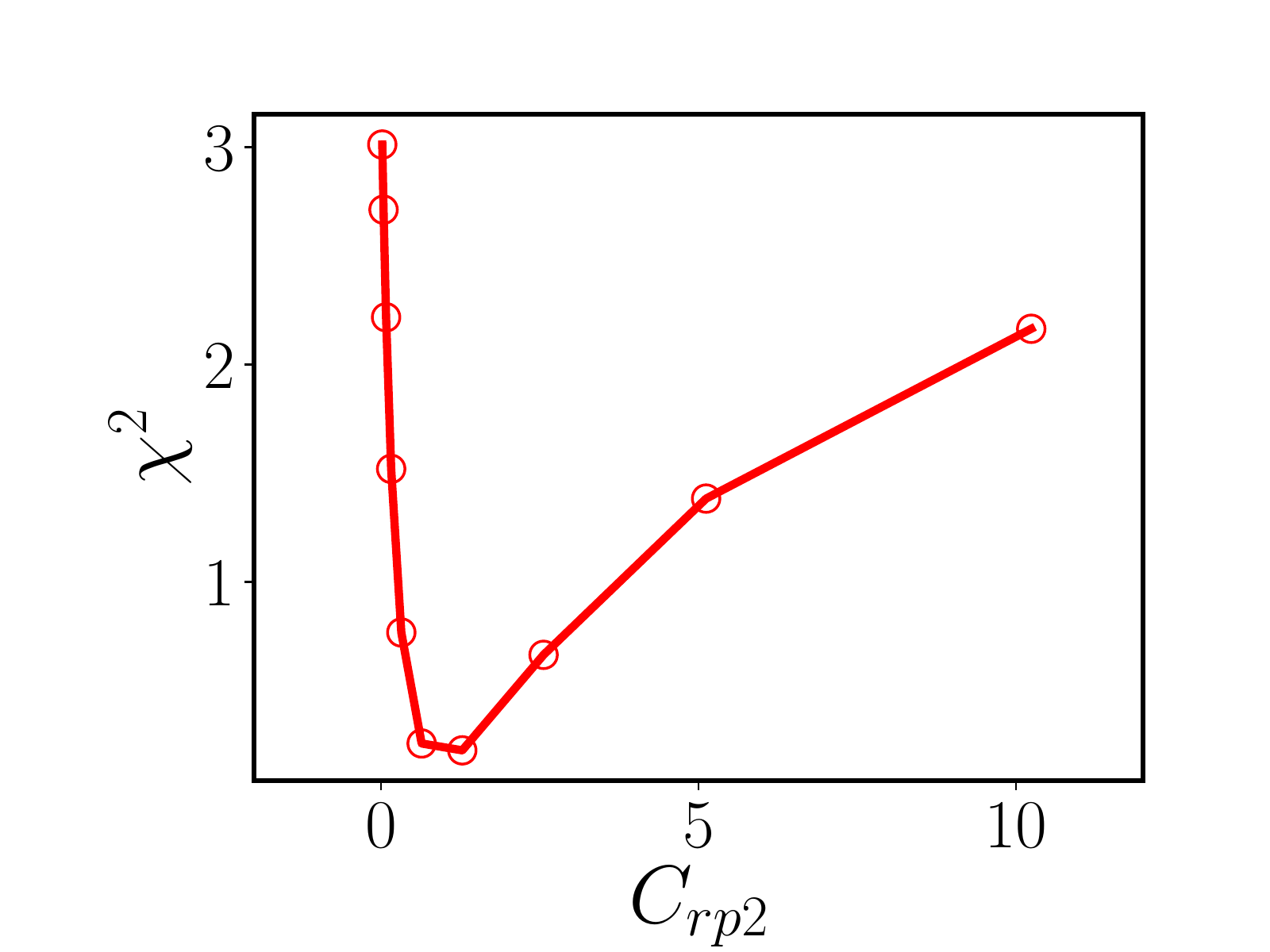}
\put(-128,100){\bf \scriptsize (b)}
\includegraphics[scale = 0.35]{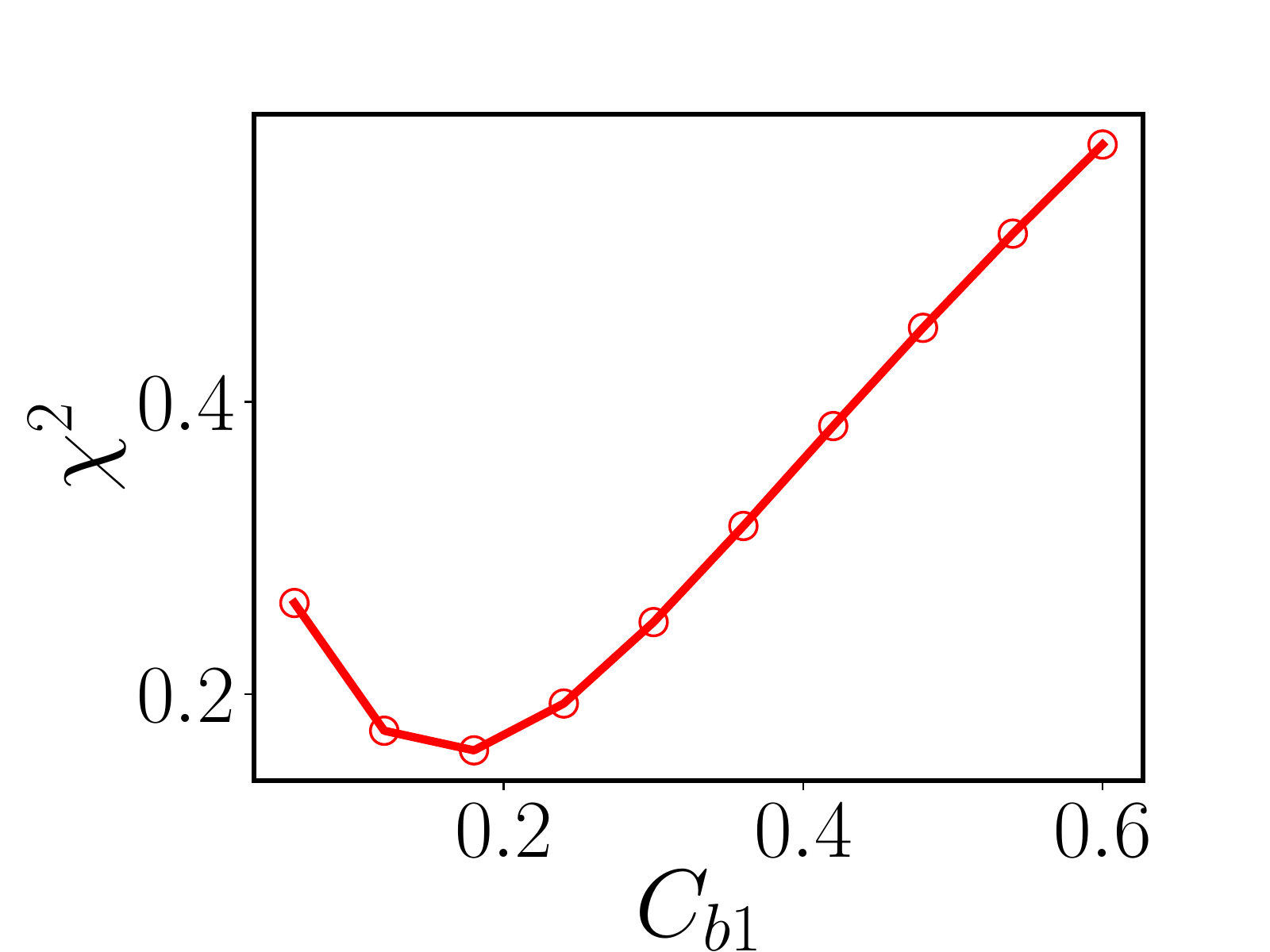}
\put(-128,100){\bf \scriptsize (c)}

\includegraphics[scale = 0.35]{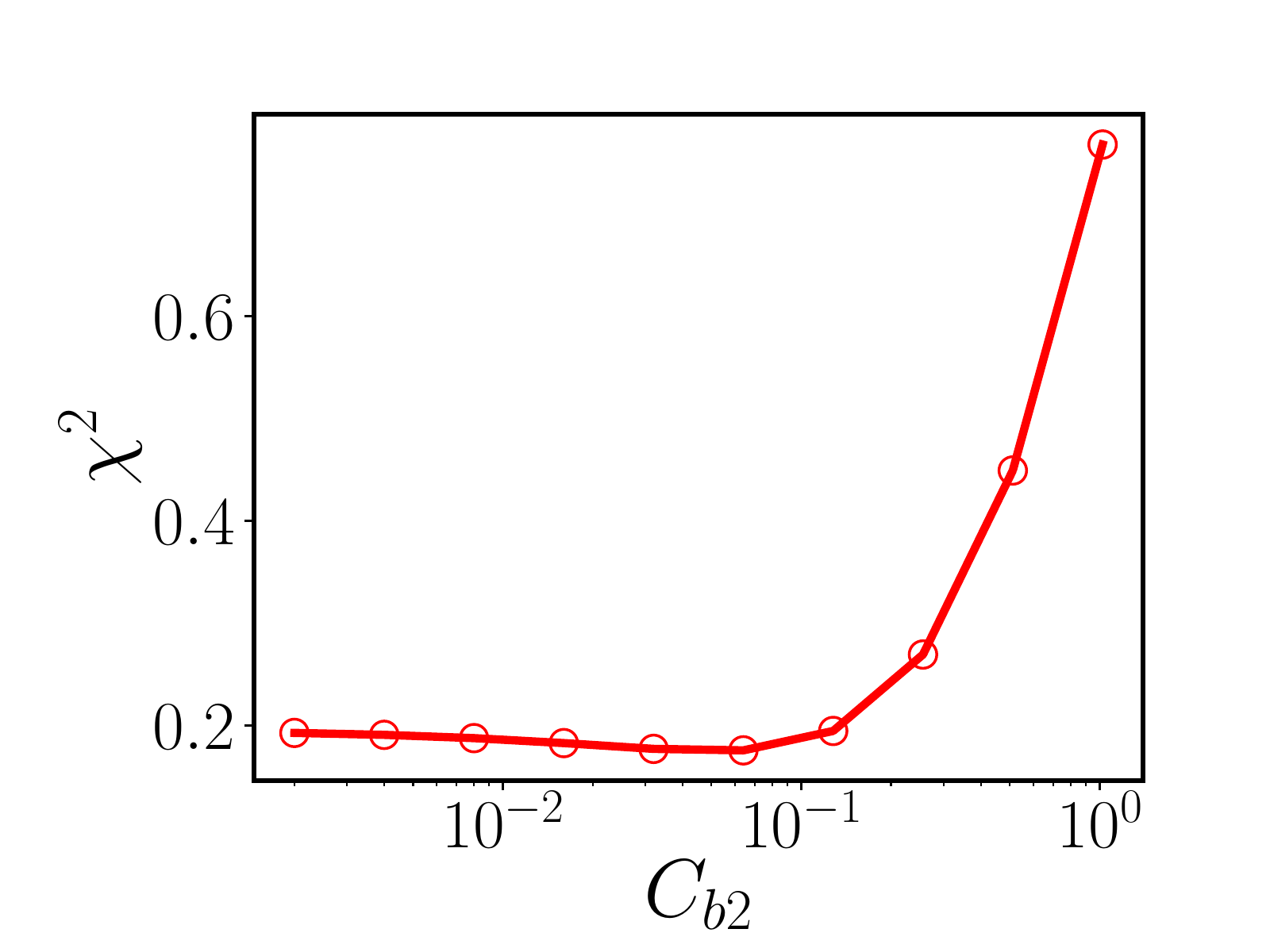}
\put(-128,100){\bf \scriptsize (d)}
\includegraphics[scale = 0.35]{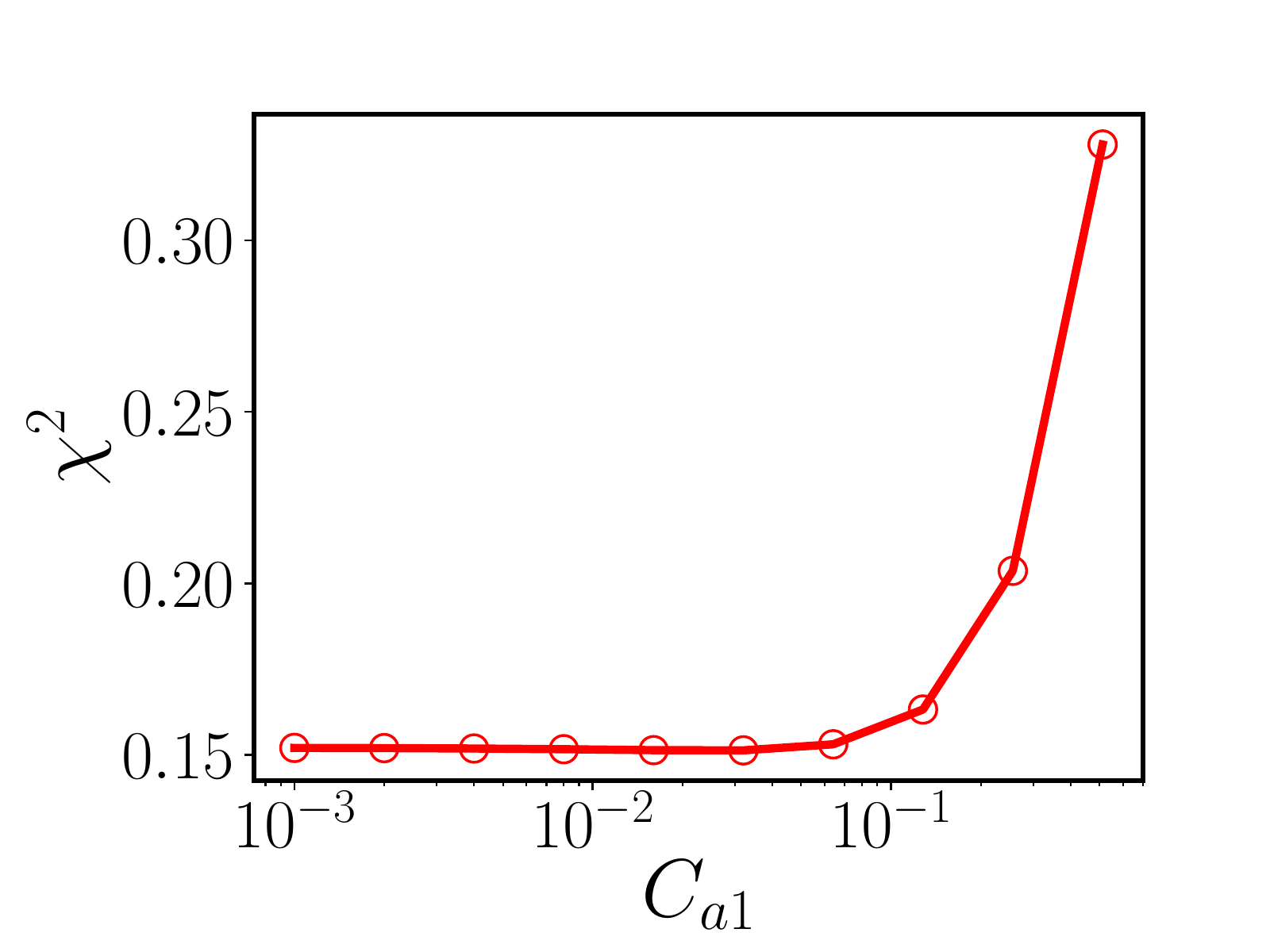}
\put(-128,100){\bf \scriptsize (e)}
\includegraphics[scale = 0.35]{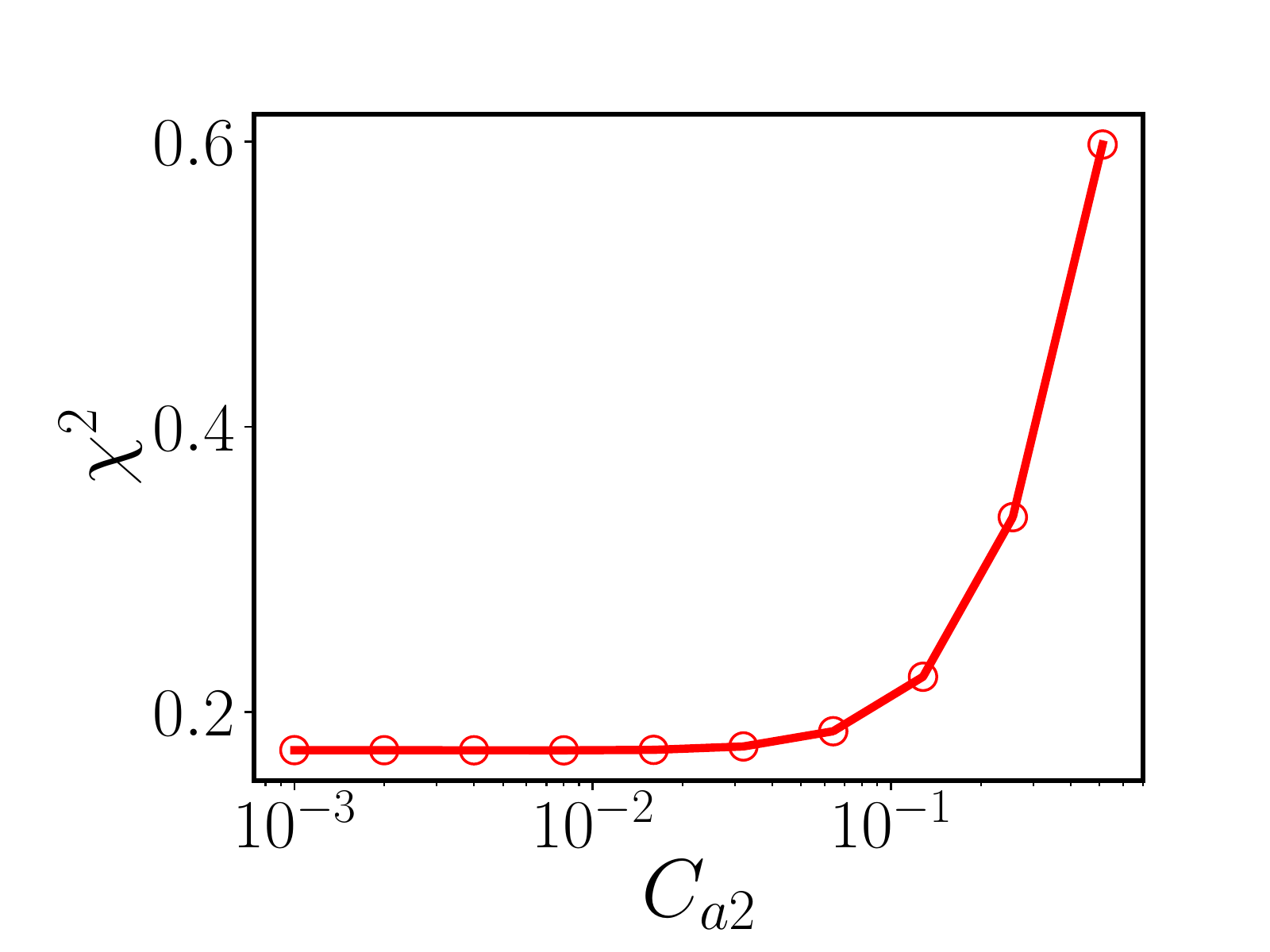}
\put(-128,100){\bf \scriptsize (f)}

\includegraphics[scale = 0.35]{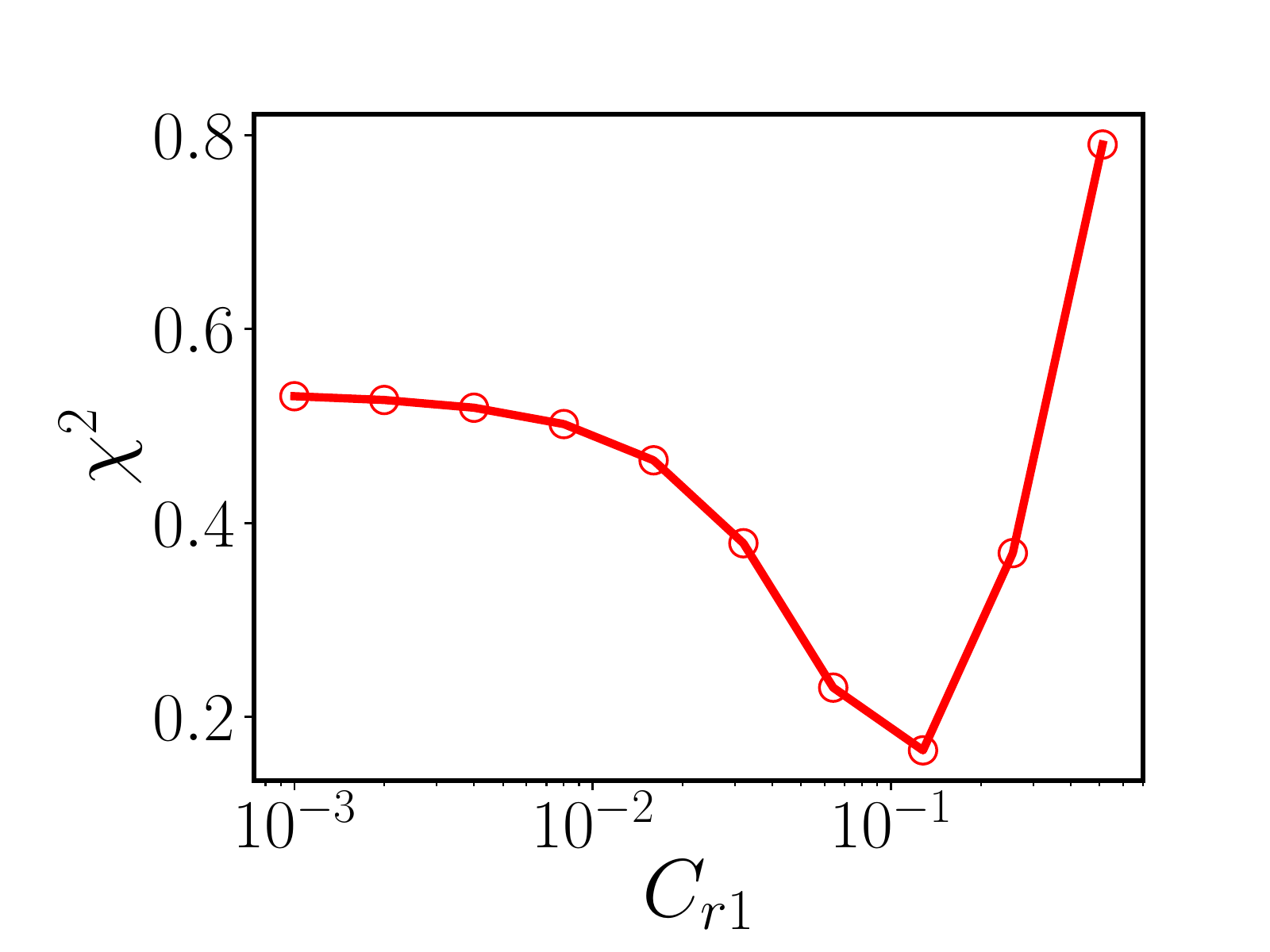}
\put(-128,100){\bf \scriptsize (g)}
\caption{[Color online] Plots of $\chi^2$ function calculated using Eq.~\ref{chisq} for different values of $C_{rp1}, C_{rp2}$, $C_{b1}$, $C_{b2}, C_{a1}, C_{a2}$ and $C_{r1}$. This minimum value of the $\chi^2$ function occurs at the optimum value of each variable.}
\label{chis}
\end{figure*}

We begin by optimizing the drag coefficient $C_{rp1}$  keeping all other constants at their nominal values. Its optimum value (and those of other coefficients subsequently) is found by minimizing the function (\ref{chisq}), as shown in the first panel of Fig. \ref{chis}. As may be seen, the error function is relatively insensitive to variation in $C_{rp1}$ and retaining a value of unity is appropriate. The $\chi^2$ as function of $C_{rp2}$ shows a minimum at $C_{rp2} \sim 1.0$. We may prescribe a conservative uncertainty estimate on the minima by specifying (roughly) the range over which the minimum $\chi^2$ is doubled. With this error specification, $C_{rp2}\simeq 1.0 \pm 0.5$ as is shown in Fig.~\ref{chis}(b), and so it is also retained as unity. 

The other coefficients are optimized in a similar manner and the quality of the optimizations are 
shown in Fig. \ref{chis}. There are two types of minima observed for the $\chi^2$ error functions shown. The first is a true parabolic minimum as for $C_{rp2}$, $C_{b1}$ and $C_{r1}$ (Fig. \ref{chis} (b), (c) and (g), and the second is an asymptotic minimum as for $C_{b2}$, $C_{a1}$ and $C_{a2}$. The extreme case is the error with respect to the already discussed $C_{rp1}$ which shows no dependence of $\chi^2$ on the coefficient value at all. 
Table \ref{table2} shows the optimized values ({\tt R1}) of all coefficients for the $At=0.05$ case along with their uncertainties. For the cases that do not have a clear minimum the nominal values from \cite{clark1995two} are retained and the range of uncertainty is taken to be all values between 0 and the first instance of the minimum. Note that the values obtained by minimizing the error over all the dynamical variables simultaneously do not depart significantly from the nominal values proposed in \cite{clark1995two}. Indeed $C_{r1}$ and $C_{r2}$ which may be derived from the Kolmogorov constant and the Lee equipartition constraint are minimized at the theoretically expected values which is a strong validation of the model and, less directly, of the assumption that the energy cascade in the variable-density mixing problem is not inconsistent with Kolmogorov dynamics.
\begin{table*}
{
\begin{tabular}{|l|l|l|l|l|l|l|l|l|}
\hline
parameters & $C_{b1}$&$C_{b2}$&$C_{a1}$&$C_{a2}$&$C_{r1}$&$C_{r2}$&$C_{rp1}$ & $C_{rp2}$ \\
\hline
{\tt R1} & $0.18 \pm 0.06$ & $0.05$ & $0.12$ & $0.06$ &$0.12\pm 0.06$&$0.06$&$1.0$&$1.0 \pm 0.5$\\
{\tt R2} &  $0.18 \pm 0.06 $ & $0.0$ & $0.0$ & $0.0$ &$0.12\pm 0.06$&$0.06$&$0.0$&$1.0\pm 0.5$\\
\hline
\end{tabular}
}
\caption{Table showing optimized values of all coefficients used for comparison with the DNS flow $At=0.05$. Those values without uncertainties quoted correspond to $\chi^2$ minima that asymptotically approach zero.}
\label{table2}
\end{table*}

Given the quality of the minima it is clear that the uncertainty in the coefficient choice may be quite large (between 30 and 50\%) either because of the shallowness of the minima or the independence of the error function to values below a certain threshold. Due to the latter feature, we may be justified in taking the values of $C_{rp1}, C_{rp2}, C_{b2}, C_{a1}$ and $C_{a2}$ to zero. The resulting sparse set of non-zero coefficients is denoted by ${\tt R2}$ in Table \ref{table2}. This set represents an attempt to assess if a mimimal number of coefficients may be extracted to yield a reasonable comparison with DNS. Note that in both rows of values in Table \ref{table2}, those which do not have uncertainties quoted correspond to the coeffiicents with asymptotic minimum $\chi^2$.
 
\begin{figure*}[ht!]
\includegraphics[width=.32\linewidth]{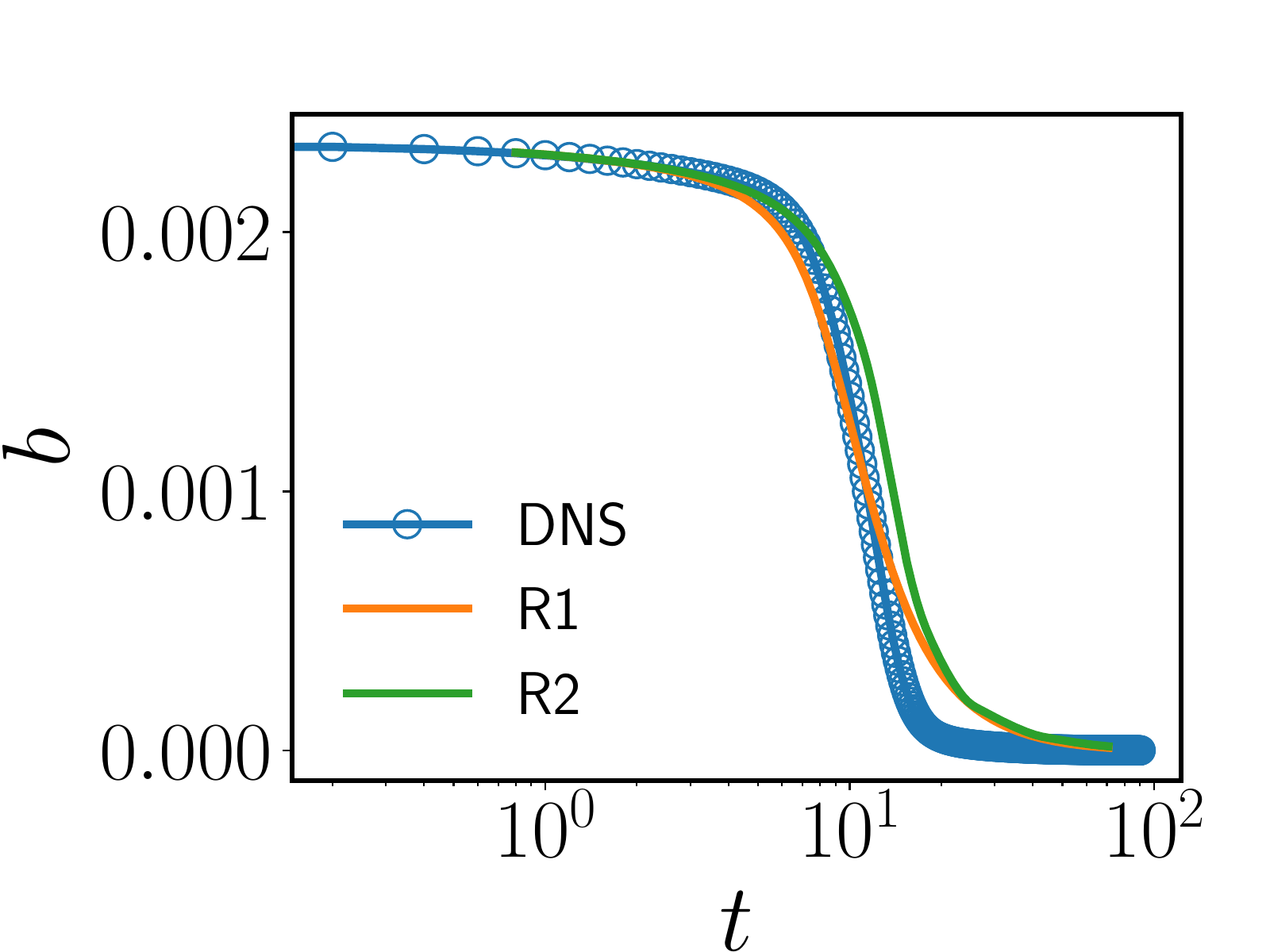}
\put(-25,100){\bf \scriptsize (a)}
\includegraphics[width=.32\linewidth]{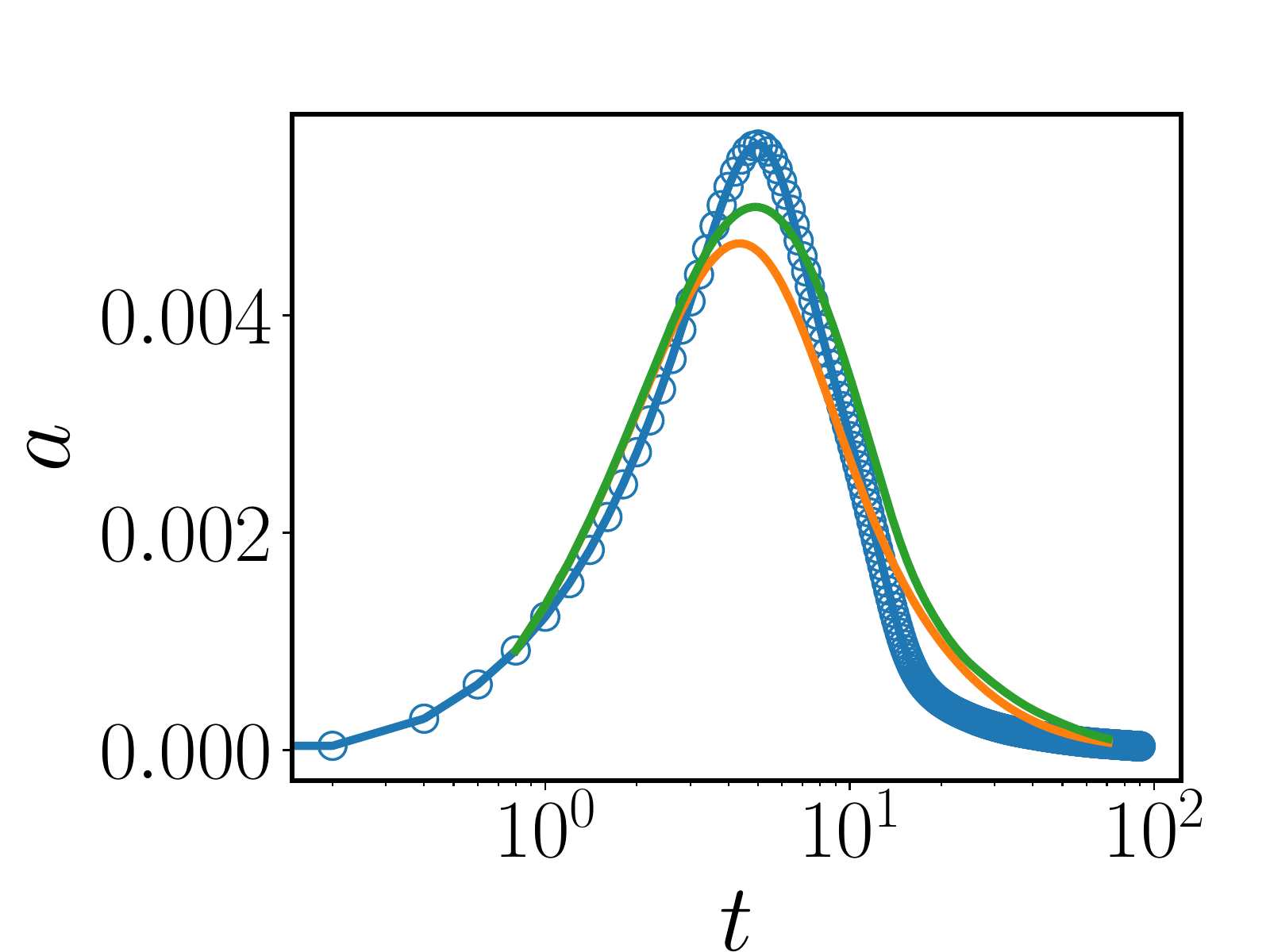}
\put(-25,100){\bf \scriptsize (b)}
\includegraphics[width=.32\linewidth]{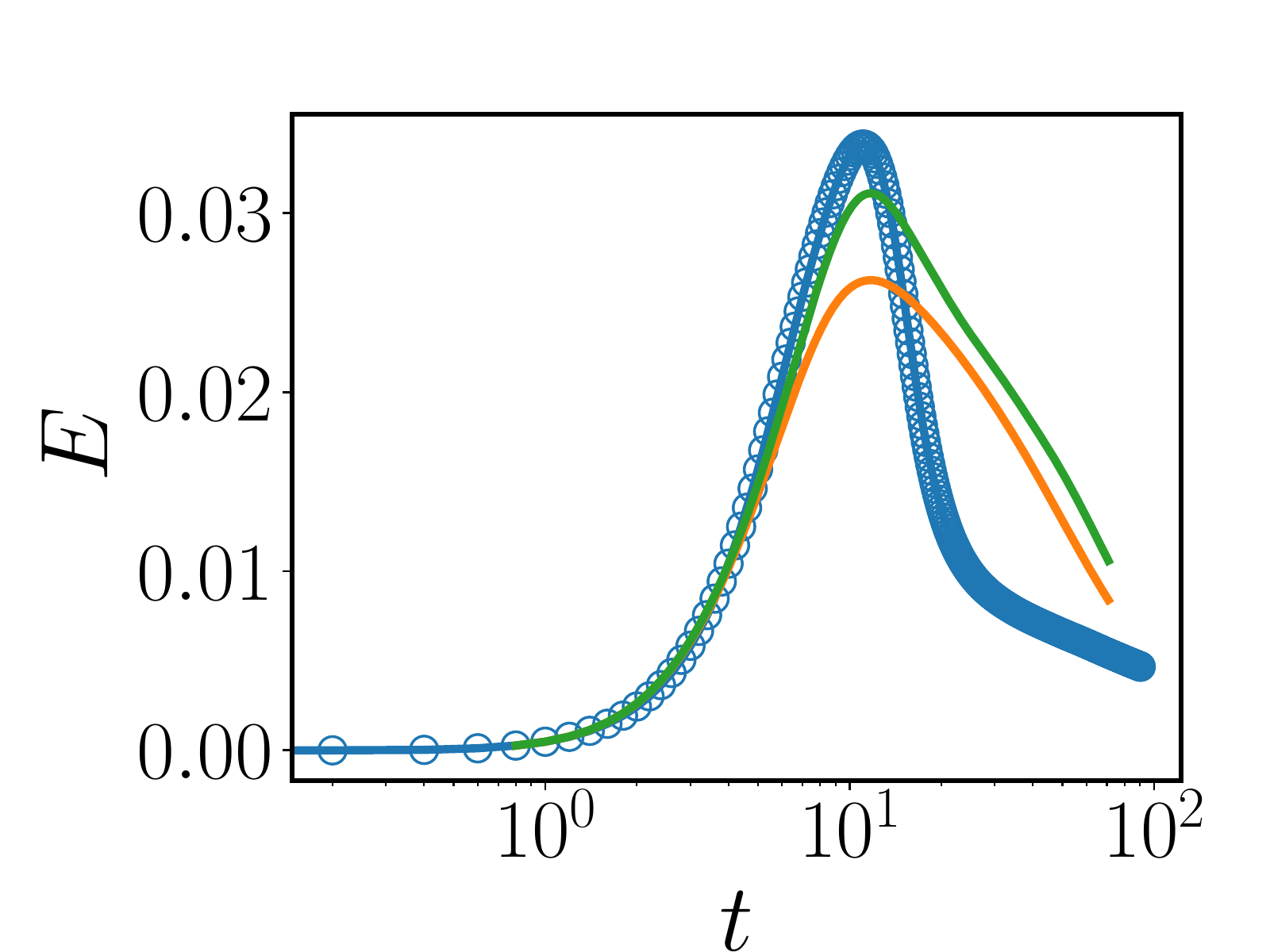}
\put(-25,100){\bf \scriptsize (c)}
\caption{[Color online] Time evolution of flow with initial $At = 0.05$. (a)Mean density-specific volume covariance $b(t)$; (b) mean mass flux $a(t)$ and (c) turbulent kinetic energy $E$ obtained from the results of the DNS calculations (blue line with circles) and from the spectral model ({\tt R1}(orange line) and {\tt R2}(green line)) with parameter values listed in Table \ref{table2}.
DNS runs in this case are at a resolution of $1024^3$ (the viscosity is $10^{-4}$).}
\label{1024_runs}
\end{figure*}

 In Fig.~\ref{1024_runs}(a)--(c) we show the comparison of the model calculations at the optimized parameters {\tt R1} (orange line) with the DNS data (blue line with circles). We observe reasonable agreement with the DNS data in the time evolution of mean $b$ (Fig.~\ref{1024_runs}(a)), the mass flux (see Fig.~\ref{1024_runs}(b)), and the kinetic energy growth stage (see Fig.~\ref{1024_runs}(c)). The magnitude of the peak of the kinetic energy is underestimated by the model, although the timing of the peak is the same as that of the DNS. The decay of the kinetic energy computed by the spectral model is slower than the decay of the kinetic energy in the DNS. Overall the model has captured the global quantities quite well, especially given that our optimization function Eq. (\ref{chisq}) requires no weighting of one quantity over another and is a fairly na\"ive choice. All three primary regimes of the dynamics, namely mix-driven growth of mass-flux followed by conversion of potential energy to turbulent kinetic energy and subsequent decaying dynamics and a fully mixed state are largely captured by the spectral model. 
 
 If we consider the minimal set of coefficients {\tt R2} we find that the comparison with DNS is very similar to that obtained by using the full set. Indeed the peak of the mass flux and energy are both in better agreement for {\tt R2}. The decay regimes are more compromised in {\tt R2} as compared to {\tt R1}. This procedure of minimization of an error function over all metrics is thus a way also to understand dominant processes and eliminate less critical contributions. Our analysis has shown that, of the modeled terms, the downscale transfer of $b$ (governed by $C_{b1}$), the downscale and upscale redistribution of energy (governed by $C_{r1}$ and $C_{r2}$) and the breakup of mass flux scales due to turbulence (governed by $C_{rp2}$) are the dominant spectral processes. The exact terms for drive and dissipation are also important but clearly not sufficient. It is particularly interesting to note that the minimal set of coefficients also seems to imply that the spectral redistribution of $\bm{a}$ is entirely subdominant in the homogeneous variable-density mixing process. 
Our systematic study of the coefficients and a fairly simple optimization procedure has thus revealed useful constraints and properties both of the model as well as of the physical processes under study. These are a significant advantage in turbulence modeling.

We next show the spectral quantities computed by the optimized model for low $At$ in Fig.~\ref{spectra}. The (a) columns shows the initial conditions of $b(k)$, $a(k)$ and $E(k)$ from the DNS in blue, and the approximation used by the model in orange. By necessity, we use a coarse representation of the spectra in the low wavenumbers because of the implementation of the surrogate coordinate $z = \ln(k)$. As time evolves the model spectra show good agreement with the DNS at the peak values (which dominate the integrals) but over-predict the spectra at both small and large $k$ for intermediate time $t = 3.2$ (column (b)). As time evolves further to $t=6.4$, both $a(k)$ and $b(k)$ model calculation show better agreement with DNS at higher $k$, but the high wavenumbers for the energy remain over-predicted. The spectra thus permit a more detailed understanding of the flow dynamics than
 do the global quantities. Indeed we re-iterate that our coefficient variation study demonstrated that spectral information was critical to developing the right time-evolution (in Fig. \ref{comp}(c), for example, very low values of the spectral transfer coefficients result in a substantially slow decay of $E$) particularly on the decay side of the process.

 
\begin{figure*}[ht!]
\includegraphics[width=.25\linewidth]{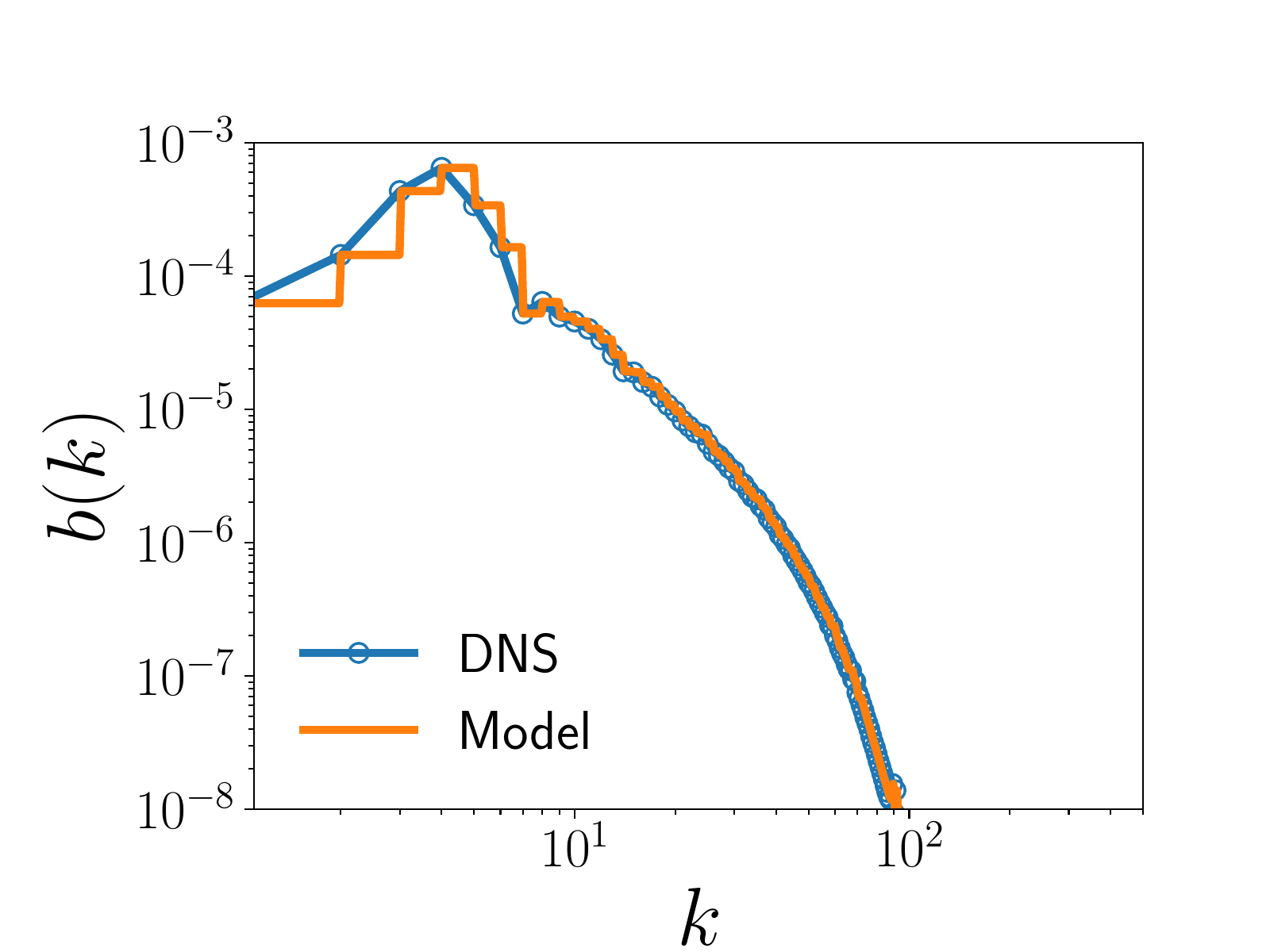}
\put(-28,73){ (a)}
\includegraphics[width=.25\linewidth]{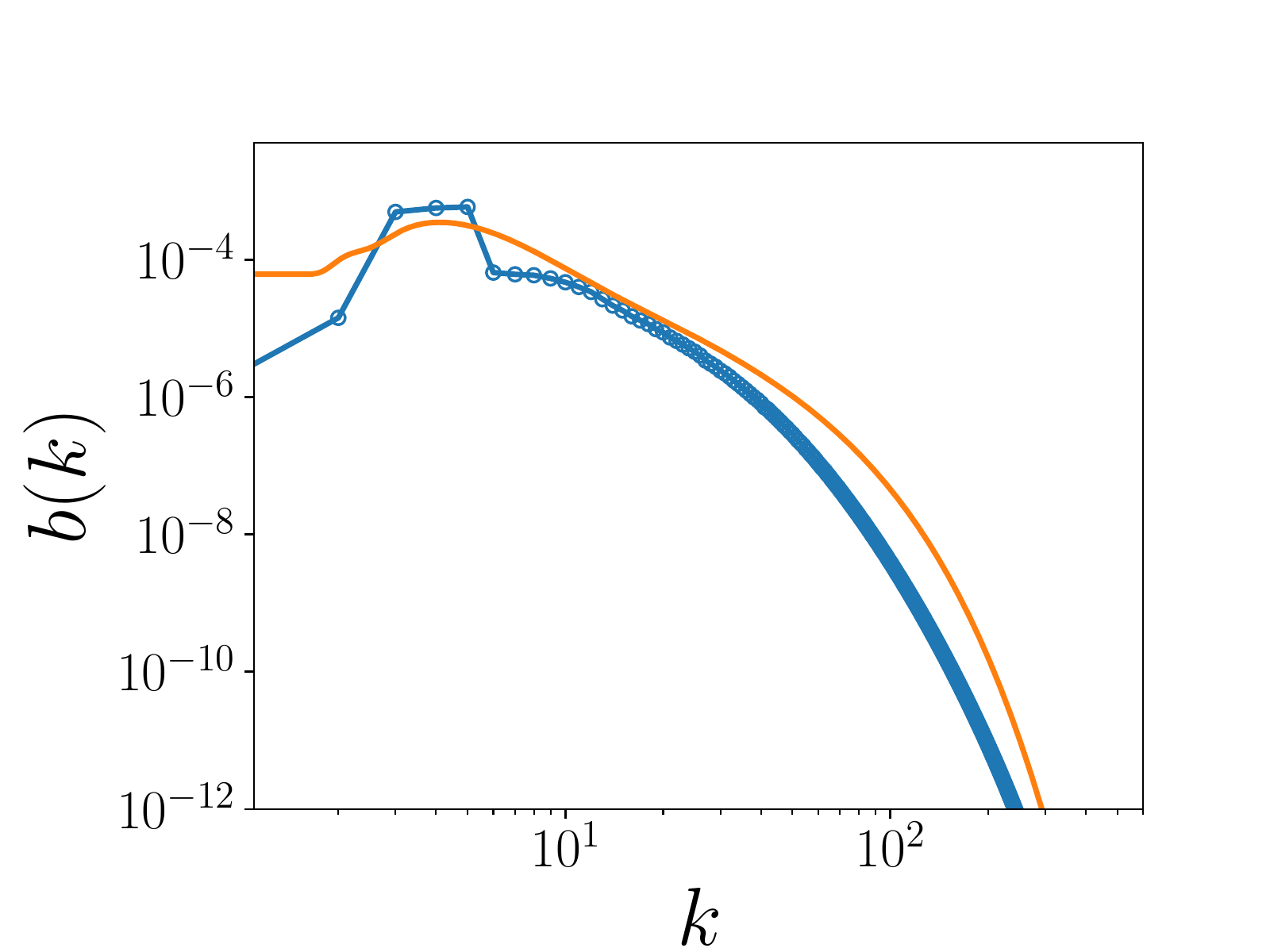}
\put(-28,73){ (b)}
\includegraphics[width=.25\linewidth]{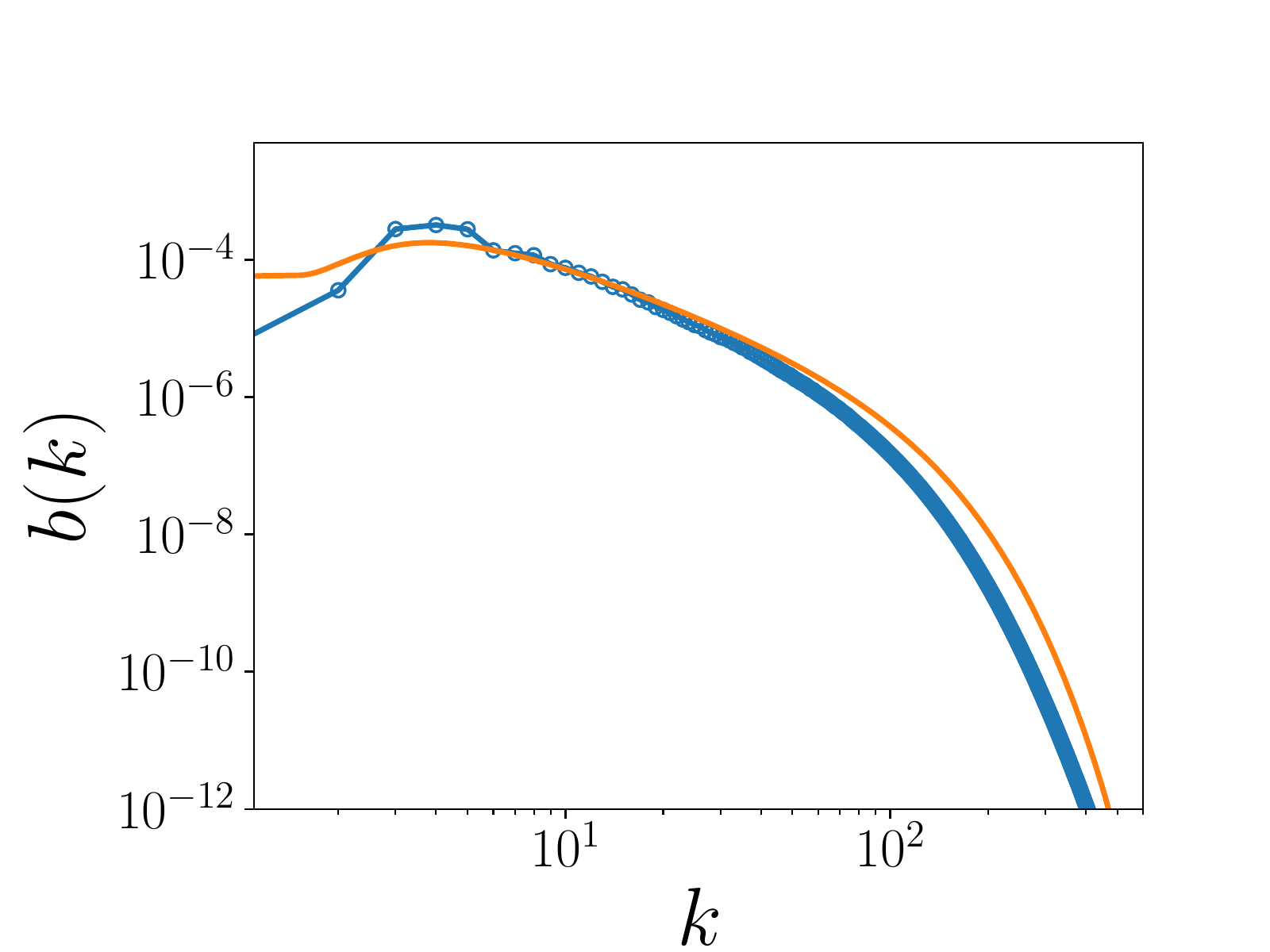}
\put(-28,73){ (c)}
\includegraphics[width=.25\linewidth]{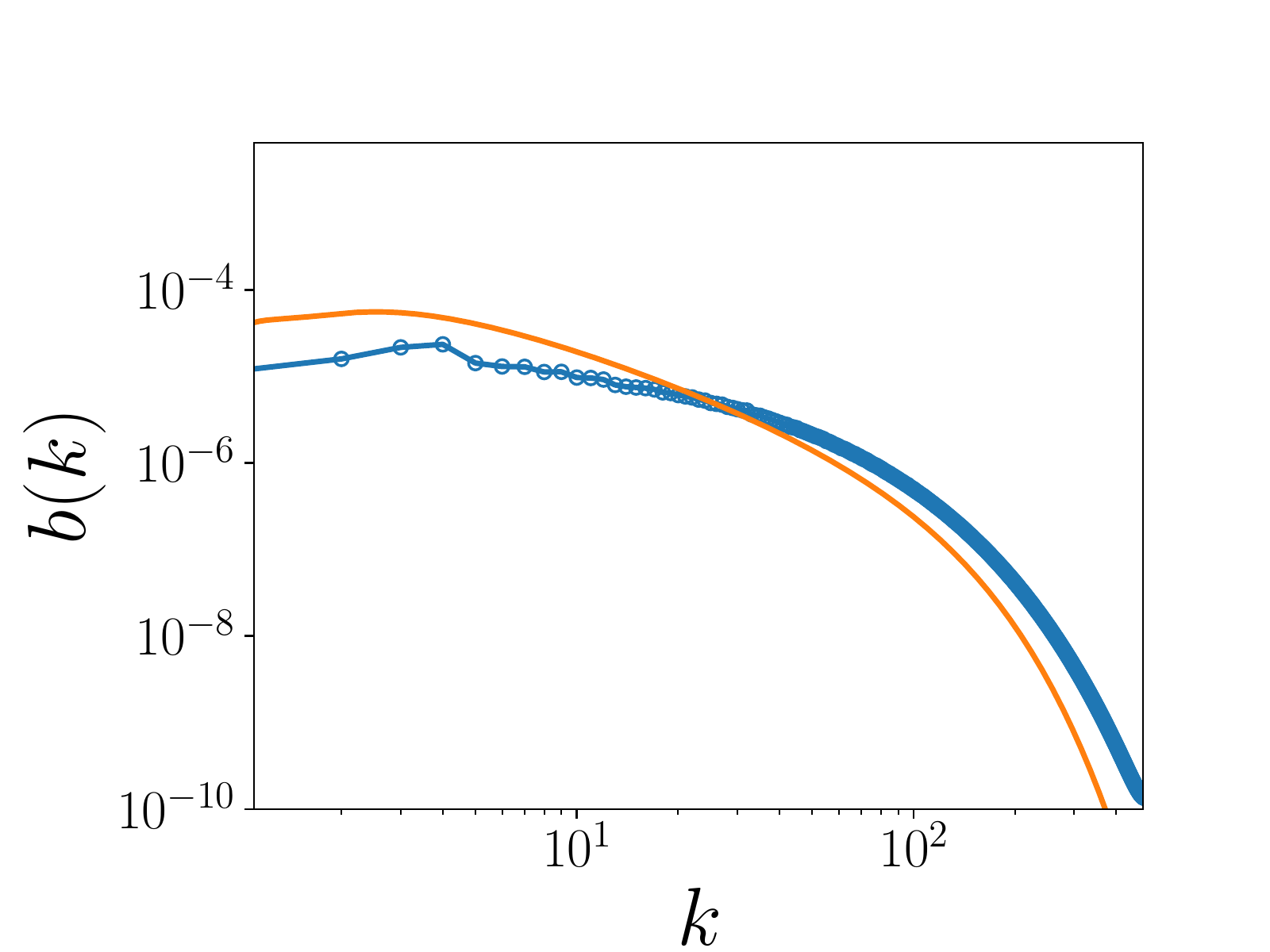}
\put(-28,73){ (d)}
\\
\includegraphics[width=.25\linewidth]{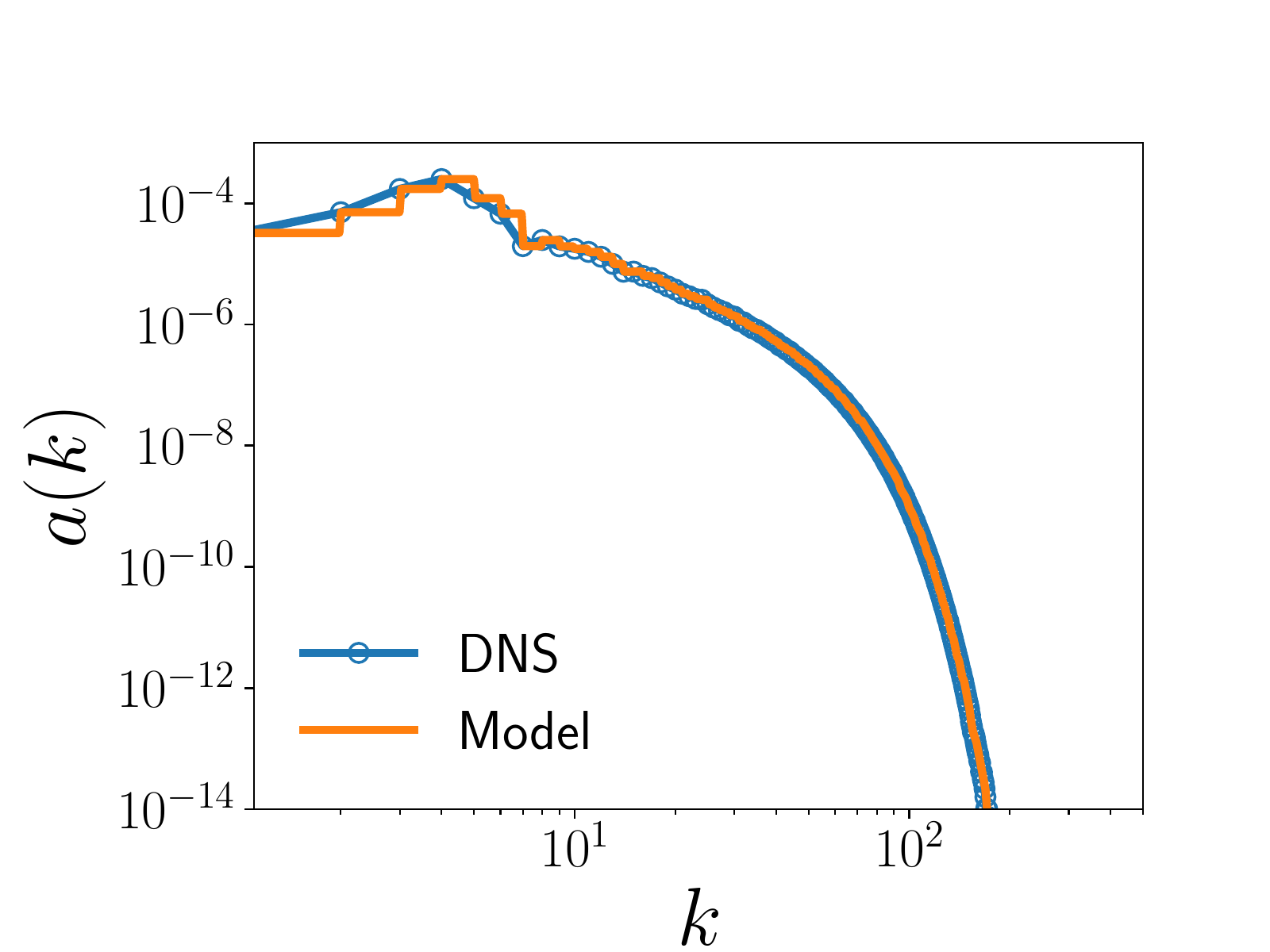}
\put(-28,73){ (e)}
\includegraphics[width=.25\linewidth]{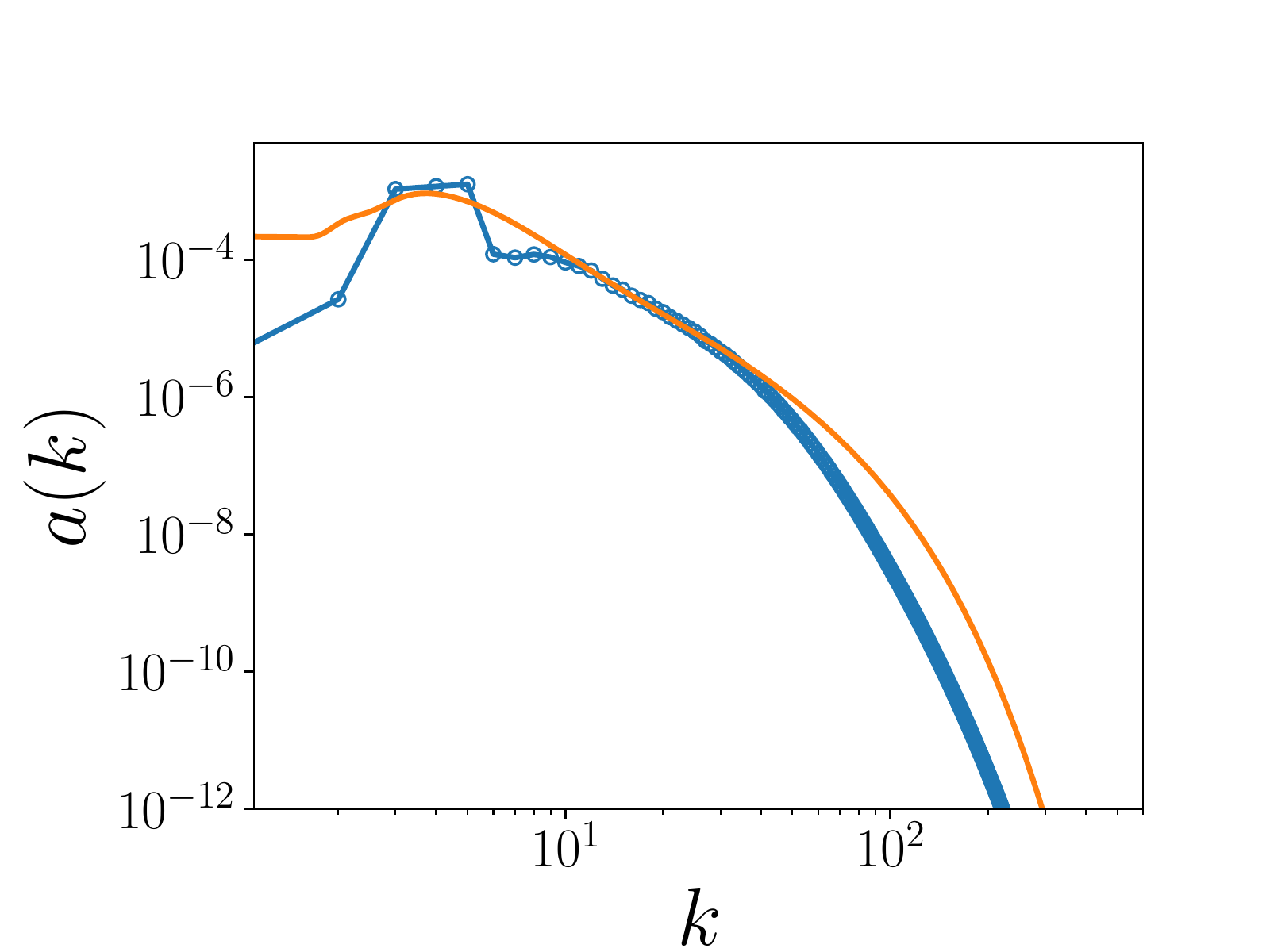}
\put(-28,73){ (f)}
\includegraphics[width=.25\linewidth]{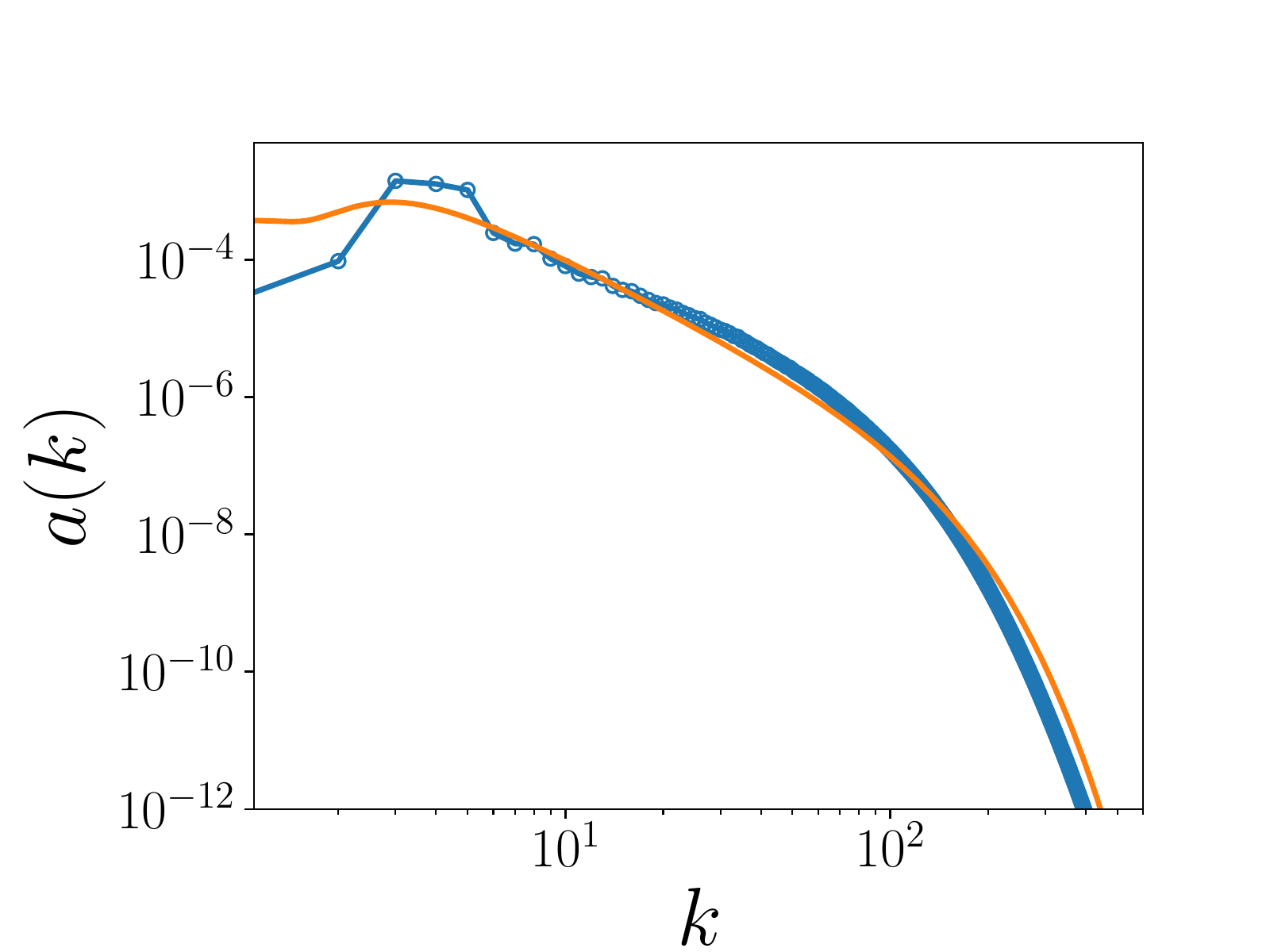}
\put(-28,73){ (g)}
\includegraphics[width=.25\linewidth]{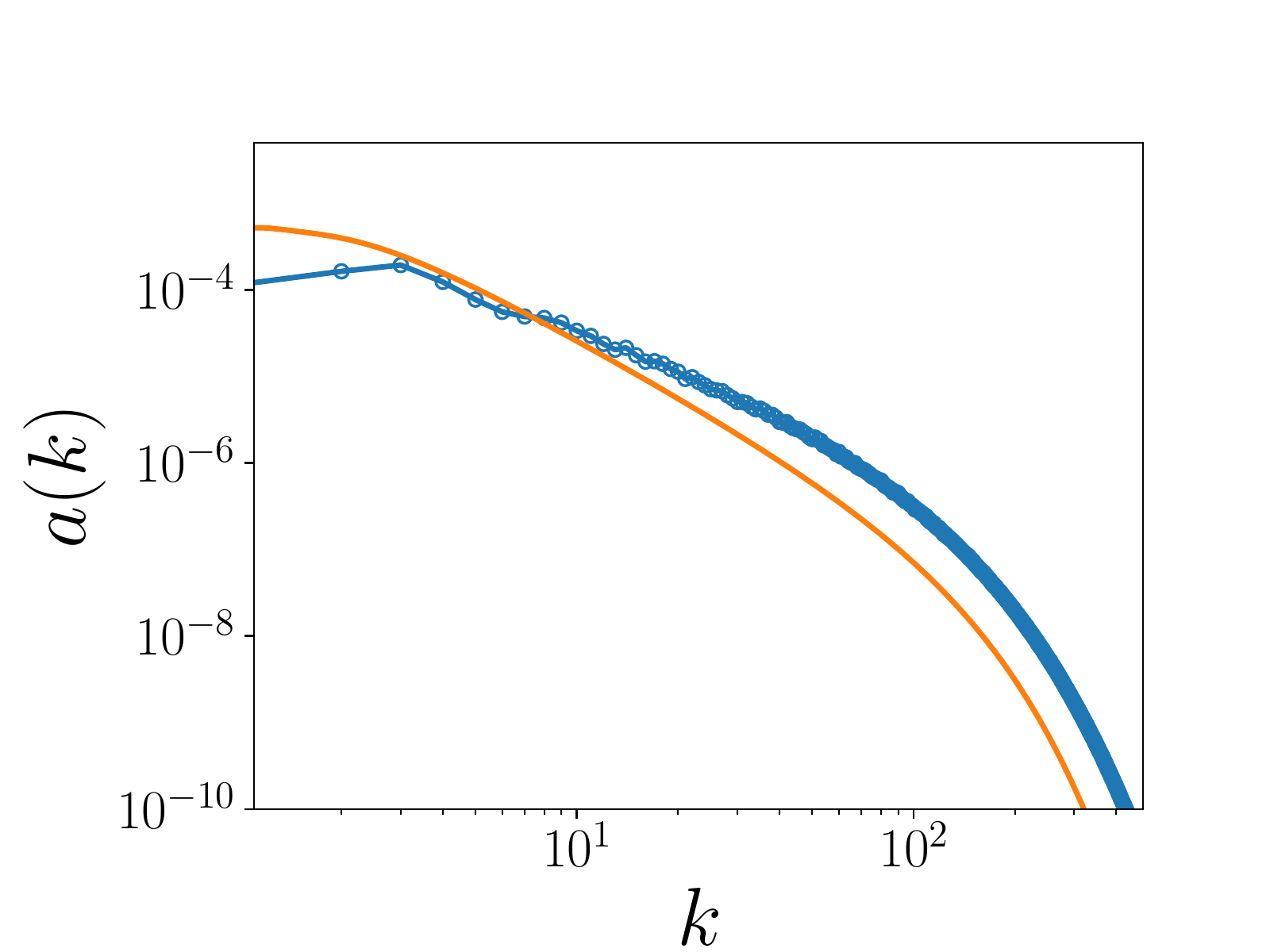}
\put(-28,73){ (h)}
\\
\includegraphics[width=.25\linewidth]{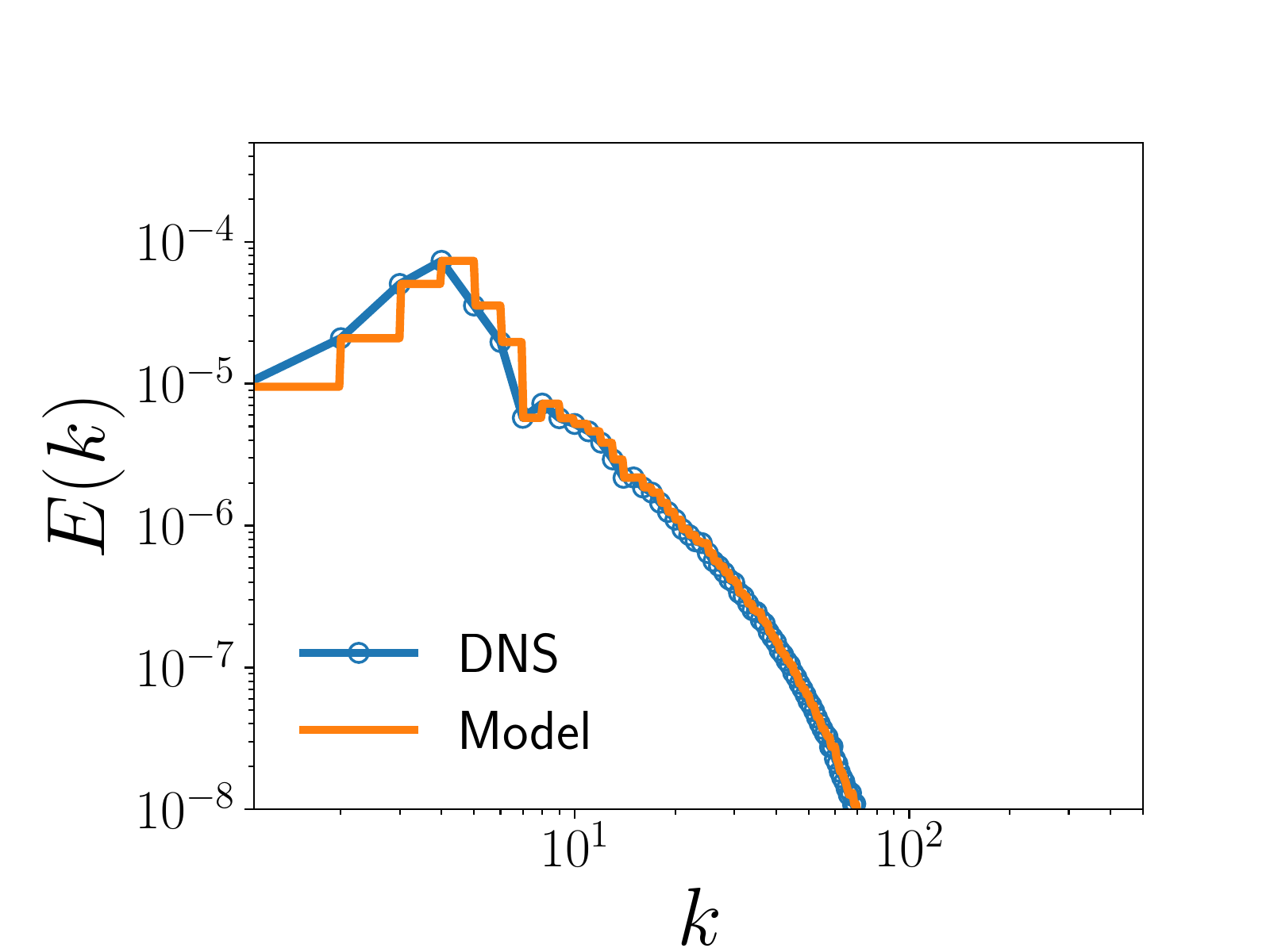}
\put(-28,73){ (i)}
\includegraphics[width=.25\linewidth]{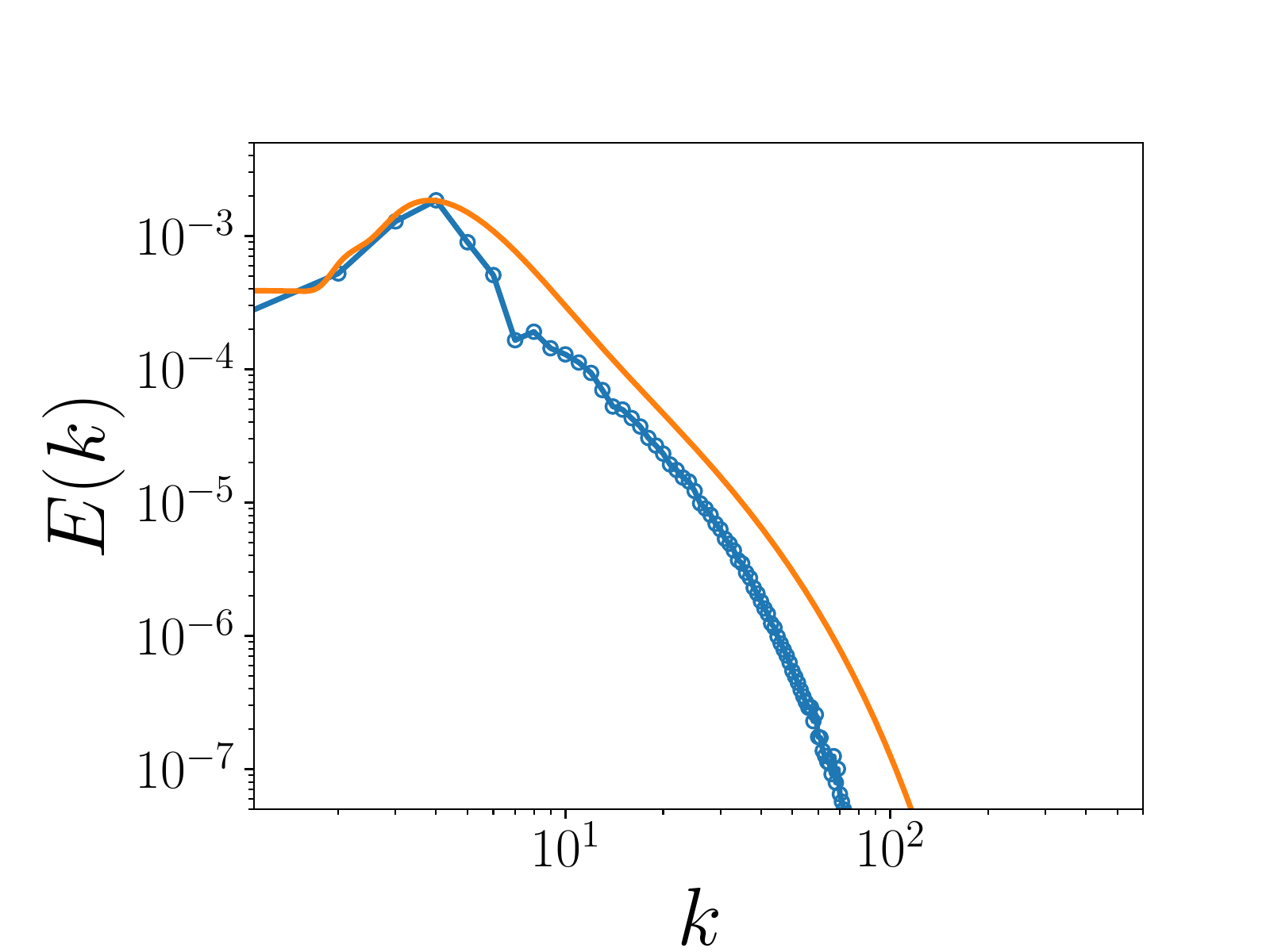}
\put(-28,73){ (j)}
\includegraphics[width=.25\linewidth]{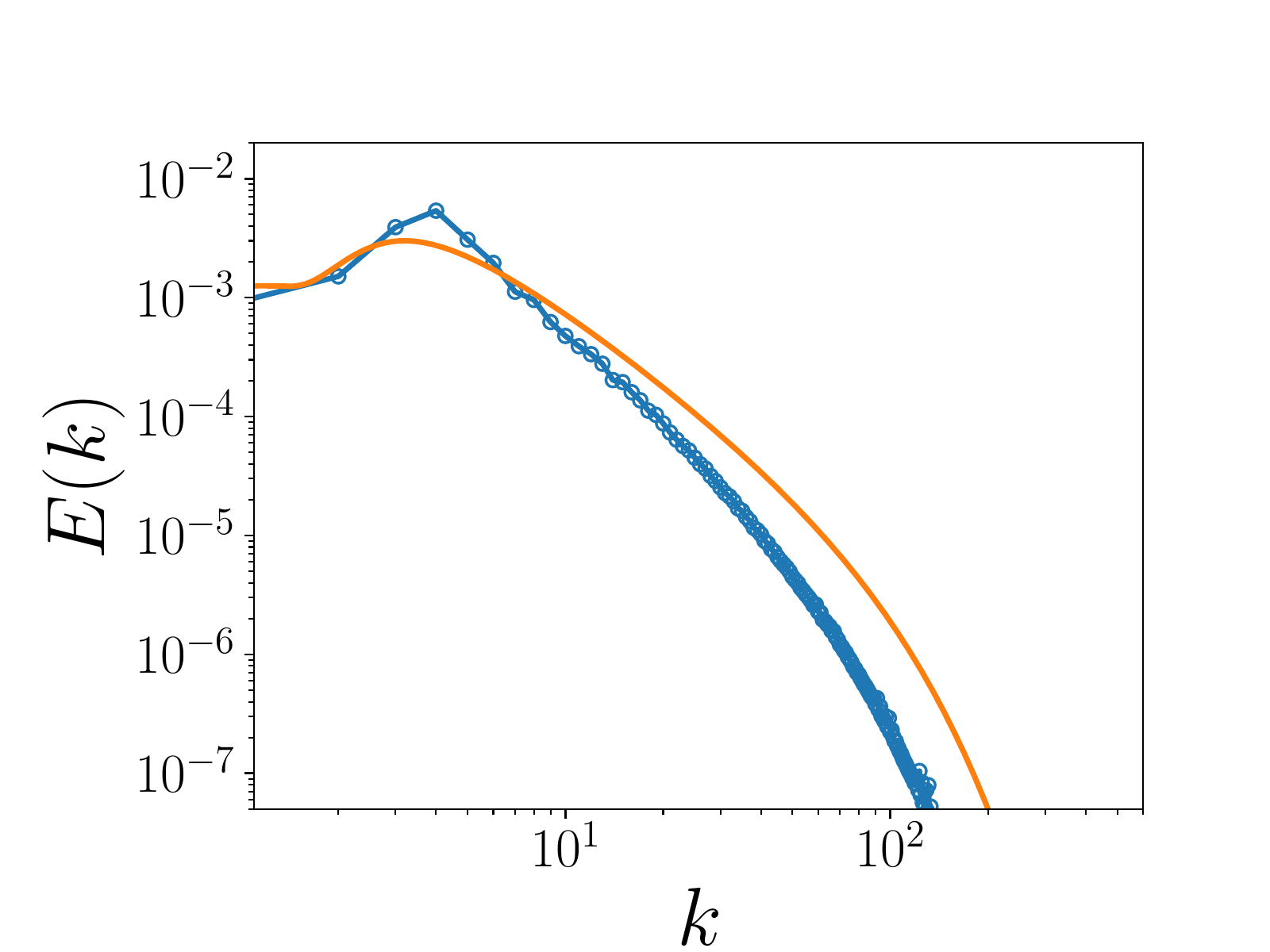}
\put(-28,73){ (k)}
\includegraphics[width=.25\linewidth]{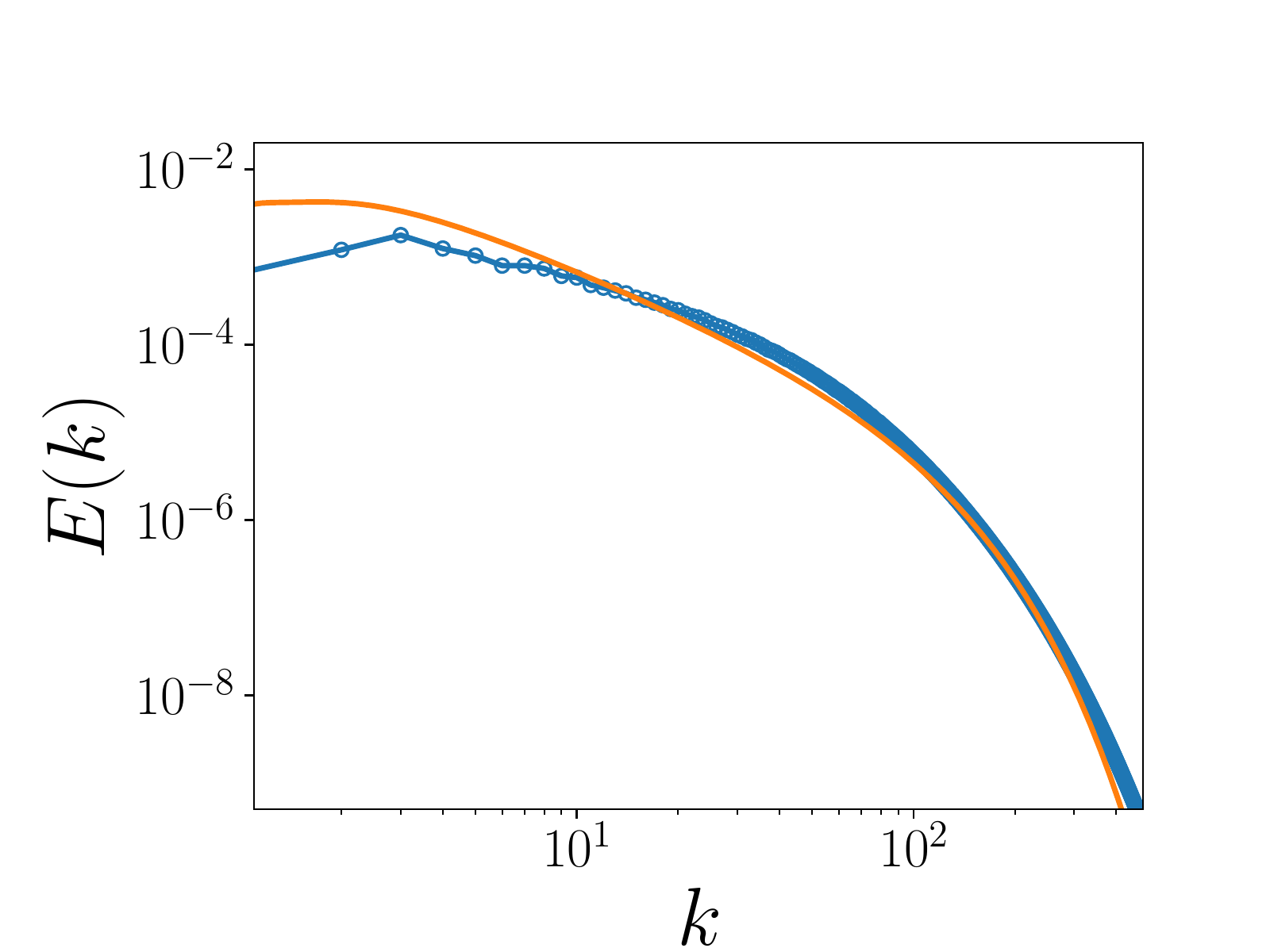}
\put(-28,73){ (l)}
\caption{[Color online] Spectra of flows with initial $At = 0.05$. Plots of $b(k)$ versus the wavenumber $k$ at times (a) $t=0.8$, (b) $t=3.2$, (c) $t=6.4$; (d)$t=14$.  Turbulent mass flux $a(k)$ versus the wavenumber $k$ at times (e) $t=0.8$, (f) $t=3.2$, (g) $t=6.4$, (h) $t=14$. Energy $E(k)$ versus the wavenumber $k$ at times (i) $t=0.8$, (j) $t=3.2$, (k) $t=6.4$, (l) $t=14$. The parameters for these data are given in Table \ref{table2}  with the corresponding DNS data resolution at $1024^3$.}
\label{spectra}
\end{figure*}

\begin{figure*}[ht!]
\includegraphics[width=.35\linewidth]{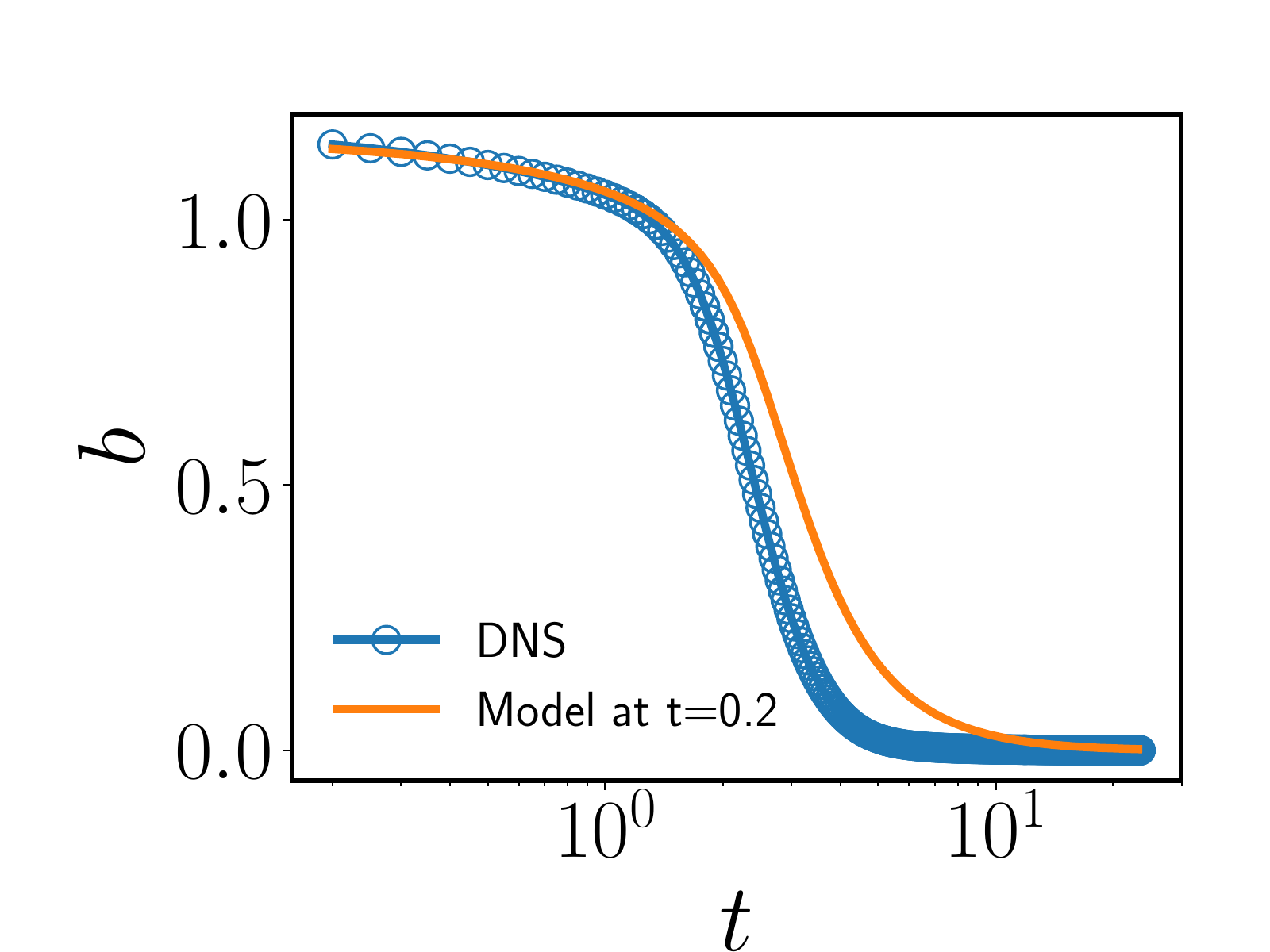}
\put(-30,105){\bf(a)}
\includegraphics[width=.35\linewidth]{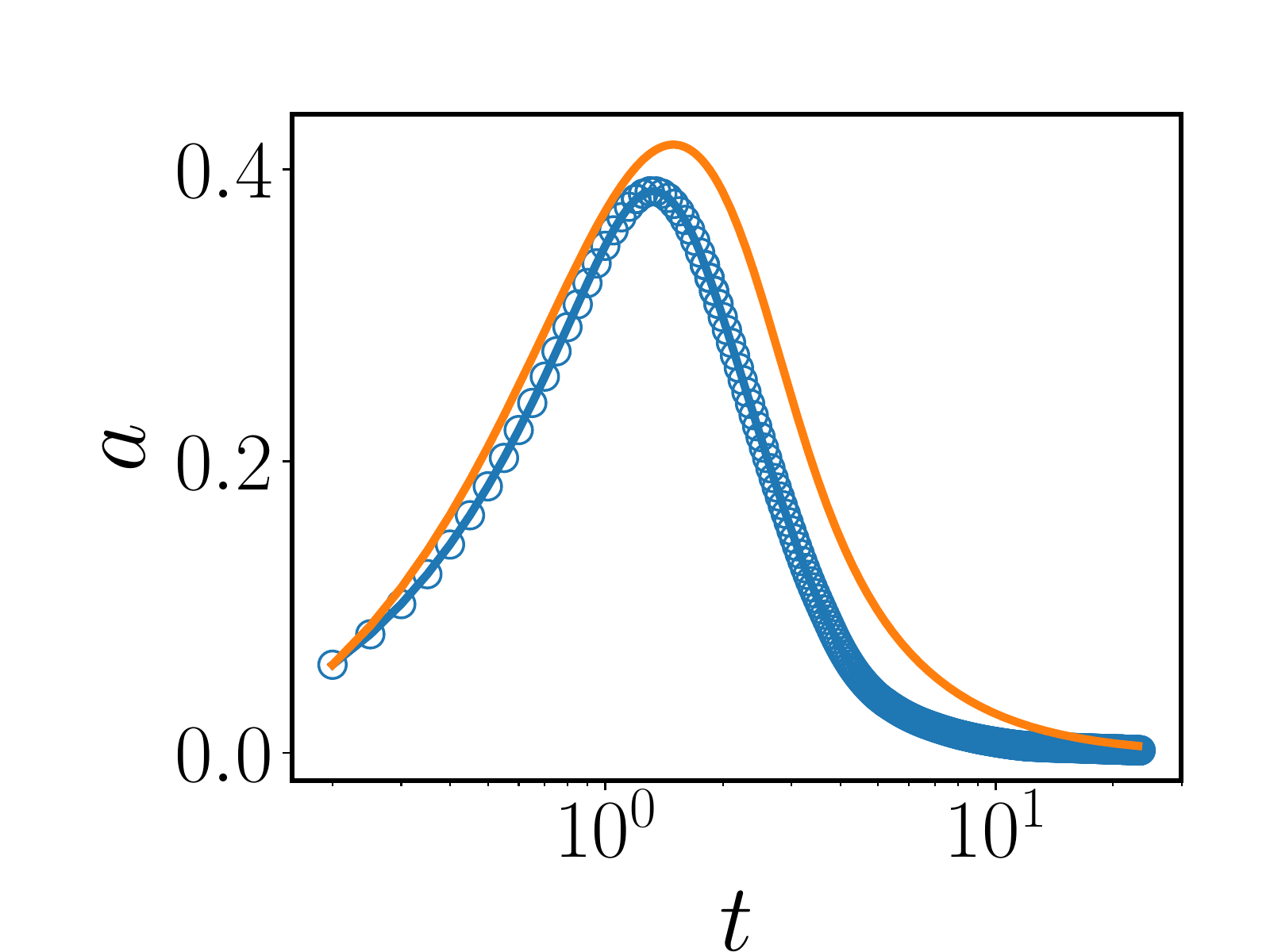}
\put(-30,105){\bf(b)}
\includegraphics[width=.35\linewidth]{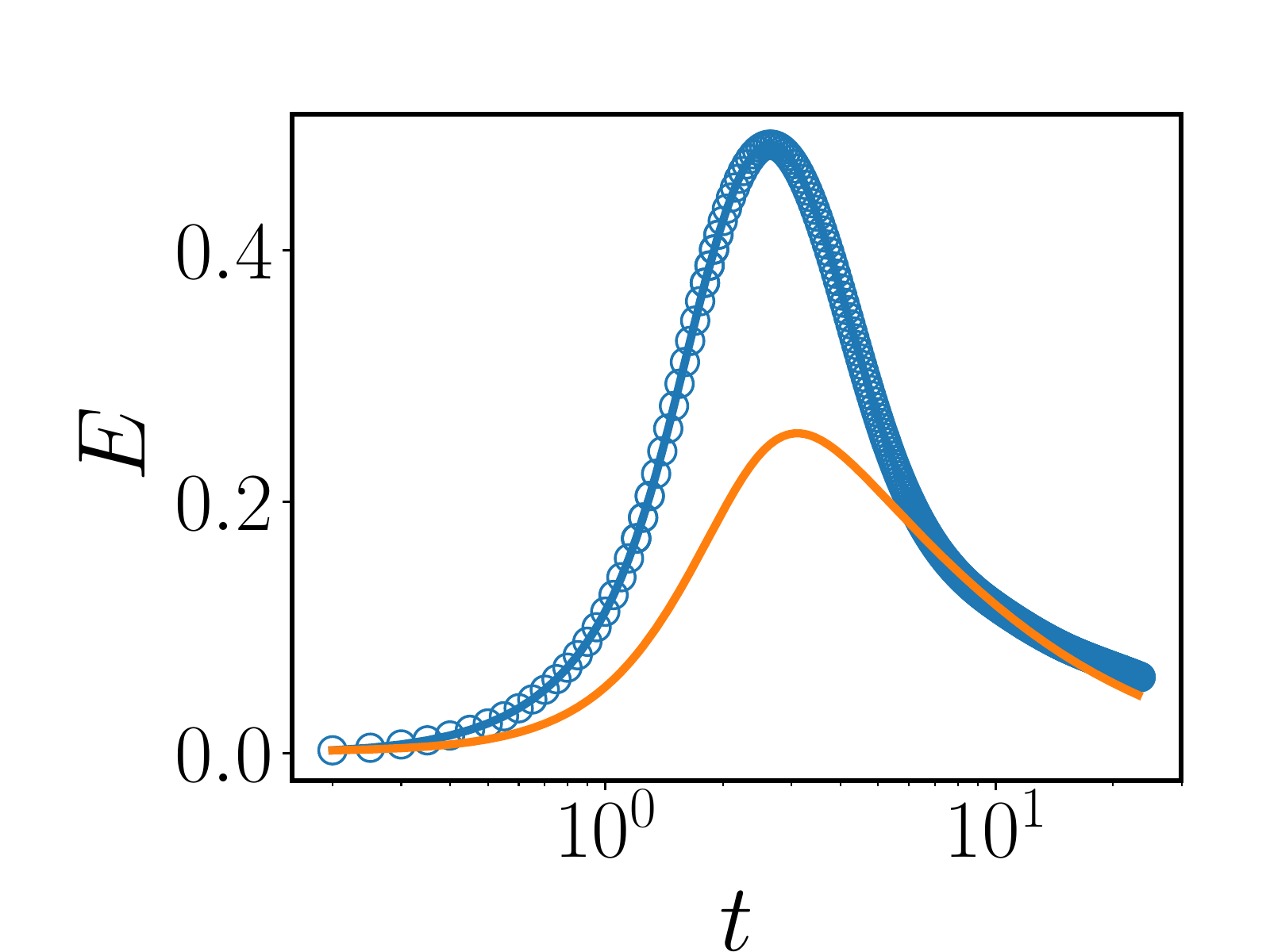}
\put(-30,105){\bf(c)}
\caption{[Color online] 
Time evolution of flow with initial $At = 0.75$.  (a) Mean density-specific volume covariance $b(t)$; (b) mean mass flux $a(t)$ and (c) turbulent kinetic energy $E$ obtained from the results of the DNS calculations (blue line with circles) and from the spectral model code (orange line) with parameter values listed in Table \ref{table2} (run {\tt R1}), and corresponding DNS resolution is $1024^3$ (the viscosity is $10^{-4}$).}
\label{1024_runs_highat}
\end{figure*}

For completeness we expand the discussion to a high Atwood number case $At = 0.75$. Atwood number in the DNS was changed by increasing the density of the heavier fluid. This corresponds, in the model to what we will call a ``density contrast'' since the Atwood number does not explicitly appear in the spectral model. This density contrast is implicit in the larger $\displaystyle \overline \rho = \frac{\rho_{max} + \rho_{min}}{2}$ that is specified in the model equations of motion. 
As a first attempt we use the coefficients optimized in the low Atwood number case shown in Table \ref{table2} {\tt R1}. The results are shown in Fig. \ref{1024_runs_highat}. We see immediately that the quantitive agreement with DNS has degraded somewhat, particularly for the turbulent kinetic energy, compared to the low At case. Thus, the optimization at low At does not hold at high At. Nevertheless, certain qualitative features are still captured quite well including
initial transition in $b$ (Fig.~\ref{1024_runs_highat}(a)) and overall shape of $a$. However, $E$ is  strongly suppressed and $a$ and $b$ are overpredicted in the decay regime. 
It must be noted that further iteration over the coefficients may well fine-tune the outcomes; but such an exercise lies beyond the scope of this paper.

\section{Discussion and Conclusion}
We have benchmarked the two-point spectral model developed by \cite{clark1995two, Steinkamp1999a, Steinkamp1999b} to study variable-density homogeneous turbulence. In summary we can observe some general trends. The main observation is that each dynamical variable is primarily affected by its spectral transfer terms governed by $C_{V1}$ and $C_{V2}$ (where $V$ is any of $b$, $a$ or $E$). The direct effect is to reduce both peak value and delay the peak timing for the respective variable. However, the indirect effect on a coupled variable may be quite different. The most striking example of this is the response of the system to variation of $C_{r1}$ and $C_{r2}$. Recall that the values of $C_{r1}$ and $C_{r2}$ were derived based on the Kolmogorov constant multiplying the spectrum, and equipartition in the invisicid case. Therefore variation away from those values implicitly permits reinterpretation of the cascade processes in the model. In Fig. \ref{comp}(c) the energy is reduced both in magnitude and in the decay rate. However, the indirect effect on $a$ and $b$, via the decrease in frequency $\Theta^{-1}$ and corresponding increase in characteristic turbulence timescale, results in delayed decay of both those quantities. Increasing this pair of variables is the only change that decreases energy while increasing both mass-flux and the mixing parameter magnitude. This is a clear demonstration of how the turbulence timescale as implemented in the model, operates as a governing parameter for the behavior of the dynamical variable.

The drag coefficients $C_{rp1}$ and $C_{rp2}$ for $\bm{a}$ govern a different process than do the spectral transfer terms.  The former have significant effect on $a$ as expected, and in turn on $E$ because of the direct coupling via the pressure gradient. The mixing parameter $b$ on the other hand is relatively insensitive to the drag on $\bm{a}$ for low to moderate values of the drag coefficients. For very large values of those coefficients, the decay rate of $b$ is slowed down. This may understood in light of the fact that the drag on $\bm{a}$ is a tertiary effect on $b$ via the turbulence energy timescale $\Theta^{-1}$. The processes governing the entire cycle of the flow may be completed before the timescale has the chance to grow enough to significantly impact $b$.

In the second part of the study we test the model against DNS data. The purpose is to offer a spectral model for variable density turbulence at high-resolution with coefficients tuned to the problem and based on well-defined error parameters. We choose to optimize the coefficients on the low $At$ case first and found that the model works well for the buoyancy production stage, as well as the decay of the different variables. In optimizing the coefficient we did not favor any one dynamical variable over another; we could envision a more targeted application which, for example, specifies greater fidelity of the turbulent kinetic energy. In that case the fidelity to the mass-flux and $b$ will be compromised and a different set of coefficients deduced. Another important factor in the outcomes of the coefficients is the time over which the functions are optimized, in this case $0 < t \leq t_{max} = 20$. If a different interval were chosen one might do better at isolating say the growth phase, or the decay phase. It is not our intention to provide a fixed set of coefficients but merely to demonstrate that the model can recover a realistic flow with satisfactory agreement to the global quantities and also to the spectral distributions. Even with these caveats, it appears that several of the eight coefficients, and hence the processes corresponding to the terms that they multiply,  become sub-dominant because of the manner in which they have asymptotically minimized errors for small values. The coefficients with true minima are $C_{b1}$, the downscale spectral transfer coefficient of $b$, $C_{r1}$ and $C_{r2}$ both of which govern upscale and downscale spectral redistribution of turbulence kinetic energy, and $C_{rp2}$ which is provides a mechanism for the breakup of fluid parcels in scale due to turbulence. It is important to note that $C_{r1}$ and $C_{r2}$ optimize to the values consistent with Kolmogorov and equipartition theory. This suggests that the cascade process in the variable density mixing problem, at least from the point of view of a second-order spectral model, is consistent with that of the constant density Kolmogorov turbulence problem. 

This emergence of four dominant coefficients leads to the understanding that, apart from the drive terms and the dissipation which are treated exactly, the model expressions for downscale transfer of $b$, the break-up of fluid blobs as they sink under gravity and couple with the turbulence, and the resulting redistribution of $E$ in spectral space are the main mechanisms at play in the homogeneous variable-density mixing problem. At the level of second-order two-point correlations therefore, the model points to and helps elucidate the dominant physical mechanisms at play.

The set of coefficients in Table \ref{table2} obtained from analysis of the low $At$ data, appear to be less suitable for a high Atwood number system At = 0.75, particularly as they relate to the energy. However, the comparison is qualitatively quite good overall considering that we only performed a first order process for determining coefficients in the low At case and the difference in $At$ for the two cases is very large. There is no approximation or assumption in the model development that requires Boussinesq or near-Boussinesq (low Atwood number) conditions. Therefore it is perhaps not surprising that one set of coefficients works quite well over a broad range of $At$.

In summary, the spectral model we have studied is able to recover the statistical and spectral outcomes from non-trivial physical processes in a mixing problem, with minimal tuning of coefficients for two widely different Atwood number flows. The tuning procedure is systematic and may be used to narrow down the space of unknown coefficients, which is always an advantage in predictive modeling. In future work, we will try to develop a more general understanding of flows with different density ratios and finally will address the canonical inhomogeneous Rayleigh-Taylor mixing problem.

\section{Acknowledgments}
NP, SK and TC were funded by the Mix and Burn project under the DOE Advanced Scientific Computing program. Work at LANL was performed under the auspices of the U.S. DOE Contract No. DE-AC52-06NA25396.
The DNS results were obtained using computational
resources provided by the Institutional Computing Program at Los Alamos
National Laboratory and the Argonne Leadership Computing Facility at Argonne National
Laboratory through a 2017 ALCC Award. 

\bibliography{var}

\begin{thebibliography}{37}%
\makeatletter
\providecommand \@ifxundefined [1]{%
 \@ifx{#1\undefined}
}%
\providecommand \@ifnum [1]{%
 \ifnum #1\expandafter \@firstoftwo
 \else \expandafter \@secondoftwo
 \fi
}%
\providecommand \@ifx [1]{%
 \ifx #1\expandafter \@firstoftwo
 \else \expandafter \@secondoftwo
 \fi
}%
\providecommand \natexlab [1]{#1}%
\providecommand \enquote  [1]{``#1''}%
\providecommand \bibnamefont  [1]{#1}%
\providecommand \bibfnamefont [1]{#1}%
\providecommand \citenamefont [1]{#1}%
\providecommand \href@noop [0]{\@secondoftwo}%
\providecommand \href [0]{\begingroup \@sanitize@url \@href}%
\providecommand \@href[1]{\@@startlink{#1}\@@href}%
\providecommand \@@href[1]{\endgroup#1\@@endlink}%
\providecommand \@sanitize@url [0]{\catcode `\\12\catcode `\$12\catcode
  `\&12\catcode `\#12\catcode `\^12\catcode `\_12\catcode `\%12\relax}%
\providecommand \@@startlink[1]{}%
\providecommand \@@endlink[0]{}%
\providecommand \url  [0]{\begingroup\@sanitize@url \@url }%
\providecommand \@url [1]{\endgroup\@href {#1}{\urlprefix }}%
\providecommand \urlprefix  [0]{URL }%
\providecommand \Eprint [0]{\href }%
\providecommand \doibase [0]{http://dx.doi.org/}%
\providecommand \selectlanguage [0]{\@gobble}%
\providecommand \bibinfo  [0]{\@secondoftwo}%
\providecommand \bibfield  [0]{\@secondoftwo}%
\providecommand \translation [1]{[#1]}%
\providecommand \BibitemOpen [0]{}%
\providecommand \bibitemStop [0]{}%
\providecommand \bibitemNoStop [0]{.\EOS\space}%
\providecommand \EOS [0]{\spacefactor3000\relax}%
\providecommand \BibitemShut  [1]{\csname bibitem#1\endcsname}%
\let\auto@bib@innerbib\@empty
\bibitem [{\citenamefont {Townsend}(1980)}]{townsend1980structure}%
  \BibitemOpen
  \bibfield  {author} {\bibinfo {author} {\bibfnamefont {A.~A.}\ \bibnamefont
  {Townsend}},\ }\href@noop {} {\emph {\bibinfo {title} {The structure of
  turbulent shear flow}}}\ (\bibinfo  {publisher} {Cambridge university
  press},\ \bibinfo {year} {1980})\BibitemShut {NoStop}%
\bibitem [{\citenamefont {Wilcox}\ \emph {et~al.}(1993)\citenamefont {Wilcox}
  \emph {et~al.}}]{wilcox}%
  \BibitemOpen
  \bibfield  {author} {\bibinfo {author} {\bibfnamefont {D.~C.}\ \bibnamefont
  {Wilcox}} \emph {et~al.},\ }\href@noop {} {\emph {\bibinfo {title}
  {Turbulence modeling for CFD}}},\ Vol.~\bibinfo {volume} {2}\ (\bibinfo
  {publisher} {DCW industries La Canada, CA},\ \bibinfo {year}
  {1993})\BibitemShut {NoStop}%
\bibitem [{\citenamefont {Hanjali{\'c}}(1994)}]{hanjalic1994advanced}%
  \BibitemOpen
  \bibfield  {author} {\bibinfo {author} {\bibfnamefont {K.}~\bibnamefont
  {Hanjali{\'c}}},\ }\href@noop {} {\bibfield  {journal} {\bibinfo  {journal}
  {International Journal of Heat and Fluid Flow}\ }\textbf {\bibinfo {volume}
  {15}},\ \bibinfo {pages} {178} (\bibinfo {year} {1994})}\BibitemShut
  {NoStop}%
\bibitem [{\citenamefont {Hanjalic}\ \emph {et~al.}(1980)\citenamefont
  {Hanjalic}, \citenamefont {Launder},\ and\ \citenamefont
  {Schiestel}}]{hanjalic1980multiple}%
  \BibitemOpen
  \bibfield  {author} {\bibinfo {author} {\bibfnamefont {K.}~\bibnamefont
  {Hanjalic}}, \bibinfo {author} {\bibfnamefont {B.}~\bibnamefont {Launder}}, \
  and\ \bibinfo {author} {\bibfnamefont {R.}~\bibnamefont {Schiestel}},\
  }\href@noop {} {\bibfield  {journal} {\bibinfo  {journal} {Turbulent Shear
  Flows}\ }\textbf {\bibinfo {volume} {2}},\ \bibinfo {pages} {36} (\bibinfo
  {year} {1980})}\BibitemShut {NoStop}%
\bibitem [{\citenamefont {Pope}(1994)}]{pope1994relationship}%
  \BibitemOpen
  \bibfield  {author} {\bibinfo {author} {\bibfnamefont {S.}~\bibnamefont
  {Pope}},\ }\href@noop {} {\bibfield  {journal} {\bibinfo  {journal} {Physics
  of Fluids}\ }\textbf {\bibinfo {volume} {6}},\ \bibinfo {pages} {973}
  (\bibinfo {year} {1994})}\BibitemShut {NoStop}%
\bibitem [{\citenamefont {Cambon}(1979)}]{cambon1979modelisation}%
  \BibitemOpen
  \bibfield  {author} {\bibinfo {author} {\bibfnamefont {C.}~\bibnamefont
  {Cambon}},\ }\emph {\bibinfo {title} {Mod{\'e}lisation spectrale en
  turbulence homogene anisotrope}},\ \href@noop {} {Ph.D. thesis} (\bibinfo
  {year} {1979})\BibitemShut {NoStop}%
\bibitem [{\citenamefont {Cambon}\ \emph {et~al.}(1981)\citenamefont {Cambon},
  \citenamefont {Jeandel},\ and\ \citenamefont {Mathieu}}]{cambon1981spectral}%
  \BibitemOpen
  \bibfield  {author} {\bibinfo {author} {\bibfnamefont {C.}~\bibnamefont
  {Cambon}}, \bibinfo {author} {\bibfnamefont {D.}~\bibnamefont {Jeandel}}, \
  and\ \bibinfo {author} {\bibfnamefont {J.}~\bibnamefont {Mathieu}},\
  }\href@noop {} {\bibfield  {journal} {\bibinfo  {journal} {Journal of Fluid
  Mechanics}\ }\textbf {\bibinfo {volume} {104}},\ \bibinfo {pages} {247}
  (\bibinfo {year} {1981})}\BibitemShut {NoStop}%
\bibitem [{\citenamefont {Godeferd}\ and\ \citenamefont
  {Cambon}(1994)}]{godeferd1994detailed}%
  \BibitemOpen
  \bibfield  {author} {\bibinfo {author} {\bibfnamefont {F.~S.}\ \bibnamefont
  {Godeferd}}\ and\ \bibinfo {author} {\bibfnamefont {C.}~\bibnamefont
  {Cambon}},\ }\href@noop {} {\bibfield  {journal} {\bibinfo  {journal}
  {Physics of Fluids}\ }\textbf {\bibinfo {volume} {6}},\ \bibinfo {pages}
  {2084} (\bibinfo {year} {1994})}\BibitemShut {NoStop}%
\bibitem [{\citenamefont {Bertoglio}\ and\ \citenamefont
  {Jeandel}(1987)}]{bertoglio1987simplified}%
  \BibitemOpen
  \bibfield  {author} {\bibinfo {author} {\bibfnamefont {J.-P.}\ \bibnamefont
  {Bertoglio}}\ and\ \bibinfo {author} {\bibfnamefont {D.}~\bibnamefont
  {Jeandel}},\ }in\ \href@noop {} {\emph {\bibinfo {booktitle} {Turbulent Shear
  Flows 5}}}\ (\bibinfo  {publisher} {Springer},\ \bibinfo {year} {1987})\ pp.\
  \bibinfo {pages} {19--30}\BibitemShut {NoStop}%
\bibitem [{\citenamefont {Gerashchenko}\ and\ \citenamefont
  {Prestridge}(2015)}]{Gerashchenko2015}%
  \BibitemOpen
  \bibfield  {author} {\bibinfo {author} {\bibfnamefont {S.}~\bibnamefont
  {Gerashchenko}}\ and\ \bibinfo {author} {\bibfnamefont {K.}~\bibnamefont
  {Prestridge}},\ }\href@noop {} {\bibfield  {journal} {\bibinfo  {journal}
  {Journal of Turbulence}\ }\textbf {\bibinfo {volume} {16}},\ \bibinfo {pages}
  {1011} (\bibinfo {year} {2015})}\BibitemShut {NoStop}%
\bibitem [{\citenamefont {Akula}\ and\ \citenamefont
  {Ranjan}(2016)}]{Ranjan2016}%
  \BibitemOpen
  \bibfield  {author} {\bibinfo {author} {\bibfnamefont {B.}~\bibnamefont
  {Akula}}\ and\ \bibinfo {author} {\bibfnamefont {D.}~\bibnamefont {Ranjan}},\
  }\href@noop {} {\bibfield  {journal} {\bibinfo  {journal} {Journal of Fluid
  Mechanics}\ }\textbf {\bibinfo {volume} {795}},\ \bibinfo {pages} {313}
  (\bibinfo {year} {2016})}\BibitemShut {NoStop}%
\bibitem [{\citenamefont {Charonko}\ and\ \citenamefont
  {Prestridge}(2017)}]{Charonko2017}%
  \BibitemOpen
  \bibfield  {author} {\bibinfo {author} {\bibfnamefont {J.~J.}\ \bibnamefont
  {Charonko}}\ and\ \bibinfo {author} {\bibfnamefont {K.}~\bibnamefont
  {Prestridge}},\ }\href@noop {} {\bibfield  {journal} {\bibinfo  {journal}
  {Journal of Fluid Mechanics}\ }\textbf {\bibinfo {volume} {825}},\ \bibinfo
  {pages} {887} (\bibinfo {year} {2017})}\BibitemShut {NoStop}%
\bibitem [{\citenamefont {Cabot}\ and\ \citenamefont
  {Cook}(2006)}]{cabot2006reynolds}%
  \BibitemOpen
  \bibfield  {author} {\bibinfo {author} {\bibfnamefont {W.~H.}\ \bibnamefont
  {Cabot}}\ and\ \bibinfo {author} {\bibfnamefont {A.~W.}\ \bibnamefont
  {Cook}},\ }\href@noop {} {\bibfield  {journal} {\bibinfo  {journal} {Nature
  Physics}\ }\textbf {\bibinfo {volume} {2}},\ \bibinfo {pages} {562} (\bibinfo
  {year} {2006})}\BibitemShut {NoStop}%
\bibitem [{\citenamefont {Livescu}\ and\ \citenamefont
  {Ristorcelli}(2008)}]{livescu2008}%
  \BibitemOpen
  \bibfield  {author} {\bibinfo {author} {\bibfnamefont {D.}~\bibnamefont
  {Livescu}}\ and\ \bibinfo {author} {\bibfnamefont {J.~R.}\ \bibnamefont
  {Ristorcelli}},\ }\href {\doibase 10.1017/S0022112008001481} {\bibfield
  {journal} {\bibinfo  {journal} {Journal of Fluid Mechanics}\ }\textbf
  {\bibinfo {volume} {605}},\ \bibinfo {pages} {145–180} (\bibinfo {year}
  {2008})}\BibitemShut {NoStop}%
\bibitem [{\citenamefont {Besnard}\ and\ \citenamefont
  {Harlow}(1985)}]{besnard1985turbulence}%
  \BibitemOpen
  \bibfield  {author} {\bibinfo {author} {\bibfnamefont {D.}~\bibnamefont
  {Besnard}}\ and\ \bibinfo {author} {\bibfnamefont {F.~H.}\ \bibnamefont
  {Harlow}},\ }\href@noop {} {\bibfield  {journal} {\bibinfo  {journal} {NASA
  STI/Recon Technical Report N}\ }\textbf {\bibinfo {volume} {86}} (\bibinfo
  {year} {1985})}\BibitemShut {NoStop}%
\bibitem [{\citenamefont {Banerjee}\ \emph {et~al.}(2010)\citenamefont
  {Banerjee}, \citenamefont {Gore},\ and\ \citenamefont
  {Andrews}}]{banerjee2010development}%
  \BibitemOpen
  \bibfield  {author} {\bibinfo {author} {\bibfnamefont {A.}~\bibnamefont
  {Banerjee}}, \bibinfo {author} {\bibfnamefont {R.~A.}\ \bibnamefont {Gore}},
  \ and\ \bibinfo {author} {\bibfnamefont {M.~J.}\ \bibnamefont {Andrews}},\
  }\href@noop {} {\bibfield  {journal} {\bibinfo  {journal} {Physical Review
  E}\ }\textbf {\bibinfo {volume} {82}},\ \bibinfo {pages} {046309} (\bibinfo
  {year} {2010})}\BibitemShut {NoStop}%
\bibitem [{\citenamefont {Schwarzkopf}\ \emph {et~al.}(2016)\citenamefont
  {Schwarzkopf}, \citenamefont {Livescu}, \citenamefont {Baltzer},
  \citenamefont {Gore},\ and\ \citenamefont {Ristorcelli}}]{Schwarzkopf2016}%
  \BibitemOpen
  \bibfield  {author} {\bibinfo {author} {\bibfnamefont {J.~D.}\ \bibnamefont
  {Schwarzkopf}}, \bibinfo {author} {\bibfnamefont {D.}~\bibnamefont
  {Livescu}}, \bibinfo {author} {\bibfnamefont {J.~R.}\ \bibnamefont
  {Baltzer}}, \bibinfo {author} {\bibfnamefont {R.~A.}\ \bibnamefont {Gore}}, \
  and\ \bibinfo {author} {\bibfnamefont {J.}~\bibnamefont {Ristorcelli}},\
  }\href@noop {} {\bibfield  {journal} {\bibinfo  {journal} {Flow, Turbulence
  and Combustion}\ }\textbf {\bibinfo {volume} {96}},\ \bibinfo {pages} {1}
  (\bibinfo {year} {2016})}\BibitemShut {NoStop}%
\bibitem [{\citenamefont {Elghobashi}\ and\ \citenamefont
  {Abou-Arab}(1983)}]{elghobashi1983two}%
  \BibitemOpen
  \bibfield  {author} {\bibinfo {author} {\bibfnamefont {S.}~\bibnamefont
  {Elghobashi}}\ and\ \bibinfo {author} {\bibfnamefont {T.}~\bibnamefont
  {Abou-Arab}},\ }\href@noop {} {\bibfield  {journal} {\bibinfo  {journal} {The
  Physics of Fluids}\ }\textbf {\bibinfo {volume} {26}},\ \bibinfo {pages}
  {931} (\bibinfo {year} {1983})}\BibitemShut {NoStop}%
\bibitem [{\citenamefont {Clark}(1999)}]{clark1999two}%
  \BibitemOpen
  \bibfield  {author} {\bibinfo {author} {\bibfnamefont {T.~T.}\ \bibnamefont
  {Clark}},\ }in\ \href@noop {} {\emph {\bibinfo {booktitle} {Modeling complex
  turbulent flows}}}\ (\bibinfo  {publisher} {Springer},\ \bibinfo {year}
  {1999})\ pp.\ \bibinfo {pages} {183--202}\BibitemShut {NoStop}%
\bibitem [{\citenamefont {Besnard}\ \emph {et~al.}(1996)\citenamefont
  {Besnard}, \citenamefont {Harlow}, \citenamefont {Rauenzahn},\ and\
  \citenamefont {Zemach}}]{besnard96}%
  \BibitemOpen
  \bibfield  {author} {\bibinfo {author} {\bibfnamefont {D.}~\bibnamefont
  {Besnard}}, \bibinfo {author} {\bibfnamefont {F.}~\bibnamefont {Harlow}},
  \bibinfo {author} {\bibfnamefont {R.}~\bibnamefont {Rauenzahn}}, \ and\
  \bibinfo {author} {\bibfnamefont {C.}~\bibnamefont {Zemach}},\ }\href@noop {}
  {\bibfield  {journal} {\bibinfo  {journal} {Theoretical and computational
  fluid dynamics}\ }\textbf {\bibinfo {volume} {8}},\ \bibinfo {pages} {1}
  (\bibinfo {year} {1996})}\BibitemShut {NoStop}%
\bibitem [{\citenamefont {Steinkamp}\ \emph
  {et~al.}(1999{\natexlab{a}})\citenamefont {Steinkamp}, \citenamefont
  {Clark},\ and\ \citenamefont {Harlow}}]{Steinkamp1999a}%
  \BibitemOpen
  \bibfield  {author} {\bibinfo {author} {\bibfnamefont {M.}~\bibnamefont
  {Steinkamp}}, \bibinfo {author} {\bibfnamefont {T.}~\bibnamefont {Clark}}, \
  and\ \bibinfo {author} {\bibfnamefont {F.}~\bibnamefont {Harlow}},\
  }\href@noop {} {\bibfield  {journal} {\bibinfo  {journal} {International
  journal of multiphase flow}\ }\textbf {\bibinfo {volume} {25}},\ \bibinfo
  {pages} {599} (\bibinfo {year} {1999}{\natexlab{a}})}\BibitemShut {NoStop}%
\bibitem [{\citenamefont {Steinkamp}\ \emph
  {et~al.}(1999{\natexlab{b}})\citenamefont {Steinkamp}, \citenamefont
  {Clark},\ and\ \citenamefont {Harlow}}]{Steinkamp1999b}%
  \BibitemOpen
  \bibfield  {author} {\bibinfo {author} {\bibfnamefont {M.}~\bibnamefont
  {Steinkamp}}, \bibinfo {author} {\bibfnamefont {T.}~\bibnamefont {Clark}}, \
  and\ \bibinfo {author} {\bibfnamefont {F.}~\bibnamefont {Harlow}},\
  }\href@noop {} {\bibfield  {journal} {\bibinfo  {journal} {International
  journal of multiphase flow}\ }\textbf {\bibinfo {volume} {25}},\ \bibinfo
  {pages} {639} (\bibinfo {year} {1999}{\natexlab{b}})}\BibitemShut {NoStop}%
\bibitem [{\citenamefont {Schwarzkopf}\ \emph {et~al.}(2011)\citenamefont
  {Schwarzkopf}, \citenamefont {Livescu}, \citenamefont {Gore}, \citenamefont
  {Rauenzahn},\ and\ \citenamefont {Ristorcelli}}]{schwarzkopf2011application}%
  \BibitemOpen
  \bibfield  {author} {\bibinfo {author} {\bibfnamefont {J.~D.}\ \bibnamefont
  {Schwarzkopf}}, \bibinfo {author} {\bibfnamefont {D.}~\bibnamefont
  {Livescu}}, \bibinfo {author} {\bibfnamefont {R.~A.}\ \bibnamefont {Gore}},
  \bibinfo {author} {\bibfnamefont {R.~M.}\ \bibnamefont {Rauenzahn}}, \ and\
  \bibinfo {author} {\bibfnamefont {J.~R.}\ \bibnamefont {Ristorcelli}},\
  }\href@noop {} {\bibfield  {journal} {\bibinfo  {journal} {Journal of
  Turbulence}\ ,\ \bibinfo {pages} {N49}} (\bibinfo {year} {2011})}\BibitemShut
  {NoStop}%
\bibitem [{\citenamefont {Dupuy}\ \emph {et~al.}(2018)\citenamefont {Dupuy},
  \citenamefont {Toutant},\ and\ \citenamefont
  {Bataille}}]{dupuy2018equations}%
  \BibitemOpen
  \bibfield  {author} {\bibinfo {author} {\bibfnamefont {D.}~\bibnamefont
  {Dupuy}}, \bibinfo {author} {\bibfnamefont {A.}~\bibnamefont {Toutant}}, \
  and\ \bibinfo {author} {\bibfnamefont {F.}~\bibnamefont {Bataille}},\
  }\href@noop {} {\bibfield  {journal} {\bibinfo  {journal} {Physics Letters
  A}\ }\textbf {\bibinfo {volume} {382}},\ \bibinfo {pages} {327} (\bibinfo
  {year} {2018})}\BibitemShut {NoStop}%
\bibitem [{\citenamefont {Youngs}(1992)}]{youngs1992two}%
  \BibitemOpen
  \bibfield  {author} {\bibinfo {author} {\bibfnamefont {D.}~\bibnamefont
  {Youngs}},\ }\href@noop {} {\bibfield  {journal} {\bibinfo  {journal} {The
  3rd Zababahin Scientific Talks, Kishtim, USSR}\ } (\bibinfo {year}
  {1992})}\BibitemShut {NoStop}%
\bibitem [{\citenamefont {Clark}\ and\ \citenamefont
  {Spitz}(1995)}]{clark1995two}%
  \BibitemOpen
  \bibfield  {author} {\bibinfo {author} {\bibfnamefont {T.}~\bibnamefont
  {Clark}}\ and\ \bibinfo {author} {\bibfnamefont {P.}~\bibnamefont {Spitz}},\
  }\href@noop {} {\emph {\bibinfo {title} {Two-point correlation equations for
  variable density turbulence}}},\ \bibinfo {type} {Tech. Rep.}\ (\bibinfo
  {institution} {Los Alamos National Lab., NM (United States)},\ \bibinfo
  {year} {1995})\BibitemShut {NoStop}%
\bibitem [{\citenamefont {Livescu}\ and\ \citenamefont
  {Ristorcelli}(2007)}]{livescu2007}%
  \BibitemOpen
  \bibfield  {author} {\bibinfo {author} {\bibfnamefont {D.}~\bibnamefont
  {Livescu}}\ and\ \bibinfo {author} {\bibfnamefont {J.~R.}\ \bibnamefont
  {Ristorcelli}},\ }\href {\doibase 10.1017/S0022112007008270} {\bibfield
  {journal} {\bibinfo  {journal} {Journal of Fluid Mechanics}\ }\textbf
  {\bibinfo {volume} {591}},\ \bibinfo {pages} {43–71} (\bibinfo {year}
  {2007})}\BibitemShut {NoStop}%
\bibitem [{\citenamefont {Leith}(1967)}]{leith67}%
  \BibitemOpen
  \bibfield  {author} {\bibinfo {author} {\bibfnamefont {C.}~\bibnamefont
  {Leith}},\ }\href@noop {} {\bibfield  {journal} {\bibinfo  {journal} {The
  Physics of Fluids}\ }\textbf {\bibinfo {volume} {10}},\ \bibinfo {pages}
  {1409} (\bibinfo {year} {1967})}\BibitemShut {NoStop}%
\bibitem [{\citenamefont {Girimaji}(1992)}]{girimaji1992modeling}%
  \BibitemOpen
  \bibfield  {author} {\bibinfo {author} {\bibfnamefont {S.~S.}\ \bibnamefont
  {Girimaji}},\ }\href@noop {} {\bibfield  {journal} {\bibinfo  {journal}
  {Physics of Fluids A: Fluid Dynamics}\ }\textbf {\bibinfo {volume} {4}},\
  \bibinfo {pages} {2529} (\bibinfo {year} {1992})}\BibitemShut {NoStop}%
\bibitem [{\citenamefont {Anderson}\ \emph {et~al.}(1984)\citenamefont
  {Anderson}, \citenamefont {Tannehill},\ and\ \citenamefont
  {Pletcher}}]{anderson1984computational}%
  \BibitemOpen
  \bibfield  {author} {\bibinfo {author} {\bibfnamefont {D.~A.}\ \bibnamefont
  {Anderson}}, \bibinfo {author} {\bibfnamefont {J.}~\bibnamefont {Tannehill}},
  \ and\ \bibinfo {author} {\bibfnamefont {R.}~\bibnamefont {Pletcher}},\
  }\href@noop {} {\bibfield  {journal} {\bibinfo  {journal} {Washington, USA}\
  } (\bibinfo {year} {1984})}\BibitemShut {NoStop}%
\bibitem [{\citenamefont {Crank}\ and\ \citenamefont
  {Nicolson}(1947)}]{crank1947practical}%
  \BibitemOpen
  \bibfield  {author} {\bibinfo {author} {\bibfnamefont {J.}~\bibnamefont
  {Crank}}\ and\ \bibinfo {author} {\bibfnamefont {P.}~\bibnamefont
  {Nicolson}},\ }in\ \href@noop {} {\emph {\bibinfo {booktitle} {Mathematical
  Proceedings of the Cambridge Philosophical Society}}},\ Vol.~\bibinfo
  {volume} {43}\ (\bibinfo {organization} {Cambridge University Press},\
  \bibinfo {year} {1947})\ pp.\ \bibinfo {pages} {50--67}\BibitemShut {NoStop}%
\bibitem [{\citenamefont {Aslangil}\ \emph {et~al.}(2018)\citenamefont
  {Aslangil}, \citenamefont {Livescu},\ and\ \citenamefont
  {Banerjee}}]{aslangil2018}%
  \BibitemOpen
  \bibfield  {author} {\bibinfo {author} {\bibfnamefont {D.}~\bibnamefont
  {Aslangil}}, \bibinfo {author} {\bibfnamefont {D.}~\bibnamefont {Livescu}}, \
  and\ \bibinfo {author} {\bibfnamefont {A.}~\bibnamefont {Banerjee}},\
  }\href@noop {} {\bibfield  {journal} {\bibinfo  {journal} {under review J.
  Fluid Mech.}\ } (\bibinfo {year} {2018})}\BibitemShut {NoStop}%
\bibitem [{\citenamefont {Lee}(1952)}]{lee1952some}%
  \BibitemOpen
  \bibfield  {author} {\bibinfo {author} {\bibfnamefont {T.}~\bibnamefont
  {Lee}},\ }\href@noop {} {\bibfield  {journal} {\bibinfo  {journal} {Quarterly
  of Applied Mathematics}\ }\textbf {\bibinfo {volume} {10}},\ \bibinfo {pages}
  {69} (\bibinfo {year} {1952})}\BibitemShut {NoStop}%
\bibitem [{\citenamefont {Livescu}\ \emph {et~al.}(2009)\citenamefont
  {Livescu}, \citenamefont {Mohd-Yusof}, \citenamefont {Petersen},\ and\
  \citenamefont {Grove}}]{livescu2009cfdns}%
  \BibitemOpen
  \bibfield  {author} {\bibinfo {author} {\bibfnamefont {D.}~\bibnamefont
  {Livescu}}, \bibinfo {author} {\bibfnamefont {J.}~\bibnamefont {Mohd-Yusof}},
  \bibinfo {author} {\bibfnamefont {M.}~\bibnamefont {Petersen}}, \ and\
  \bibinfo {author} {\bibfnamefont {J.}~\bibnamefont {Grove}},\ }\href@noop {}
  {\bibfield  {journal} {\bibinfo  {journal} {Los Alamos National Laboratory
  Technical Report No. LA-CC-09-100}\ } (\bibinfo {year} {2009})}\BibitemShut
  {NoStop}%
\bibitem [{\citenamefont {Livescu}(2013)}]{livescu2013numerical}%
  \BibitemOpen
  \bibfield  {author} {\bibinfo {author} {\bibfnamefont {D.}~\bibnamefont
  {Livescu}},\ }\href@noop {} {\bibfield  {journal} {\bibinfo  {journal} {Phil.
  Trans. R. Soc. A}\ }\textbf {\bibinfo {volume} {371}},\ \bibinfo {pages}
  {20120185} (\bibinfo {year} {2013})}\BibitemShut {NoStop}%
\bibitem [{\citenamefont {Clark}\ \emph {et~al.}(1997)\citenamefont {Clark},
  \citenamefont {Chen}, \citenamefont {Turner},\ and\ \citenamefont
  {Zemach}}]{Clarkthesis}%
  \BibitemOpen
  \bibfield  {author} {\bibinfo {author} {\bibfnamefont {T.}~\bibnamefont
  {Clark}}, \bibinfo {author} {\bibfnamefont {S.-Y.}\ \bibnamefont {Chen}},
  \bibinfo {author} {\bibfnamefont {L.}~\bibnamefont {Turner}}, \ and\ \bibinfo
  {author} {\bibfnamefont {C.}~\bibnamefont {Zemach}},\ }\href@noop {} {\emph
  {\bibinfo {title} {Turbulence and turbulence spectra in complex fluid
  flows}}},\ \bibinfo {type} {Tech. Rep.}\ (\bibinfo  {institution} {Los Alamos
  National Lab., NM (United States)},\ \bibinfo {year} {1997})\BibitemShut
  {NoStop}%
\bibitem [{\citenamefont {Pearson}(1956)}]{pearson1956karl}%
  \BibitemOpen
  \bibfield  {author} {\bibinfo {author} {\bibfnamefont {K.}~\bibnamefont
  {Pearson}},\ }\href@noop {} {\emph {\bibinfo {title} {Karl Pearson's early
  statistical papers}}}\ (\bibinfo  {publisher} {University Press},\ \bibinfo
  {year} {1956})\BibitemShut {NoStop}%
\end{thebibliography}%
\end{document}